\documentclass[%
 aip,
 amsmath,amssymb,
preprint,%
]{revtex4-1}
\usepackage{graphicx}
\usepackage{dcolumn}
\usepackage{bm}
\usepackage{caption,subcaption}
\begin{document}

\preprint{AIP/123-QED}

\title{Performance evaluation of high-order reconstruction for discrete unified gas-kinetics scheme in tracking fluid interfaces}

\author{Zeren Yang}
\email[Yang, Z. R.: ]{zeren@mail.nwpu.edu.cn}
\author{Chengwen Zhong}
\email[Corresponding author: Zhong, C. W., ]{zhongcw@nwpu.edu.cn}
\author{Congshan Zhuo}%
\email[Zhuo, C. S.: ]{zhuocs@nwpu.edu.cn}
\affiliation{National Key Laboratory of Science and Technology on Aerodynamic Design and Research, Northwestern Polytechnical University, Xi'an, Shaanxi 710072, China.}
\date{\today}
\begin{abstract}
With a noticeable increase in research centered on modeling micro fluid interfaces in the framework of mesoscopic methods, we conduct an exhaustive study of discrete unified gas-kinetics scheme (DUGKS) in handling complicated interface deformations. High-order isotropic finite-difference schemes are first utilized in DUGKS to improve its capability in tracking interfaces. The performance of third-stage third-order DUGKS where source term is incorporated has also been assessed for the first time and a series of numerical tests have been conducted to investigate their capability. The comparative analysis have revealed the reason why the performance of lattice Boltzmann method is superior to that of discrete velocity method and DUGKS in general condition from an informed perspective. The mechanism behind the performance distinction between the central scheme and upwind scheme utilized in meso-flux construction in DUGKS have also been clarified. Numerical results have shown that the employment of high-order schemes in DUGKS does have an effect on the reduction of numerical dissipation, but the overall accuracy of this method is limited by the precision of prediction of source terms on mesh interface. The capability of third-stage third-order DUGKS is severely inhibited by its intrinsic limitation of the ratio of time step to particle collision time. Among the various kinds of DUGKS employed with different reconstruction methods, the most promising scheme is the one with third-order isotropic reconstruction and upwind-based meso-flux evaluation, which is able to ensure an unique balance between efficiency and accuracy.
\end{abstract}
\maketitle
\clearpage
\section{\label{sec:introduction}INTRODUCTION}
Multi-phase flow at micro scales have drawn the focus of numerous researchers due to its specific physical mechanisms as well as wide applications in engineering and technology \cite{JMReese2003,Axel2006,Martin2012,Li2015}. One of the fundamental issues on modeling multi-phase flow is the description of complex interfacial behavior. Over the past decades, several efficient methods have been proposed for conveniently depicting the evolution of interfaces \cite{HIRT1981JCP,OSHER1988JCP,Rudman1998,SUSSMAN1998CF,Yuan2018POF}, among which the diffusive interface approach has attracted considerable attention owing to its distinctive feature that the interface is captured implicitly and irregular topology changes are handled naturally without any special procedures \cite{Anderson1998ARFM,ZHENG2006JCP,Zhang2016POF}. Usually different phases in the fluid domain is identified by a continuous variable (or order parameter) and the physical properties are distributed smoothly across the interface. The order parameter used to describe different phases is generally governed by Cahn-Hilliard (C-H) equation \cite{Cahn1958} or Allen-Cahn (A-C) equation \cite{ALLEN1976}. The analytical solutions for those partial differential equations are difficult to derive, hence plenty of numerical methods, including finite-difference method \cite{JACQMIN1999JCP,KIM2014IJES,ZHAI2015IJHMT}, finite-volume method \cite{DING2007JCP,CAI2016CT}, finite-element method \cite{ZHANG2010JCP}, and spectral method \cite{Yue2004JFM,LIU2003PD}, have been applied to solve the A-C equation or C-H equation.
\\
With the rapid development of kinetic schemes, more and more interfacial dynamic problems have been investigated by this type of method \cite{Swift1995PRL,He1999JCP,Chen2009POF,Biferale2010POF,Gan2015SM}. Compared to the traditional methods, the kinetic schemes are capable to model complex multi-phase flows at the mesoscopic level, which fills the gap between the macroscopic descriptions of the interfacial dynamics and microscopic intermolecular interactions appeared in multi-phase systems \cite{Guo2011PRE}. Amongst various kinds of kinetic schemes, lattice Boltzmann method (LBM) has received particular attention due to the distinct way in depicting phase interactions. Plenty of models with excellent performance have been devised \cite{INAMURO2004JCP,Zheng2008IJNMF,Li2012PRE,Chen2013POF,LIU2017IJHMT}, among which the phase-field-based models, designed in a pattern of solving C-H equation or A-C equation within the framework of LBM, have made great progress \cite{Niu2018POF,Zhang2019POF,Liu2019POF,Liang2019POF}. While they share the advantages of simplicity and efficiency rooted in LBM, the required uniformity of the lattice structure has posed a challenge on the application of complex boundaries. Another issue is multi-phase models based on lattice Boltzmann equation lack the ability to accurately predict the non-equilibrium effects at micro scales.
\\
As a newly developed kinetic scheme, discrete unified gas-kinetic scheme (DUGKS) has proven its excellent ability in a range of fields including microflows \cite{Guo2013PRE,Liu2018CF}, binary gas flows \cite{Zhang2018PRE,Zhang2019YuePOF}, phonon transportation \cite{Luo2017IJHMT} and radiative heat transfer \cite{Guo2016IJHMT}. Compared to lattice Boltzmann method, DUGKS is implemented within the framework of finite volume method and thus is no longer limited to uniform grid. With the collision effect taken into consideration in the solution reconstruction at cell interface, DUGKS preserves second-order accuracy in both continuum and free-molecular regimes \cite{Guo2013PRE}. Based on the aforementioned advantages, DUGKS is a promising choice to resolve the non-equilibrium effects emerged in multi-phase flow at micro scales. However, studies on the performance of DUGKS in coping with multi-phase flows is still limited \cite{Zhang2018IJHMT}. Also it has been verified that original DUGKS \cite{Guo2013PRE}, where central-based meso-flux construction is utilized, fails to match LBM when the flow is dominated by convection \cite{Yang2019PRE}. Before any further investigation, it is necessary to seek out ways to improve the capability of DUGKS in tackling with interface dynamics. As high-order interpolation templates are frequently employed to improve scheme fidelity in traditional methods \cite{Macdonald2008JSC,WANGY2015JCP}, it is worth trying high-order schemes for the desired variable reconstruction. In this work, we apply several high-order finite-difference schemes for the reconstruction of local characteristic solution in DUGKS and evaluate their performance on tracking interface evolution. To objectively assess the capability of DUGKS with high-order reconstruction, lattice Boltzmann method and discrete velocity method (DVM) \cite{YANG2016JCP} has been introduced for reference results.
\\
The rest of the paper is organized as follows. In Sec.~\ref{sec:sec2}, the methodology of two kinetic schemes for the conservative Allen-Cahn equation would be introduced. In Sec.~\ref{sec:sec3}, four representative numerical cases are carried out to compare their capabilities and brief discussions are presented. Final discussions and conclusions are given in Sec.~\ref{sec:sec4}.

\section{\label{sec:sec2}ALLEN-CAHN EQUATION AND KINETIC MODELS}
\subsection{Conservative Allen-Cahn equation}\label{subsec:sec2A}
The conservative Allen-Cahn equation was first introduced by Geier et al. \cite{Geier2015PRE}. A more general version is given by Ren et al. \cite{Ren2016PRE} in the form of
\begin{equation}
\partial_t{\phi}+\nabla\cdot(\phi\bm{u}) = M_\phi[\nabla^2\phi-\nabla\cdot(\theta\bm{n})],
\label{Eq:ACE}
\end{equation}
where $\phi$ short for $\phi(\bm{x},t)$ is the order parameter used to indicate the phase ($\bm{x}$ stands for position and $t$ stands for time), $\bm{u}$ is the flow velocity, $M_\phi$ is the mobility coefficient and $\bm{n}$ is the unit vector normal to the interface, with the expression given by
\begin{equation}
\bm{n}=\nabla{\phi}/\vert\nabla{\phi}\vert.
\label{Eq:AC:Normal}
\end{equation}

The parameter $\theta$ is the norm of the order parameter's gradient when the interface comes to an equilibrium state, at which $\phi$ satisfies the one-dimensional profile equation \cite{Kim2012CiCP},
\begin{equation}
\phi(z)=\frac{\phi_A+\phi_B}{2} + \frac{\phi_A-\phi_B}{2}\text{tanh}\Big(\frac{2z}{W}\Big),
\end{equation}
where $W$ is the width of interface and z is along the direction normal to the interface. $\phi_A$ and $\phi_B$ stand for the phase indicators for phase A and phase B respectively. Then the general form for $\theta$ can be easily derived from the above equilibrium profile equation,
\begin{equation}
\theta = \frac{-4(\phi-\phi_A)(\phi-\phi_B)}{W(\phi_A-\phi_B)}.
\end{equation}
It is worth mentioning that the divergence-free velocity condition, $\nabla{\cdot}\bm{u} = 0$, is adopted in the derivation of conservative Allen-Cahn equation.

The reason why we choose conservative A-C equation instead of C-H equation is that A-C equation imposes less requirements on the numerical scheme compared to that of C-H equation. The highest-order derivative C-H equation contains is fourth-order while in A-C equation it is only second-order. The price of solving a second-order partial differential equation (PDE) is surely much less expensive than that of dealing with a fourth-order one. Another reason is that C-H equation cannot be recovered precisely from the kinetics methods through the second-order Chapman-Enskog expansion \cite{Wang2016PRE}. Since the purpose of current study focus mainly on the numerical properties of various kinetic schemes, our conclusion would be more reliable if we select the benchmark tests involving fewer irrelevant impacts.
\subsection{\label{sec2B}Kinetic methods for Allen-Cahn equation}
The Allen-Cahn equation implemented by lattice Boltzmann method has been widely investigated by former researchers \cite{Geier2015PRE,Ren2016PRE,Chai2018IJHMT,Liu2019POF}. Also, the Allen-Cahn equation solved by the primitive DUGKS can be found in literature \cite{Yang2019PRE}. Here we will not waste our energy to explain those mature methods again but concentrate on interpreting Allen-Cahn equation in the language of streaming and collision DVM and third-order DUGKS, respectively.

In general, the Boltzmann equation with BGK collision model can be expressed as
\begin{equation}
\frac{\partial{f}}{\partial{t}}+\bm{\xi}\cdot\nabla_{x}{f} + \bm{a}\cdot\nabla_{\bm{\xi}}f = -\frac{f-f^{eq}}{\tau},
\label{Eq:Boltzmann-BGK}
\end{equation}
where $f$, short for $f(\bm{x},\bm{\xi},t)$, is the distribution function which depends on space $\bm{x}$, particle velocity $\bm{\xi}$, and time $t$, $\bm{a}$ is the particle acceleration, ${\tau}$ is the relaxation time and $f^{eq}$ stands for the equilibrium state approached by $f$ within each collision.
The conservative variable $\phi$ is the zeroth moment of distribution function $f$, i.e.,
\begin{equation}
\phi = {\int}f(\bm{x},\bm{\xi},t)d\bm{x}d\bm{\xi}.
\end{equation}
After discretizing the continuous velocity space into finite ones, Eq.~(\ref{Eq:Boltzmann-BGK}) turns into
\begin{equation}
\frac{\partial{f_i}}{\partial{t}}+\bm{\xi}_i\cdot\nabla_{x}{f_i} + \bm{a}\cdot\nabla_{\bm{\xi}_i}f_i = -\frac{f_i-f_i^{eq}}{\tau},
\label{Eq:Discrete-BGK}
\end{equation}
where $\bm{\xi}_i$ denotes the $i$th discretized velocity and $f_i = f(\bm{x},\bm{\xi}_i,t)$ is the distribution function with velocity $\bm{\xi}_i$. The main endeavor in this paper is try to solve conservative Allen-Cahn equation written in the form of Eq.~(\ref{Eq:Discrete-BGK}) via various of schemes.

\subsubsection{\label{sec2B.1}Streaming and collision discrete velocity method}
The original DVM with streaming and collision process is a multi-scale scheme devised by Yang et al. \cite{YANG2016JCP}. It is more like a semi-Lagrangian scheme due to the incorporation of streaming and collision process.
 Integrate Eq.~(\ref{Eq:Discrete-BGK}) in space and time within a single time step $\Delta{t}$, we can obtain
\begin{equation}
\begin{split}
f_i(\bm{x}_c,t+\Delta{t})-f_i(\bm{x}_c-\bm{\xi}_i{\Delta}t,t) = \frac{\Delta{t}}{2}[\Omega_i(\bm{x}_c,t+{\Delta}t)+\Omega_i(\bm{x}_c-\bm{\xi}_i{\Delta}t,t)
\\
+ S_i(\bm{x}_c,t+{\Delta}t) + S_i(\bm{x}_c-\bm{\xi}_i{\Delta}t,t)],
 \end{split}
 \label{Eq:DVM}
\end{equation}
where $\bm{x}_c$ is the cell center in the discretized physical space, $\Omega_i=-(f_i-f_{i}^{eq})/{\tau}$ is the collision term and $S_i = -\bm{a}\cdot\nabla_{\bm{\xi}_i}{f_i}$ is the source term. Here trapezoidal rule is utilized for the integration of collision and source term, which leads to the implicitness during the update of $f_i(\bm{x}_c)$. To remove this implicit treatment, an auxiliary distribution function is introduced,
\begin{equation}
\tilde{f_i}^{+}=f_i + \frac{\Delta{t}}{2}\Omega_{i}+\frac{\Delta{t}}{2}S_i.
\label{Eq:AuxiliaryDF}
\end{equation}
Substituting Eq.~(\ref{Eq:AuxiliaryDF}) into Eq.~(\ref{Eq:DVM}), we have
\begin{equation}
\tilde{f_i}^{+}(\bm{x}_c,t+{\Delta}t) = \frac{2\tau-\Delta{t}}{2\tau+\Delta{t}}\tilde{f_i}^{+}(\bm{x}_c-\bm{\xi}_i{\Delta}t,t)+\frac{2\Delta{t}}{2\tau+\Delta{t}}[f_i^{eq}(\bm{x}_c,t+{\Delta}t) + {\tau}S_i(\bm{x}_c,t+{\Delta}t)].
\label{Eq:SCDVM}
\end{equation}
The evolution of $\tilde{f_i}^{+}(\bm{x}_c,t+{\Delta}t)$ in Eq.~(\ref{Eq:SCDVM}) can be decomposed into two processes, i.e.,\\
Streaming step:
\begin{subequations}
\begin{equation}
\tilde{f_i}^{'}(\bm{x}_c,t+{\Delta}t) = \tilde{f_i}^{+}(\bm{x}_c-\bm{\xi}_i{\Delta}t,t),
\label{Eq:SCDVM-streaming}
\end{equation}
\text{Collision step:}
\begin{equation}
\tilde{f_i}^{+}(\bm{x}_c,t+{\Delta}t) = \tilde{f_i}^{'}(\bm{x}_c,t+{\Delta}t)+\frac{2\Delta{t}}{2\tau+\Delta{t}}[f_i^{eq}(\bm{x}_c,t+{\Delta}t)-\tilde{f_i}^{'}(\bm{x}_c,t+{\Delta}t) + {\tau}S_i(\bm{x}_c,t+{\Delta}t)],
\label{Eq:SCDVM-collision}
\end{equation}
\end{subequations}
where $\tilde{f_i}^{'}(\bm{x}_c,t+{\Delta}t)$, evolved from surrounding points, is the temporary distribution function located at the cell center. To determine the value of $\tilde{f_i}^{+}(\bm{x}_c-\bm{\xi}_i{\Delta}t,t)$, general interpolation techniques can be employed. Here the distribution function and its first and second order derivatives at the cell center are utilized to do this job, i.e,
\begin{equation}
\tilde{f_i}^{+}(\bm{x}_c-\bm{\xi}_i{\Delta}t,t) = \tilde{f_i}^{+}(\bm{x}_n,t)+(\bm{x}_c-\bm{\xi}_i{\Delta}t-\bm{{x}_n})\cdot\nabla\tilde{f_i}^{+}(\bm{x}_n,t)+(\bm{x}_c-\bm{\xi}_i{\Delta}t-\bm{{x}_n})^2:\nabla^2\tilde{f_i}^{+}(\bm{x}_n,t).
\label{Eq:DVMDerivatives}
\end{equation}
Here $\bm{x}_n$ denotes the center of a neighbor cell of which the distribution function streams out. If the time step ${\Delta}t$ is set to be small enough, the whole streaming process would happen within a single cell. In such a condition, $\bm{x}_n$ would be identical with $\bm{x}_c$ and information across cells would transmit at a slow pace, resulting in a relatively large dissipation. This is the reason why Monotone Upstream-centered Schemes for Conservation Laws (MUSCL) approach was utilized in the work of Yang et al. \cite{YANG2016JCP}. By choosing an appropriate time step in this work, it can be guaranteed that the starting cell from which particles migrated is just adjacent to the targeting cell. In this way, the influence region of any point in the flow field will be scaled to such a suitable size that it would neither get too small to keep a low dissipation nor become too large to be physical.
\\
Since the simulations in current work focus on various static and dynamic shapes, the isotropic finite-difference scheme \cite{Kumar2004JCP} is utilized to calculate the derivatives of $\tilde{f_i}^{+}(\bm{x}_c,t)$. Thus with the determination of $\tilde{f_i}^{+}(\bm{x}_c-\bm{\xi}_i{\Delta}t,t)$ comes the evaluation of $\tilde{f_i}^{'}(\bm{x}_c,t+{\Delta}t)$, which is the streaming step depicted by Eq.~(\ref{Eq:SCDVM-streaming}). When comes to the collision step, the evolution process meets a bit of difficulty due to that the equilibrium distribution function $f_i^{eq}$ and external source term $S_i$ at time $t+{\Delta}t$ needs to be known before the update of $\tilde{f_i}^{+}(\bm{x}_c,t+{\Delta}t)$. Hence we should shift our focus to obtain the conservative variables at time $t+{\Delta}t$. Notice that the moments of $\tilde{f}_i^{+}$ are identical to ${f}_i$, so the conservative variable used in this work can be updated by
\begin{equation}
\phi(\bm{x}_c,t+{\Delta}t)=\sum_{i}\omega_i\tilde{f}_i^{+}(\bm{x}_c-\bm{\xi}_i\Delta{t},t),
\label{Eq:DVM:Moments}
\end{equation}
where $\omega_i$ stand for the weights associated with each discretized velocity $\bm{\xi}_i$.
After we get the newest conservative variable $\phi$, the latest equilibrium distribution function and source term can be calculated by
\begin{subequations}
\label{Eq:DVM:Eq-Source}
\begin{equation}
f_i^{eq} = \omega_i\phi(1+\bm{\xi}_i\cdot\bm{u}/RT),
\label{Eq:DVM:Eq}
\end{equation}
\text{and}
\begin{equation}
S_i=\omega_i\theta\bm{\xi}_i\cdot\bm{n}+\omega_i\bm{\xi}_i\partial_t(\phi\bm{u})/RT,
\label{Eq:DVM:Source}
\end{equation}
\end{subequations}
where $\bm{u}$ is the flow velocity and $\sqrt{RT}$ is the reference velocity. The others symbols $\phi$, $\theta$, and $\bm{n}$ have the same meaning as those in Sec.~\ref{subsec:sec2A}. After all of the information needed at time $t+{\Delta}t$ has been updated, the collision step is performed and so we obtain the newest $\tilde{f}^{+}$. The details of evolution process are shown as follows,
\\
\textit{Streaming step.}
\\
\begin{equation*}
\tilde{f}^{+}(\bm{x}_n,\bm{\xi}_i,t)\xrightarrow{(\ref{Eq:DVMDerivatives})\ }\tilde{f}^{+}(\bm{x}_c-\bm{\xi}_i{\Delta}t,\bm{\xi}_i,t)\xrightarrow{(\ref{Eq:SCDVM-streaming})}\tilde{f}^{'}(\bm{x}_c,\bm{\xi}_i,t+\Delta{t}).
\end{equation*}
\\
\textit{Collision step.}
\\
\begin{equation*}
\tilde{f}^{'}(\bm{x}_c,\bm{\xi}_i,t+\Delta{t})\stackrel{(\ref{Eq:DVM:Moments})\ }{\longrightarrow}\phi(\bm{x}_c,t+\Delta{t})\xrightarrow[(\ref{Eq:DVM:Source})\ ]{(\ref{Eq:DVM:Eq})\ }
\left\{
\begin{aligned}
f^{eq}(\bm{x}_c,\bm{\xi}_i,t+{\Delta}t)&
\\
S(\bm{x}_c,\bm{\xi}_i,t+{\Delta}t)&
\end{aligned}
\right\}\xrightarrow{(\ref{Eq:SCDVM-collision})}\tilde{f}^{+}(\bm{x}_n,\bm{\xi}_i,t+{\Delta}t).
\end{equation*}
\\
\subsubsection{\label{sec2B.2}Third-Order DUGKS}
Inspired by the work of Li et al. \cite{Li2016}, Wu et al. \cite{Wu2018} developed the third-stage third-order discrete unified gas-kinetic scheme (DUGKS-T3S3) for low speed isothermal flows without source terms. Here we applied this method in the simulation of interfacial dynamic problems driven by a predefined velocity field and conducted an exhaustive investigation on its performance. For the cumbersome equations appeared during the derivation of DUGKS-T3S3, readers are recommended to refer to the original literature \cite{Wu2018}. Here we begin with the derived evolution equation in its discretized form, i.e.,
\begin{equation}
\tilde{f}_i^{n+\frac{1}{3}} = \tilde{f}_i^{+,{n}} + \frac{1}{3}\Delta{t}L(f_i^{*})
\label{Eq:DUGKS:T2S2}
\end{equation}
and
\begin{equation}
\hat{f}_i^{n+1} = {f_i}^{n}+\frac{3}{4}\Delta{t}[\Omega(f_i')+S(f_i')]+\frac{1}{7}\Delta{t}[3L(f_i^*)+4L(f_i^{**})].
\label{Eq:DUGKS-T3S3}
\end{equation}
The two above equations introduce several new abbreviations. It is worth mentioning that $\tilde{f}_i^{+,n}$, ${f_i}^{n}$, $\hat{f_i}^{n+1}$, $\Omega$ and $S$ are all cell-averaged values since DUGKS is implemented by finite volume method. The full form of abbreviated distribution function ${f_i}^{n}$ is
\begin{equation}
{f_i}^{n} = \frac{1}{\vert{V_c}\vert}{\int}_{V_c}f(\bm{x}_c,\bm{\xi}_i,t_n),
\end{equation}
where $i$ indicates the $i$th discretized velocity $\bm{\xi}_i$, $n$ denotes the time at $t_n$ and $V_c$ is the control volume centered at $\bm{x}_c$.
\\
The pair of auxiliary distribution function $\tilde{f}_i^{n+\frac{1}{3}}$ and $\tilde{f}_i^{+,{n}}$ in Eq.~(\ref{Eq:DUGKS:T2S2}) are defined as
\begin{subequations}
\begin{equation}
\tilde{f}_i^{n+\frac{1}{3}} = \tilde{f}_i(\bm{x}_c,t_n + \frac{\Delta{t}}{3})=f_i^{n+\frac{1}{3}}-\frac{\Delta{t}}{6}\Omega_i^{n+\frac{1}{3}}-\frac{\Delta{t}}{6}S_i^{n+\frac{1}{3}}
\end{equation}
\begin{equation}
\tilde{f}_i^{n,+} = \tilde{f}_i^{+}(\bm{x}_c,t_n)= f_i^{n}+\frac{\Delta{t}}{6}\Omega_i^{n}+\frac{\Delta{t}}{6}S_i^{n}
\end{equation}
\end{subequations}
The auxiliary distribution function $\hat{f_i}^{n+1}$, presented as a shorthand for $\hat{f}(\bm{x}_c,\bm{\xi}_i,t_n+\Delta{t})$, have the following definition,
\begin{equation}
\hat{f_i} = f_i-\frac{\Delta{t}}{4}\Omega_i-\frac{\Delta{t}}{4}S_i,
\end{equation}
where $\Omega_i$ and $S_i$ have the same meaning as those in Sec.~\ref{sec2B.1}.
\\
The others symbols with different superscripts including $f^{'}_i$, $f^{*}_i$, and $f^{**}_i$ are abbreviations of original distribution function in the form of $f(\bm{x}_c,\bm{\xi}_i,t_n^{'})$, $f(\bm{x}_c,\bm{\xi}_i,t_n^{*})$, $f(\bm{x}_c,\bm{\xi}_i,t_n^{**})$, separately. And the intermediate various time steps have values of
\begin{equation}
t_n^{'} = t_n + \frac{\Delta{t}}{3},\ t_n^{*} = t_n + \frac{\Delta{t}}{6},\ t_n^{**} = t_n + \frac{3\Delta{t}}{4}.
\end{equation}
Now the only remaining unknown symbol in Eq.~(\ref{Eq:DUGKS-T3S3}) is $L$, which is the meso-flux operator with the expression of
\begin{equation}
L(f_i) = \frac{1}{{\vert}V_c{\vert}}{\int}_{\partial{V_c}}(\bm{\xi}_i\cdot\bm{n})f(\bm{x}_c,\bm{\xi}_i,t)d\bm{S}.
\label{Eq:DUGKS:MicroFlux}
\end{equation}
Here $\partial{V_c}$ is the surface of cell ${V_c}$ and $\bm{n}$ is the inward unit vector normal to the surface.
To evaluate the meso-flux at different intermediate time steps with sufficient precision, the key point is to get the original distribution function on cell interfaces $x_b$ with enough accuracy. Luckily, the reconstruction procedure in originalDUGKS \cite{Guo2013PRE} preserves the third-order temporal accuracy and therefore can be used without any modification.
\\
Integrating Eq.~(\ref{Eq:Discrete-BGK}) along its characteristic line with a small time step $h$ and applying the trapezoidal rule to the collision term and source term, we have
\begin{equation}
\begin{split}
f_i(\bm{x}_b,t+h)-f_i(\bm{x}_b-\bm{\xi}_ih,t) = \frac{h}{2}[\Omega_i(\bm{x}_b,t+h)+\Omega_i(\bm{x}_b-\bm{\xi}_ih,t)
\\
+ S_i(\bm{x}_b,t+h) + S_i(\bm{x}_b-\bm{\xi}_ih,t)].
\label{Eq:Characteristic-Integral-Solution}
\end{split}
\end{equation}
By introducing the following auxiliary distribution functions,
\begin{subequations}
\begin{equation}
\bar{f}_i = f_i - \frac{h}{2}{\Omega_i}-\frac{h}{2}S_i,
\end{equation}
\begin{equation}
\bar{f}_i^{+,h} = f_i + \frac{h}{2}{\Omega_i}+\frac{h}{2}S_i,
\end{equation}
\end{subequations}
Eq.~(\ref{Eq:Characteristic-Integral-Solution}) turns into
\begin{equation}
\bar{f}_i(\bm{x}_b,t+h) = \bar{f}_i^{+,h}(\bm{x}_b-\bm{\xi}_i{h},t).
\label{Eq:Auxiliary-Characteristic}
\end{equation}
In the following the superscript $h$ is omitted when unnecessary. Since there exits no information at time $t+h$, to estimate $\bar{f}_i(\bm{x}_b,t+h)$, we should first evaluate $\bar{f}_i^{+}(\bm{x}_b-\bm{\xi}_i{h},t)$.  Two kinds of interpolation approaches, central scheme and upwind scheme, were put forward by Guo et al. \cite{Guo2013PRE,Guo2015PRE} successively during the development of DUGKS.
\\
For the central scheme, the value of $\bar{f}_i^{+}(\bm{x}_b-\bm{\xi}_i{h},t)$ should be updated by
\begin{equation}
\bar{f}_i^{+}(\bm{x}_b-\bm{\xi}_i{h},t) = \bar{f}_i^{+}(\bm{x}_b,t) - (\bm{\xi}_i{h})\cdot{\nabla}\bar{f}_i^{+}(\bm{x}_b,t),
\label{Eq:BarPlus-Central}
\end{equation}
where $\bar{f}_i^{+}(\bm{x}_b,t)$ as well as its gradient is estimated by the surrounding values of $\bar{f}_i^{+}$ at cell center. For the upwind scheme, the value of $\bar{f}_i^{+}(\bm{x}_b-\bm{\xi}_i{h},t)$ should be updated by
\begin{equation}
\bar{f}_i^{+}(\bm{x}_b-\bm{\xi}_i{h},t) = \bar{f}_i^{+}(\bm{x}_n,t) + (\bm{x}_b-\bm{\xi}_i{h}-\bm{x}_n)\cdot{\nabla{\bar{f}_i^{+}(\bm{x}_n,t)}} + (\bm{x}_b-\bm{\xi}_i{h}-\bm{x}_n)^2:{\nabla^2{\bar{f}_i^{+}(\bm{x}_n,t)}},
\label{Eq:BarPlus-Upwind}
\end{equation}
where the coordinate $\bm{x}_n$ indicates the cell center which is nearest to the back-traced position $\bm{x}_b-\bm{\xi}_i{h}$. Again, the isotropic finite-difference method \cite{Kumar2004JCP} is utilized to calculate the spatial derivatives of $\bar{f}_i^{+}$.
\\
The value of $\bar{f}_i(\bm{x}_b,t+h)$ can be achieved by Eq.~(\ref{Eq:Auxiliary-Characteristic}) after the update of $\bar{f}_i^{+}(\bm{x}_b-\bm{\xi}_i{h},t)$. To determine the original distribution function, the equilibrium distribution function needs to be updated first due to the following relation,
\begin{equation}
f_i = \frac{2\tau}{2\tau+h}\bar{f}_i+\frac{h}{2\tau+h}f_i^{eq} + \frac{\tau{h}}{2\tau + h}S_i.
\label{Eq:DUGKS:PrimitiveDFonInterface}
\end{equation}
The expression of $f_i^{eq}$ and $S_i$ takes the same form as those in Section~\ref{sec2B.1}. The conservative variable $\phi$ on cell interface can be calculated by
\begin{equation}
\phi(\bm{x}_b,t+h) = \sum_{i}{\omega}_i\bar{f}_i(\bm{x}_b,t+h).
\label{Eq:CV:Interface}
\end{equation}
\\
To obtain the summation of collision term and source term at $t_n^{'} = t_n + \Delta{t}/3$, the following relation is needed:
\begin{equation}
\Omega_i(\bm{x}_c,t_n^{'}) + S_i(\bm{x}_c,t_n^{'}) = -\frac{6{\Delta}t}{6\tau+{\Delta}t}[\tilde{f}_i(\bm{x}_c,t_n^{'})-f_i^{eq}(\bm{x}_c,t_n^{'})-{\tau}S_i(\bm{x}_c,t_n^{'})].
\label{Eq:DUGKS:Sum-Omega-S}
\end{equation}
The conservative variable $\phi$ at cell center is calculated by
\begin{subequations}
\begin{equation}
\phi(\bm{x}_c,t_n) = \sum_i\hat{f}_i(\bm{x}_c,t_n),
\end{equation}
\begin{equation}
\phi(\bm{x}_c,t_n^{'}) = \sum_i\tilde{f}_i(\bm{x}_c,t_n^{'}).
\label{Eq:CV:Cell:DtA}
\end{equation}
\end{subequations}
Other useful relations needed during this evolution process are
\begin{equation}
\bar{f}_i^{+,\frac{\Delta{t}}{6}}(\bm{x}_c,t_n) = \frac{12\tau-\Delta{t}}{12\tau+3\Delta{t}}\hat{f}_i(\bm{x}_c,t_n) + \frac{4\Delta{t}}{12\tau+3\Delta{t}}[f_i^{eq}(\bm{x}_c,t_n) + {\tau}S_i(\bm{x}_c,t_n)],
\label{Eq:DUGKS-T3S3-BarPlusA}
\end{equation}
\begin{equation}
f_i(\bm{x}_c,t_n) = \frac{3}{4}\bar{f}_i^{+,\frac{\Delta{t}}{6}}(\bm{x}_c,t_n) + \frac{1}{4}\hat{f}_i(\bm{x}_c,t_n),
\label{Eq:DUGKS-T3S3-Prime}
\end{equation}
\begin{equation}
\tilde{f}_i^{+}(\bm{x}_c,t_n) = \frac{5}{4}\bar{f}_i^{+,\frac{\Delta{t}}{6}}(\bm{x}_c,t_n) - \frac{1}{4}\hat{f}_i(\bm{x}_c,t_n),
\label{Eq:DUGKS-T3S3-TildePlus}
\end{equation}
\begin{equation}
\bar{f}_i^{+,\frac{5}{12}\Delta{t}}(\bm{x}_c,t_n^{'}) = \frac{24\tau-5\Delta{t}}{24\tau+4\Delta{t}}\tilde{f}_i(\bm{x}_c,t_n^{'}) + \frac{9\Delta{t}}{24\tau+4\Delta{t}}[f_i^{eq}(\bm{x}_c,t_n^{'}) + {\tau}S_i(\bm{x}_c,t_n^{'})].
\label{Eq:DUGKS-T3S3-BarPlusK}
\end{equation}

By far we have defined all the symbols and built all the relations used in the process of solving Eq.~(\ref{Eq:DUGKS-T3S3}). Base on these information, it is possible to provide an exhaustive clarification on the evolution process.
\\
\textit{Step (i)}. Construct meso-flux at $t_n^{*}$,
\begin{equation}
\hat{f}_i(\bm{x}_c,t_n)\xrightarrow{(\ref{Eq:DVM:Eq-Source}),(\ref{Eq:DUGKS-T3S3-BarPlusA})}\bar{f}_i^{+,\frac{\Delta{t}}{6}}(\bm{x}_c,t_n)\xrightarrow{(\ref{Eq:BarPlus-Upwind}),(\ref{Eq:Auxiliary-Characteristic})}\bar{f}_i(\bm{x}_c,t_n^{*})\xrightarrow{(\ref{Eq:CV:Interface}),(\ref{Eq:DVM:Eq-Source}),(\ref{Eq:DUGKS:PrimitiveDFonInterface}),(\ref{Eq:DUGKS:MicroFlux})}L_i(t_n^{*}).
\end{equation}
\textit{Step (ii)}. Update $\tilde{f}_i(\bm{x}_c,t_n^{'})$ by originalDUGKS and compute $\Omega_i(\bm{x}_c,t_n^{'})+S_i(\bm{x}_c,t_n^{'})$,
\begin{equation}
L(t_n^{*})\xrightarrow{(\ref{Eq:DUGKS-T3S3-TildePlus}),(\ref{Eq:DUGKS:T2S2})}\tilde{f}(\bm{x}_c,t_n^{'})\xrightarrow{(\ref{Eq:CV:Cell:DtA})}\phi(\bm{x}_c,t_n^{'})\xrightarrow{(\ref{Eq:DVM:Eq-Source}),(\ref{Eq:DUGKS:Sum-Omega-S})}[\Omega_i(\bm{x}_c,t_n^{'})+S_i(\bm{x}_c,t_n^{'})].
\end{equation}
\textit{Step (iii)}. Construct meso-flux at $t_n^{**}$,
\begin{equation}
\tilde{f}_i(\bm{x}_c,t_n^{'})\xrightarrow{(\ref{Eq:DUGKS-T3S3-BarPlusK})}\bar{f}_i^{+,\frac{5}{12}\Delta{t}}(\bm{x}_c,t_n^{'})\xrightarrow{(\ref{Eq:BarPlus-Upwind}),(\ref{Eq:Auxiliary-Characteristic})}\bar{f}_i(\bm{x}_c,t_n^{**})\xrightarrow{(\ref{Eq:CV:Interface}),(\ref{Eq:DVM:Eq-Source}),(\ref{Eq:DUGKS:PrimitiveDFonInterface}),(\ref{Eq:DUGKS:MicroFlux})}L_i(t_n^{**}).
\end{equation}
\textit{Step (iv)}. Update $\hat{f}_i^{n+1}$ by the known terms,
\begin{equation}
\hat{f}_i(\bm{x}_c,t_n)\xrightarrow{(\ref{Eq:DUGKS-T3S3-Prime})}f_i{(\bm{x}_c,t_n)}\xrightarrow{(\ref{Eq:DUGKS-T3S3})}\hat{f}_i(\bm{x}_c,t_{n+1}).
\end{equation}
With the comprehensive elaboration of the evolution process of DVM and DUGKS-T3S3 in this section, there should not be much difficulty in comprehending those two methods. Since the distribution function is discrete in its velocity space as well as spatial space during calculation, it is indispensable to present the discretization method applied to it. Here the three-point Gauss-Hermite quadrature is employed to discretize the velocity space in each single dimension. The discrete velocities and associated weights used in this study are
\begin{equation*}
\begin{aligned}
\bm{\xi}&=\sqrt{3RT}
\Bigg{[}
\begin{aligned}
0&&1&&1&&0&&-1&&-1&&-1&&0&&1
\\
0&&0&&1&&1&&1&&0&&-1&&-1&&-1
\end{aligned}
\Bigg{]},
\\
\omega_i &=
	\begin{cases}
	\frac{4}{9},&i = 0\\
	\frac{1}{9},&i = 1,3,5,7\\
	\frac{1}{36},&i = 2,4,6,8
	\end{cases},
\end{aligned}
\end{equation*}
where $RT$ is the dimensionless velocity with a fixed value of 1/3.
The time step is determined by the Courant-Friedrichs-Lewy (CFL) condition as follows
\begin{equation}
\Delta{t} = C \frac{\Delta{x}}{\sqrt{3RT}},
\label{Eq:CFL}
\end{equation}
where $C $ is the CFL number and $\Delta{x}$ is the size of the shortest mesh interface.
\\
\section{\label{sec:sec3}NUMERICAL TESTS}
In this section, four standard benchmark tests, including interface diagonal translation, Zalesak's disk rotation, interface extension and interface deformation, are simulated to assess the performance of LBM, DVM and DUGKS in capturing interfaces. The velocity of flow field is specified in advance. Therefore, there is no need to solve the hydrodynamic equation. The dimensionless parameters used in this paper are P\'{e}clet number and Cahn number with the definitions of
\begin{equation}
\begin{aligned}
Pe &=\frac{U_0L_0}{M_\phi},\\
Cn &= \frac{W}{L_0},
\end{aligned}
\end{equation}
where $U_0$ is the reference velocity and $L_0$ is the side length of computational domain. Uniform Cartesian mesh with a constant cell size of unity has been used among these four simulations. To conduct an quantitative evaluation on the performance of the relating methods, the $L_2$-norm-based error of the order parameter $\phi$ is used \cite{Liang2014PRE}:
\begin{equation}
E_{\phi}(t) = \sqrt{\frac{\sum_{\bm{x}}\vert\phi(\bm{x},t)-\phi(\bm{x},0)\vert^2}{\sum_{\bm{x}}\vert\phi(\bm{x},0)\vert^2}}.
\end{equation}
The time is scaled by
\begin{equation}
T_f=\frac{L_0}{U_0}.
\end{equation}
\subsection{\label{sec:sec3.A}Interface diagonal translation}
\begin{figure}[htbp]
    \centering
    \begin{minipage}[b]{0.5 \columnwidth}
    \begin{subfigure}{0.9 \columnwidth}
      \centering
      \includegraphics[width=1\linewidth]{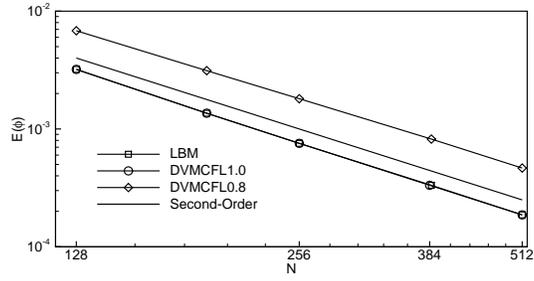}
      \caption{LBM, DVM}
      \label{FIG:TD:LBMDVML2Phi}
    \end{subfigure}
    \end{minipage}
    \begin{minipage}[b]{0.9 \columnwidth}
    \begin{subfigure}{0.5 \columnwidth}
    \centering
      \centering
      \includegraphics[width=1\linewidth]{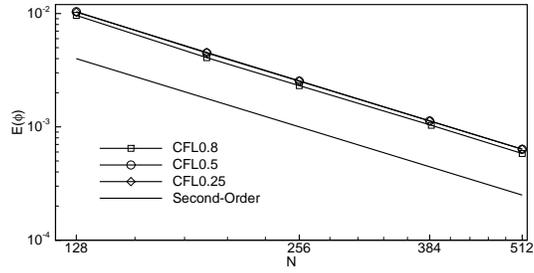}
      \caption{DUGKS-T2S2CD}
      \label{FIG:TD:DUGKST2S2CDL2Phi}
    \end{subfigure}
    \end{minipage}
    \begin{minipage}[b]{0.9 \columnwidth}
    \begin{subfigure}{0.5 \columnwidth}
    \centering
      \centering
      \includegraphics[width=1\linewidth]{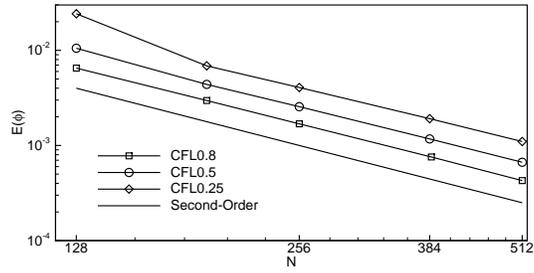}
      \caption{DUGKS-T2S3}
      \label{FIG:TD:DUGKST2S3L2Phi}
    \end{subfigure}
    \end{minipage}
    \begin{minipage}[b]{0.9 \columnwidth}
    \begin{subfigure}{0.5 \columnwidth}
    \centering
      \centering
      \includegraphics[width=1\linewidth]{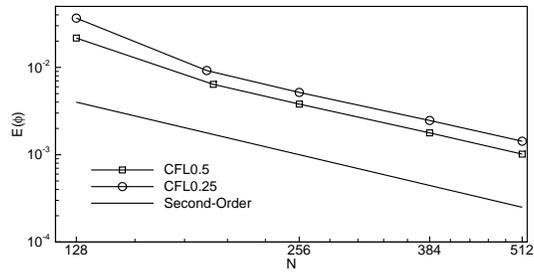}
      \caption{DUGKS-T3S3}
      \label{FIG:TD:DUGKST3S3L2Phi}
    \end{subfigure}
    \end{minipage}
    \begin{minipage}[b]{0.9 \columnwidth}
    \begin{subfigure}{0.5 \columnwidth}
    \centering
      \centering
      \includegraphics[width=1\linewidth]{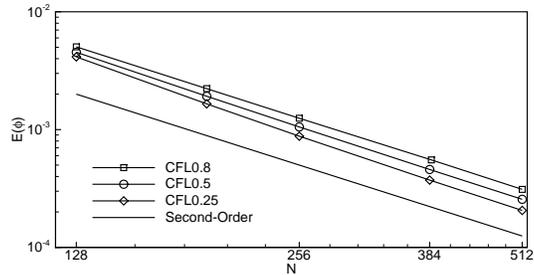}
      \caption{DUGKS-T2S5}
      \label{FIG:TD:DUGKST2S5L2Phi}
    \end{subfigure}
    \end{minipage}
\caption{Convergence rate of $\phi$ obtained by multiple methods with $Pe = 128$, $Cn = 1/32$, $M_\phi = 0.02$.}
\label{FIG:TD:L2Phi}
\end{figure}

Here we investigated the translation of a circular interface moving along the diagonal direction of the computational domain, which is driven by a constant velocity field $\bm{u} = \textrm{(}U_0,U_0\textrm{)}$. Initially, a circular interface with radius $R = L_0/5$ is placed in the center of a periodic $L_0{\times}L_0$ domain. After a $T_f$ of elapsed time, the interface will transport along the diagonal direction to the original location. A comparison between the initial and the final profile of interface would be able to validate the performance of multiple methods quantitatively.
\\
The comparative results of convergence rate of each method are illustrated in Fig.~\ref{FIG:TD:L2Phi}  and the corresponding detailed results are presented in Table~\ref{tab:TD:L2Phi}. The mesh number along each side of the square domain is refined from 128 to 512. The reference velocity $U_0$ and interface width $W$ are adjusted along with the variation of mesh number in order to keep the P\'{e}clet number, Cahn number, and mobility coefficient $M_\phi$ consistent.
\\
\begin{table}
\caption
{
  \label{tab:TD:L2Phi}
  $L_2$-norm error of $\phi$ for diagonal translation obtained by various methods with $Pe = 128$, $Cn = 1/32$, $M_\phi = 0.02$.
}
	\begin{minipage}{1.0\textwidth}
	\begin{subtable}{1.0\textwidth}
	\caption{\label{tab:TD:LBMDVML2Phi}LBM, DVM}
		\begin{ruledtabular}
		\begin{tabular}{cccccc}
		N&128&192&256&384&512\\
		\colrule
		LBM & $3.20\times10^{-3}$ & $1.36\times10^{-3}$ & $7.55\times10^{-4}$ & $3.33\times10^{-4}$ & $1.87\times10^{-4}$\\
		DVM-CFL1.0 & $3.20\times10^{-3}$ & $1.36\times10^{-3}$ & $7.55\times10^{-4}$ & $3.33\times10^{-4}$ & $1.87\times10^{-4}$\\
		DVM-CFL0.8 & $6.82\times10^{-3}$ & $3.12\times10^{-3}$ & $1.81\times10^{-3}$ & $8.21\times10^{-4}$ & $4.66\times10^{-4}$\\
		\end{tabular}
		\end{ruledtabular}
	\end{subtable}%
	\end{minipage}
	\par\medskip
	\begin{minipage}{1.0\textwidth}
	\begin{subtable}{1.0\textwidth}
	\caption{\label{tab:TD:DUGKS-T2S2CD-L2Phi}DUGKS-T2S2CD}
		\begin{ruledtabular}
		\begin{tabular}{cccccc}
		N&128&192&256&384&512\\
		\colrule
		DVM-CFL0.25 & $1.03\times10^{-2}$ & $4.54\times10^{-3}$ & $2.55\times10^{-3}$ & $1.13\times10^{-3}$ & $6.35\times10^{-4}$\\
		DVM-CFL0.5 & $1.04\times10^{-2}$ & $4.49\times10^{-3}$ & $2.53\times10^{-3}$ & $1.13\times10^{-3}$ & $6.35\times10^{-4}$\\
		DVM-CFL0.8 & $9.63\times10^{-3}$ & $4.08\times10^{-3}$ & $2.31\times10^{-3}$ & $1.03\times10^{-3}$ & $5.82\times10^{-4}$\\
		\end{tabular}
		\end{ruledtabular}
	\end{subtable}%
	\end{minipage}
    \par\medskip
	\begin{minipage}{1.0\textwidth}
	\begin{subtable}{1.0\textwidth}
	\caption{\label{tab:TD:DUGKS-T2S3-L2Phi}DUGKS-T2S3}
		\begin{ruledtabular}
		\begin{tabular}{cccccc}
		N&128&192&256&384&512\\
		\colrule
		DVM-CFL0.25 & $2.43\times10^{-2}$ & $6.87\times10^{-3}$ & $4.07\times10^{-3}$ & $1.91\times10^{-3}$ & $1.10\times10^{-3}$\\
		DVM-CFL0.5 & $1.05\times10^{-2}$ & $4.39\times10^{-3}$ & $2.56\times10^{-3}$ & $1.17\times10^{-3}$ & $6.67\times10^{-4}$\\
		DVM-CFL0.8 & $6.51\times10^{-3}$ & $2.97\times10^{-3}$ & $1.69\times10^{-3}$ & $7.57\times10^{-4}$ & $4.27\times10^{-4}$\\
		\end{tabular}
		\end{ruledtabular}
	\end{subtable}%
	\end{minipage}
	\par\medskip
	\begin{minipage}{1.0\textwidth}
	\begin{subtable}{1.0\textwidth}
	\caption{\label{tab:TD:DUGKS-T3S3-L2Phi}DUGKS-T3S3}
		\begin{ruledtabular}
		\begin{tabular}{cccccc}
		N&128&192&256&384&512\\
		\colrule
		DVM-CFL0.25 & $3.67\times10^{-2}$ & $9.21\times10^{-3}$ & $5.16\times10^{-3}$ & $2.46\times10^{-3}$ & $1.43\times10^{-3}$\\
		DVM-CFL0.5 & $2.17\times10^{-2}$ & $6.40\times10^{-3}$ & $3.80\times10^{-3}$ & $1.78\times10^{-3}$ & $1.01\times10^{-3}$
		\end{tabular}
		\end{ruledtabular}
	\end{subtable}%
	\end{minipage}
	\par\medskip
	\begin{minipage}{1.0\textwidth}
	\begin{subtable}{1.0\textwidth}
	\caption{\label{tab:TD:DUGKS-T2S5-L2Phi}DUGKS-T2S5}
		\begin{ruledtabular}
		\begin{tabular}{cccccc}
		N&128&192&256&384&512\\
		\colrule
		DVM-CFL0.25 & $4.14\times10^{-3}$ & $1.65\times10^{-3}$ & $8.79\times10^{-4}$ & $3.73\times10^{-4}$ & $2.06\times10^{-4}$\\
		DVM-CFL0.5 & $4.53\times10^{-3}$ & $1.92\times10^{-3}$ & $1.05\times10^{-3}$ & $4.59\times10^{-4}$ & $2.56\times10^{-4}$\\
		DVM-CFL0.8 & $5.04\times10^{-3}$ & $2.23\times10^{-3}$ & $1.25\times10^{-3}$ & $5.53\times10^{-4}$ & $3.11\times10^{-4}$\\
		\end{tabular}
		\end{ruledtabular}
	\end{subtable}%
	\end{minipage}
\end{table}

\begin{figure}[htbp]
    \centering
    \begin{minipage}[b]{0.9 \columnwidth}
    \begin{subfigure}{0.45 \columnwidth}
      \centering
      \includegraphics[width=1\linewidth]{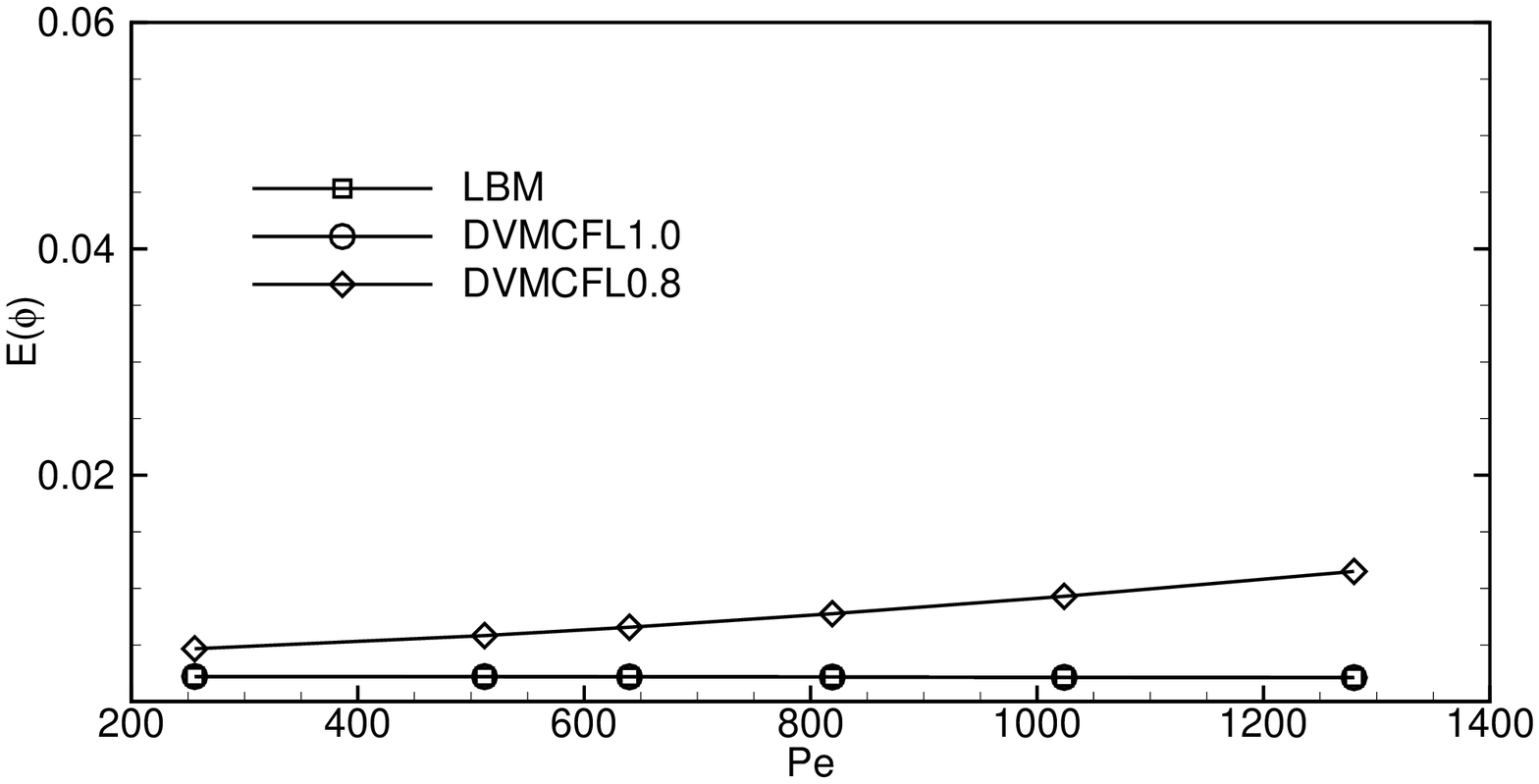}
      \caption{LBM, DVM}
      \label{FIG:TD:LBMDVM-Pe-L2Phi}
    \end{subfigure}
    \end{minipage}
    \begin{minipage}[b]{0.9 \columnwidth}
    \begin{subfigure}{0.45 \columnwidth}
    \centering
      \centering
      \includegraphics[width=1\linewidth]{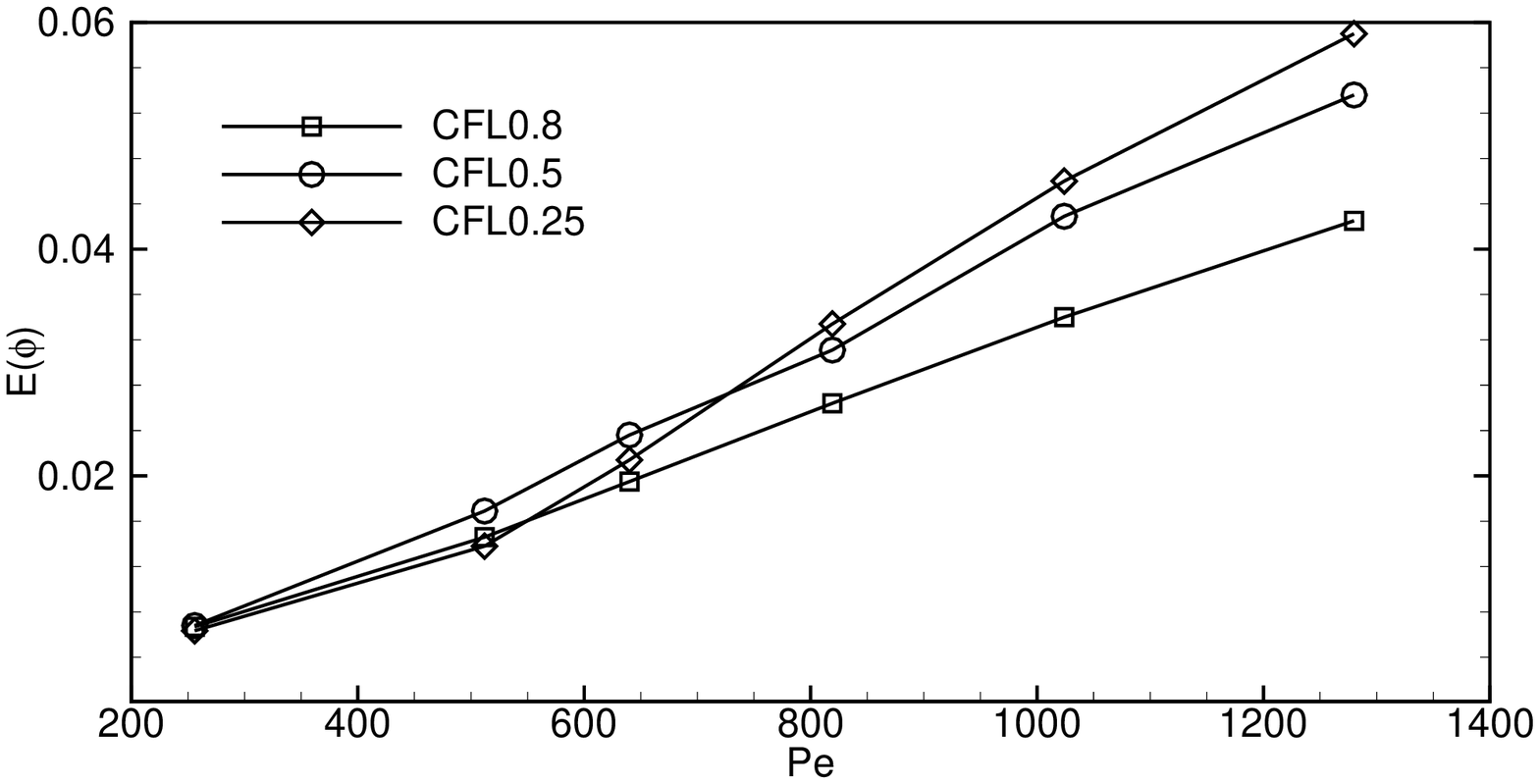}
      \caption{DUGKS-T2S2CD}
      \label{FIG:TD:DUGKST2S2CD-Pe-L2Phi}
    \end{subfigure}
    \end{minipage}
    \begin{minipage}[b]{0.9 \columnwidth}
    \begin{subfigure}{0.45 \columnwidth}
    \centering
      \centering
      \includegraphics[width=1\linewidth]{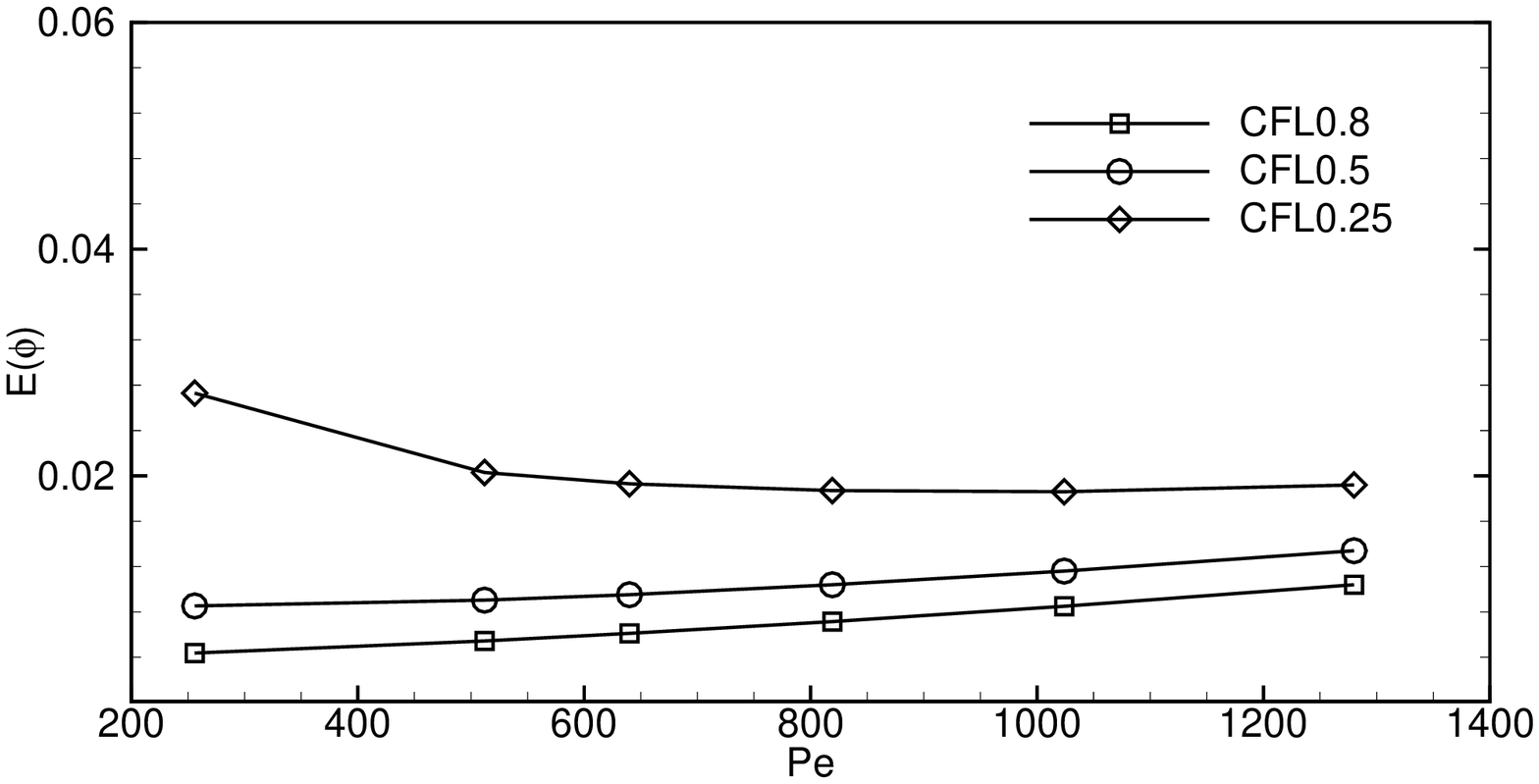}
      \caption{DUGKS-T2S3}
      \label{FIG:TD:DUGKST2S3-Pe-L2Phi}
    \end{subfigure}
    \end{minipage}
    \begin{minipage}[b]{0.9 \columnwidth}
    \begin{subfigure}{0.45 \columnwidth}
    \centering
      \centering
      \includegraphics[width=1\linewidth]{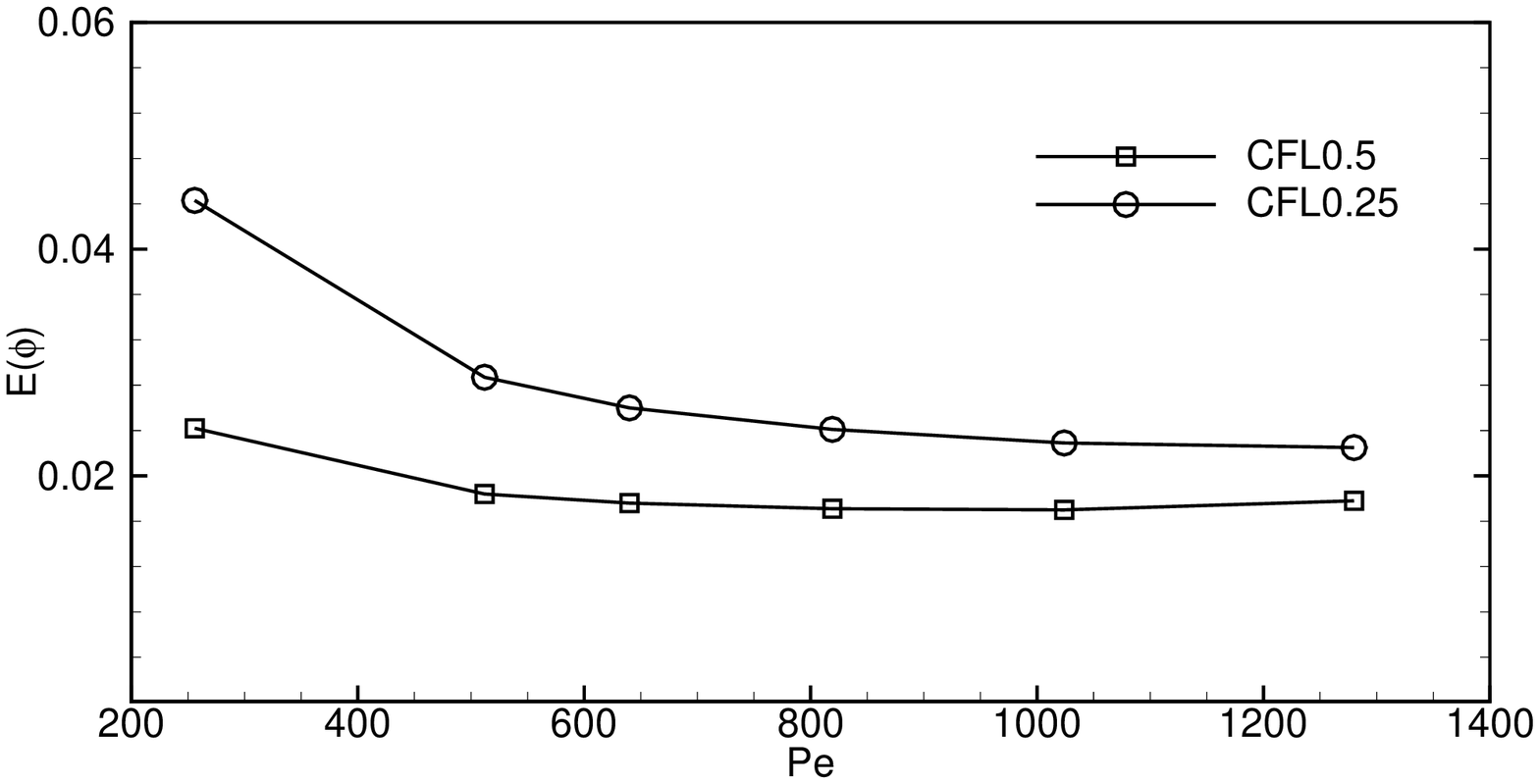}
      \caption{DUGKS-T3S3}
      \label{FIG:TD:DUGKST3S3-Pe-L2Phi}
    \end{subfigure}
    \end{minipage}
    \begin{minipage}[b]{0.9 \columnwidth}
    \begin{subfigure}{0.45 \columnwidth}
    \centering
      \centering
      \includegraphics[width=1\linewidth]{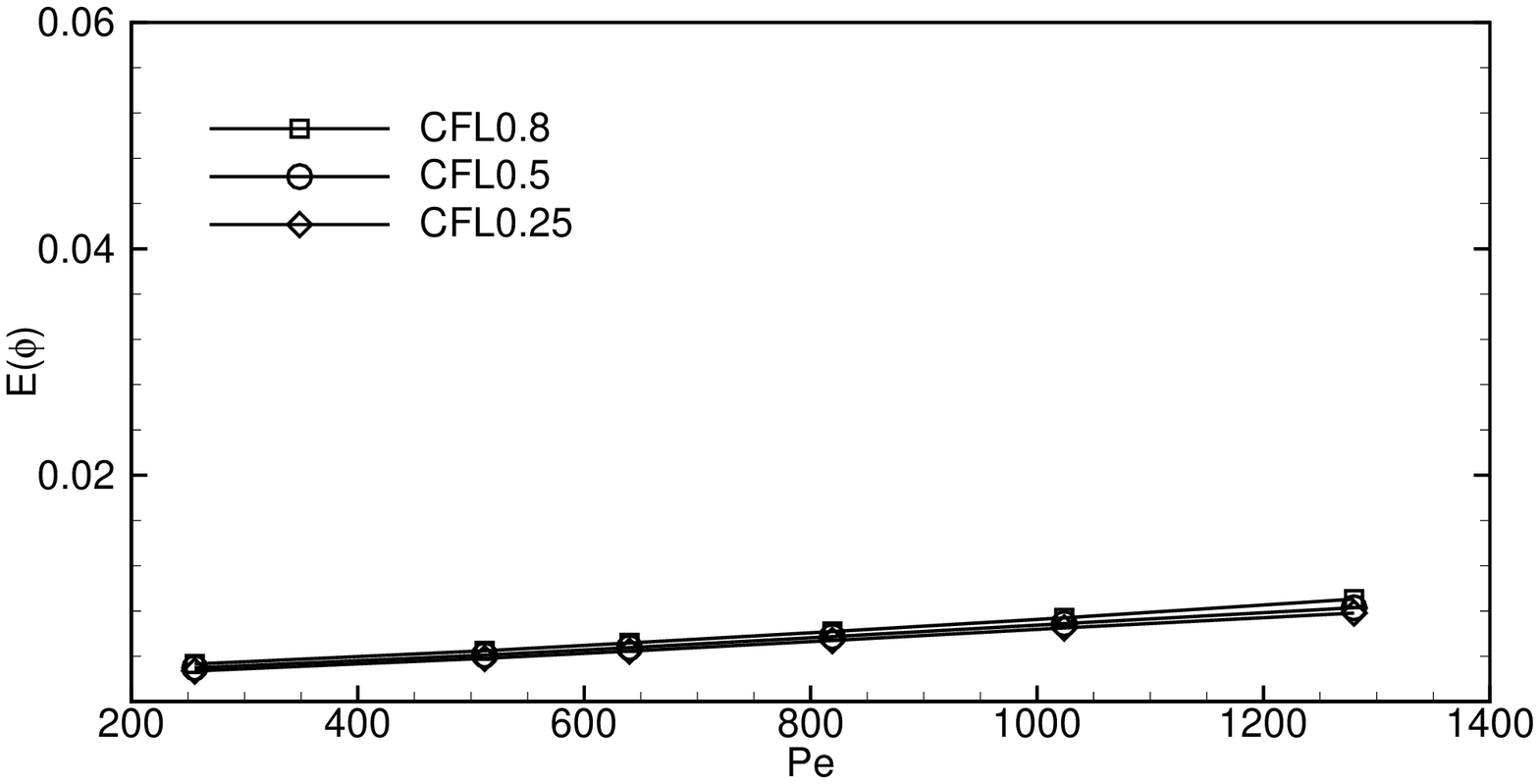}
      \caption{DUGKS-T2S5}
      \label{FIG:TD:DUGKST2S5-Pe-L2Phi}
    \end{subfigure}
    \end{minipage}
\caption{$L_2$-norm error of $\phi$ for diagonal translation obtained by multiple methods with various $Pe$, $Cn = 4/256$, $M_\phi = 0.02$.}
\label{FIG:TD:Pe-L2Phi}
\end{figure}

\begin{figure}[htbp]
    \centering
    \begin{minipage}[b]{0.9 \columnwidth}
    \begin{subfigure}{0.45 \columnwidth}
      \centering
      \includegraphics[width=1\linewidth]{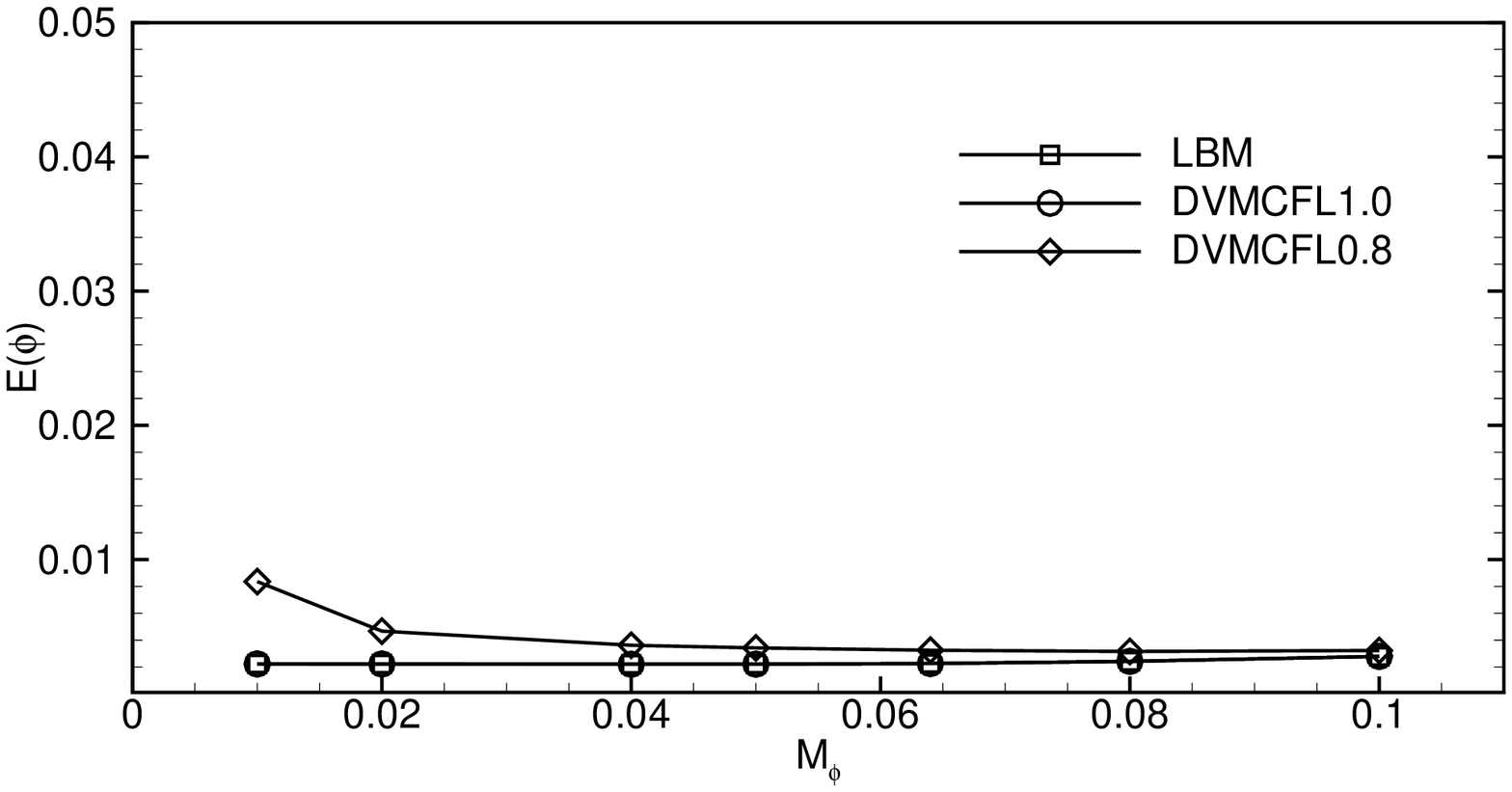}
      \caption{LBM, DVM}
      \label{FIG:TD:LBMDVM-M-L2Phi}
    \end{subfigure}
    \end{minipage}
    \begin{minipage}[b]{0.9 \columnwidth}
    \begin{subfigure}{0.45 \columnwidth}
    \centering
      \centering
      \includegraphics[width=1\linewidth]{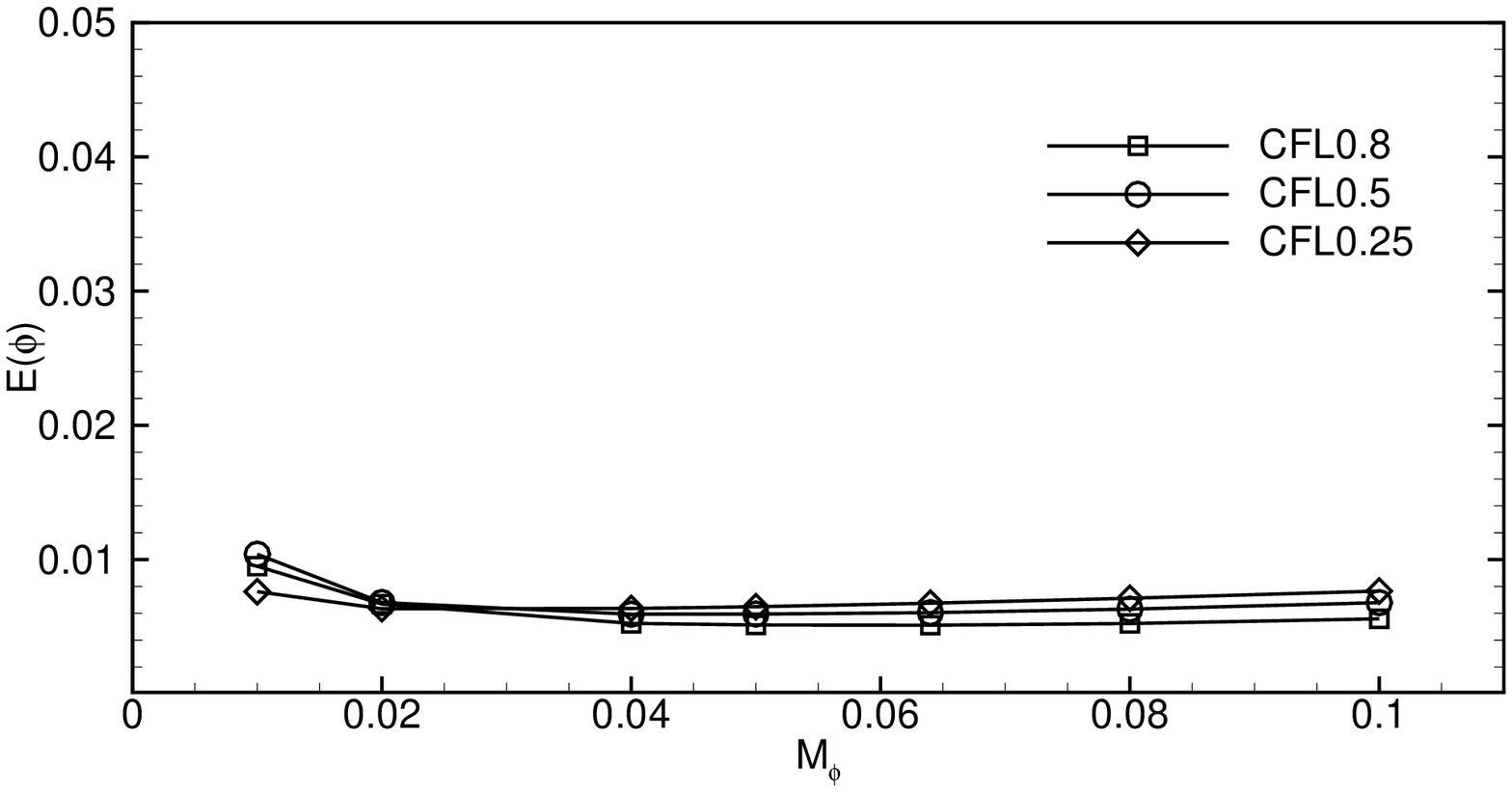}
      \caption{DUGKST2S2CD}
      \label{FIG:TD:DUGKST2S2CD-M-L2Phi}
    \end{subfigure}
    \end{minipage}
    \begin{minipage}[b]{0.9 \columnwidth}
    \begin{subfigure}{0.45 \columnwidth}
      \centering
      \includegraphics[width=1\linewidth]{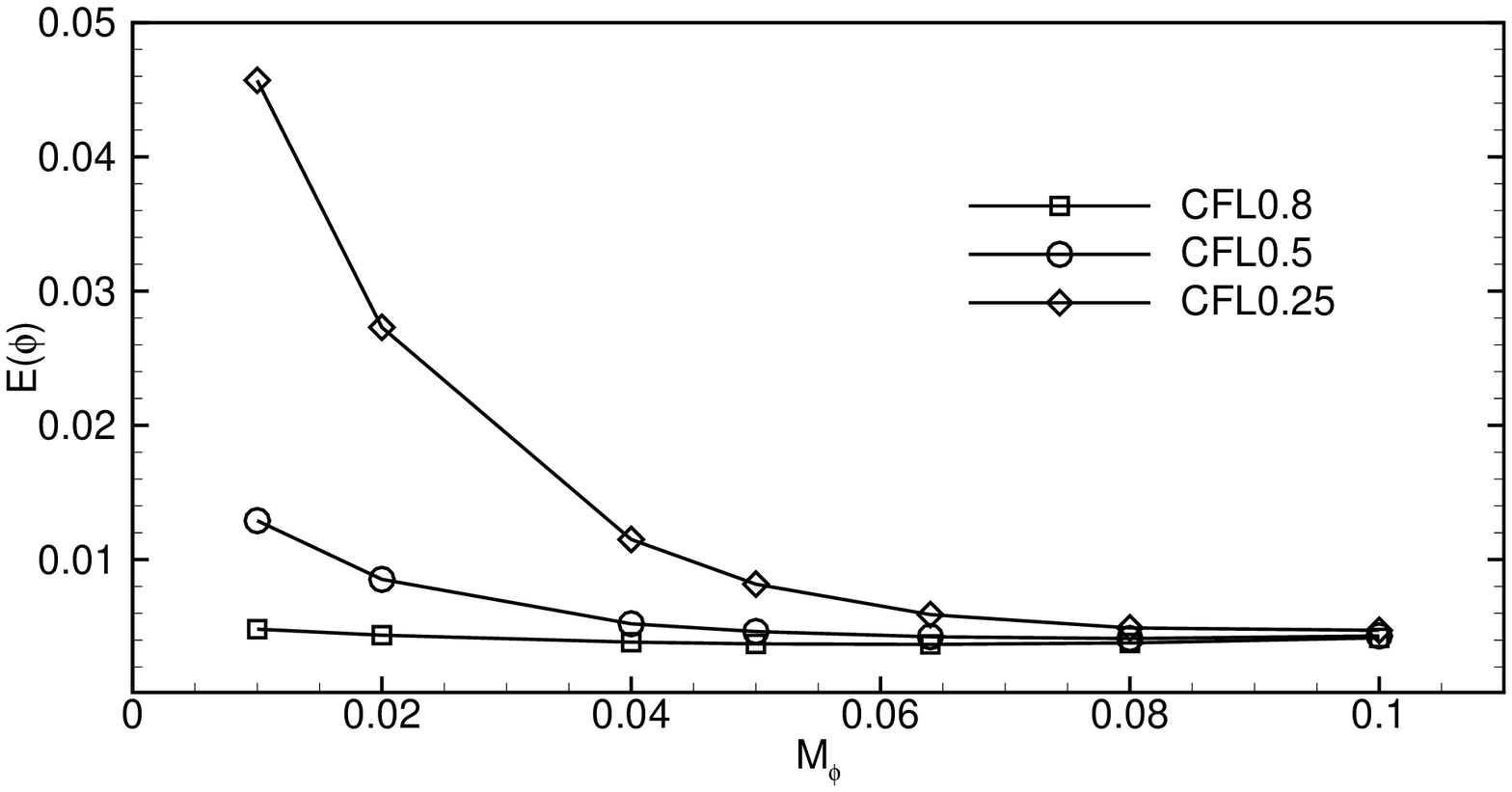}
      \caption{DUGKST2S3}
      \label{FIG:TD:DUGKST2S3-M-L2Phi}
    \end{subfigure}
    \end{minipage}
    \begin{minipage}[b]{0.9 \columnwidth}
    \begin{subfigure}{0.45 \columnwidth}
      \centering
      \includegraphics[width=1\linewidth]{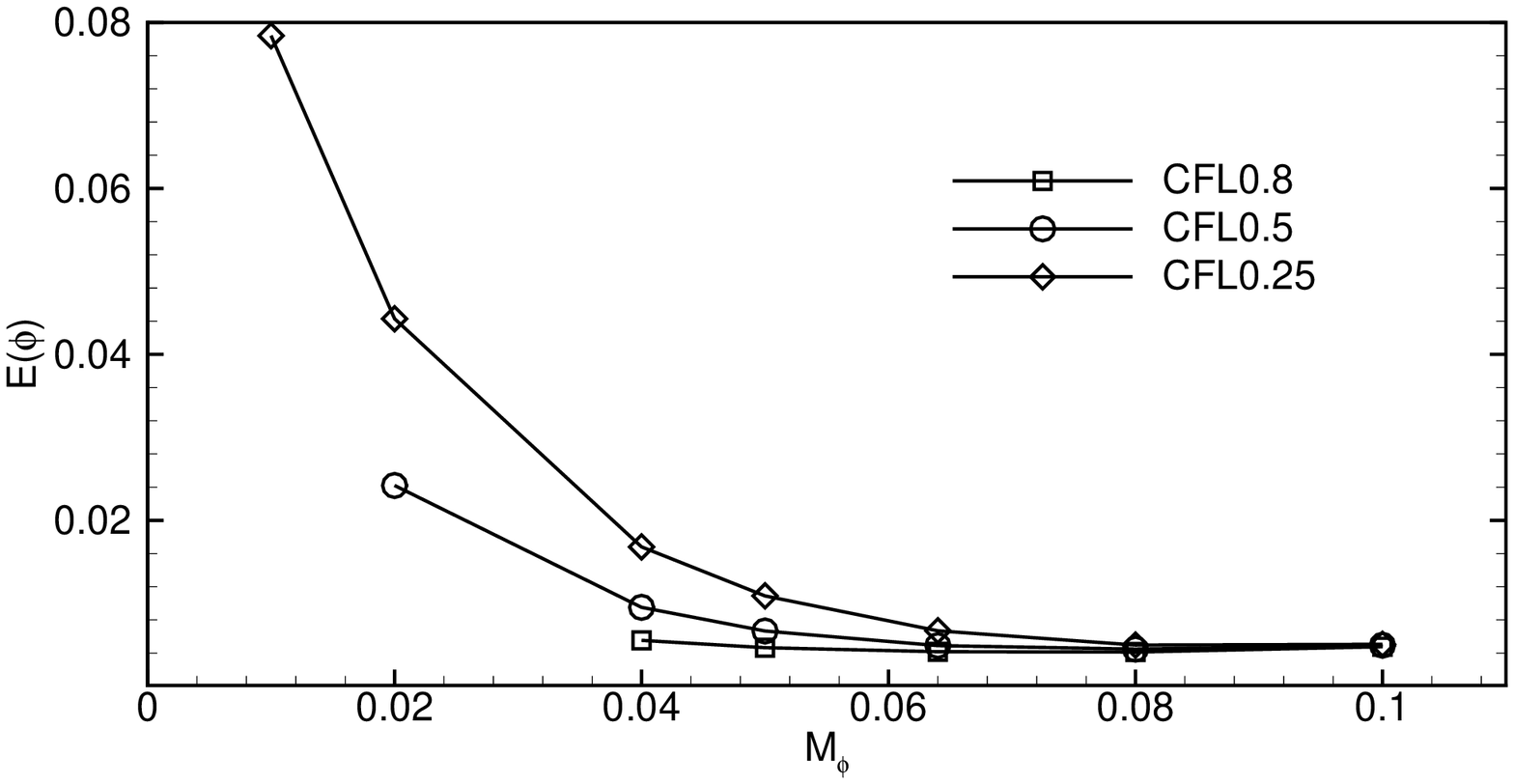}
      \caption{DUGKST3S3}
      \label{FIG:TD:DUGKST3S3-M-L2Phi}
    \end{subfigure}
    \end{minipage}
    \begin{minipage}[b]{0.9 \columnwidth}
    \begin{subfigure}{0.45 \columnwidth}
      \centering
      \includegraphics[width=1\linewidth]{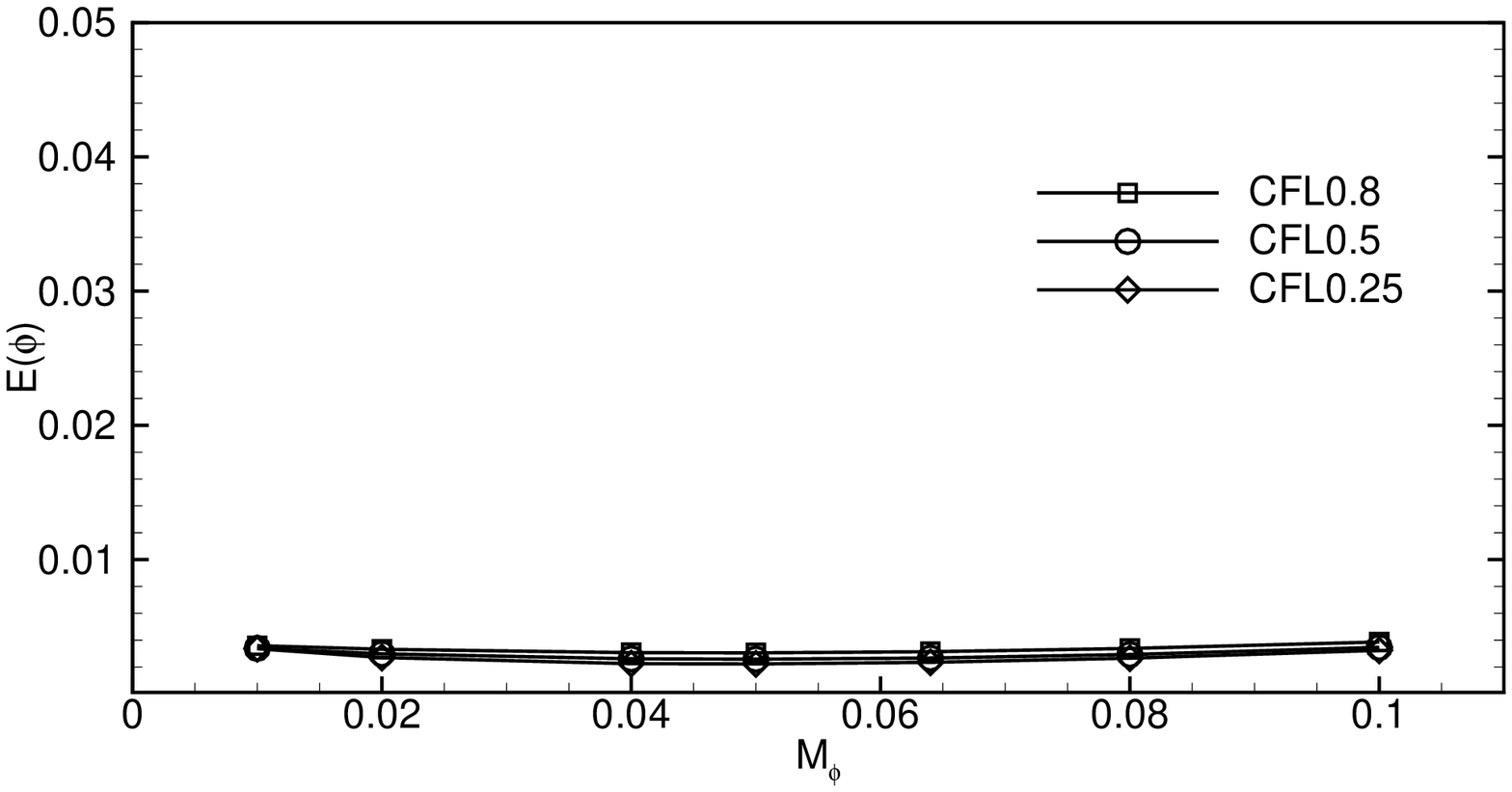}
      \caption{DUGKST2S5}
      \label{FIG:TD:DUGKST2S5-M-L2Phi}
    \end{subfigure}
    \end{minipage}
\caption{$L_2$-norm error of $\phi$ for diagonal translation obtained by multiple methods with various $M_\phi$, $Cn = 4/256$, $Pe = 256$.}
\label{FIG:TD:M-L2Phi}
\end{figure}

\begin{figure}[htbp]
    \begin{subfigure}{0.2 \columnwidth}
      \centering
      \includegraphics[width=1\linewidth]{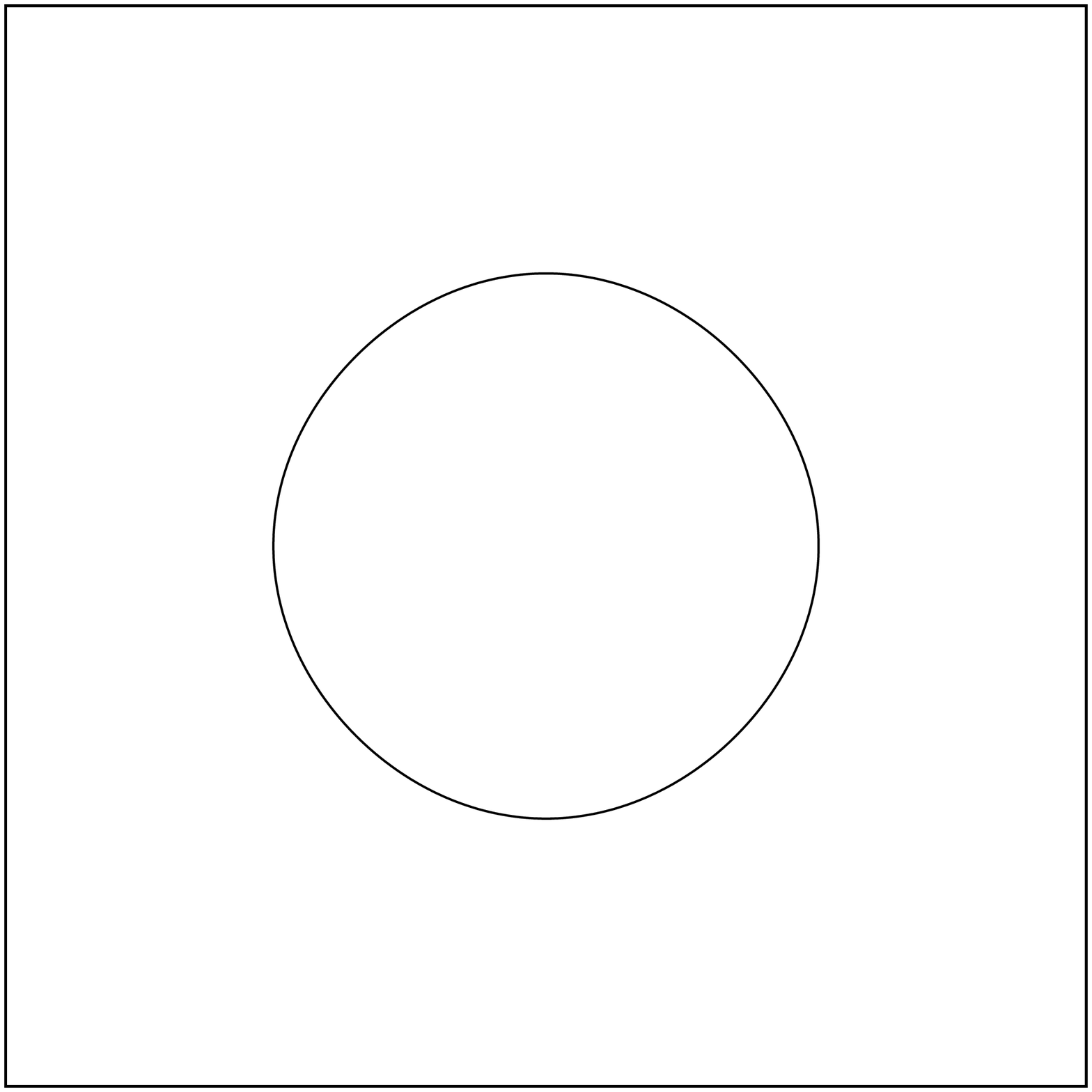}
      \caption{LBM-Isotropy}
      \label{FIG:TD:LBMDVM-Isotropy}
    \end{subfigure}
    \begin{subfigure}{0.2 \columnwidth}
      \centering
      \includegraphics[width=1\linewidth]{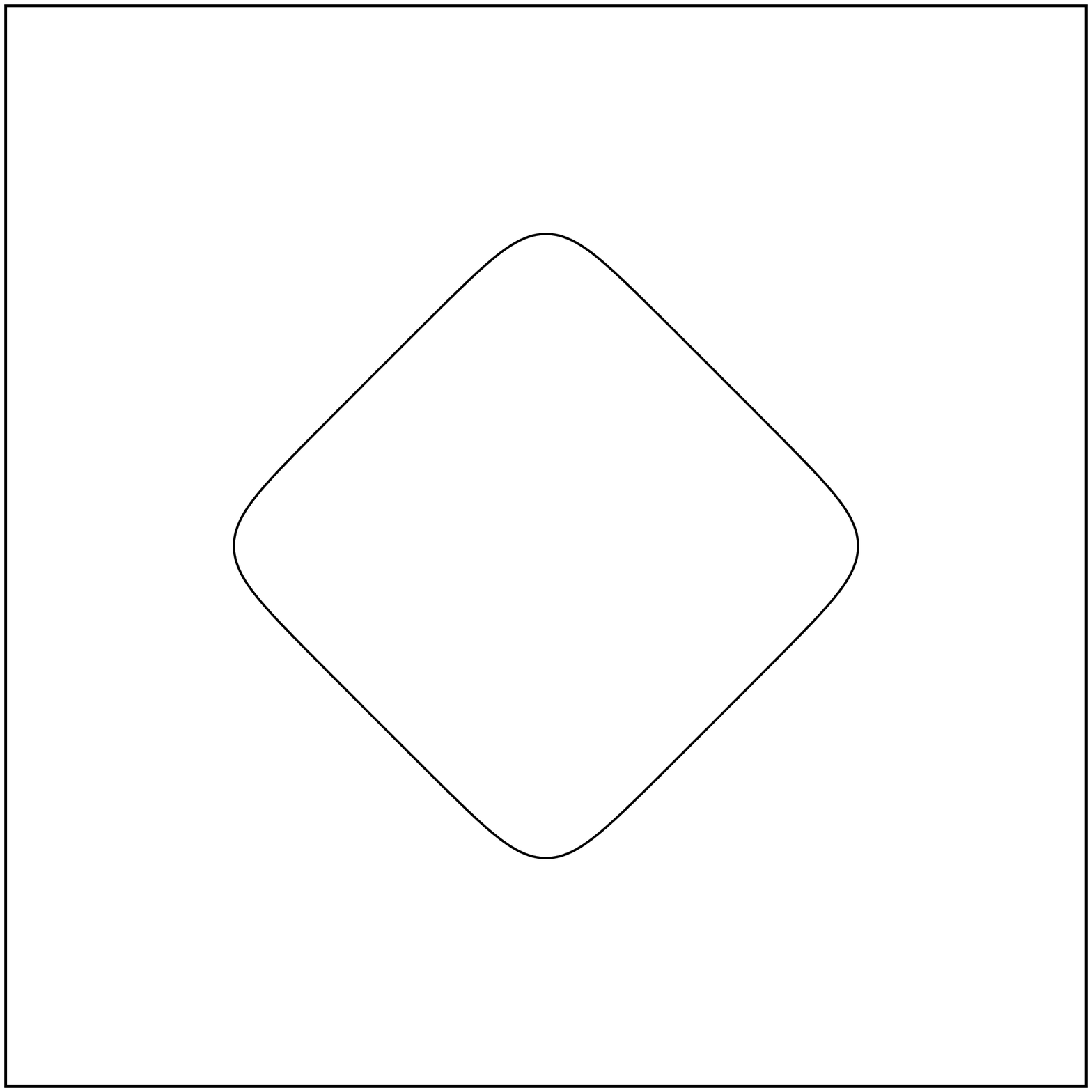}
      \caption{LBM-Anisotropy}
      \label{FIG:TD:LBMDVM-Anisotropy}
    \end{subfigure}
    \begin{subfigure}{0.2 \columnwidth}
      \centering
      \includegraphics[width=1\linewidth]{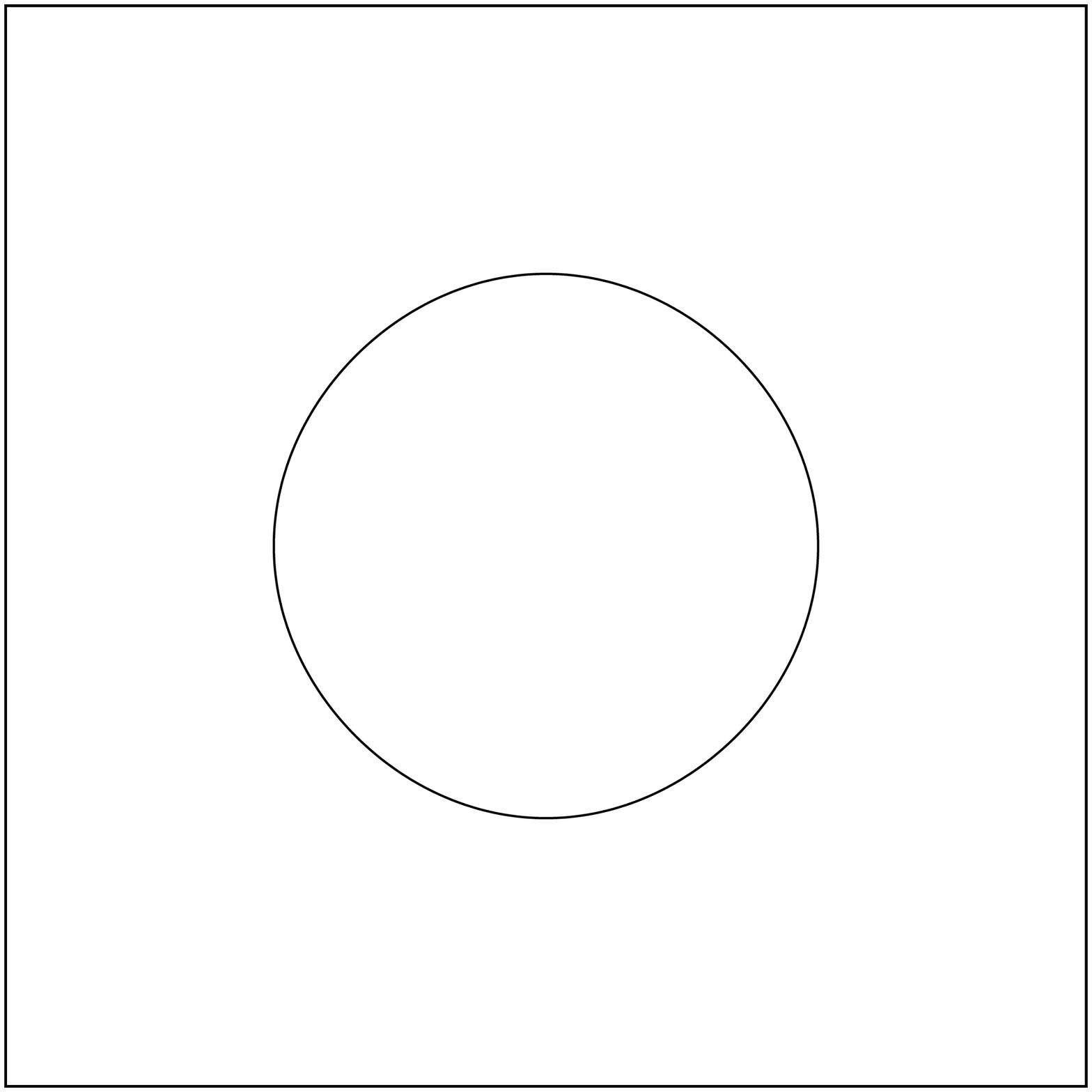}
      \caption{DUGKS-Isotropy}
      \label{FIG:TD:DUGKST2S3-Isotropy}
    \end{subfigure}
    \begin{subfigure}{0.2 \columnwidth}
      \centering
      \includegraphics[width=1\linewidth]{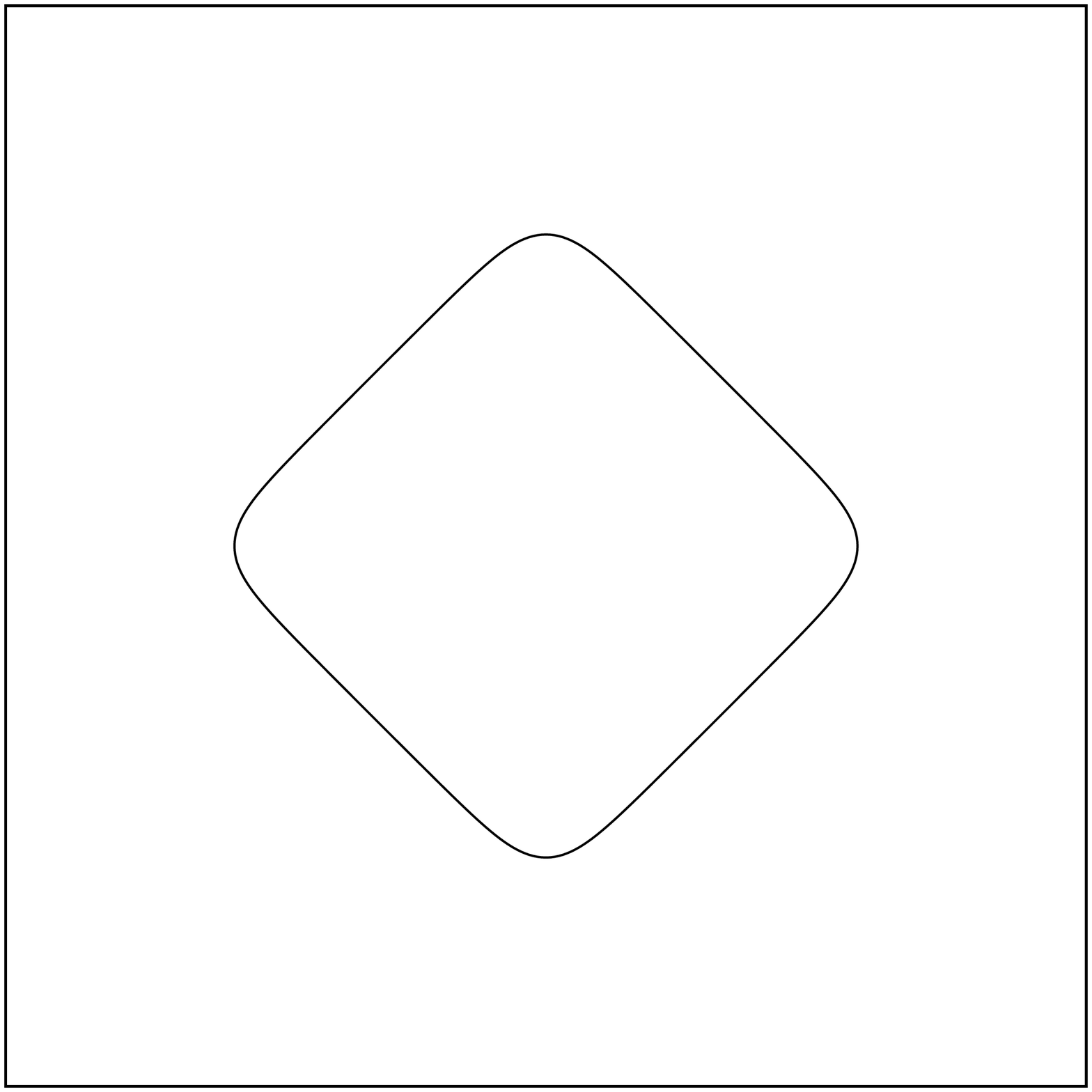}
      \caption{DUGKS-Anisotropy}
      \label{FIG:TD:DUGKST2S3-Anisotropy}
    \end{subfigure}
\caption{Interface shapes for LBM and DUGKS with different finite-difference schemes.}
\label{FIG:TD:Isotropy}
\end{figure}

It can be seen that no matter which spatial interpolation scheme is utilized, the overall convergence rate of any method maintains a second-order accuracy. The results got with LBM are in perfect consistence with previous results in literature \cite{Yang2019PRE}, where second-order convergence rate of LBM for interface translation test is validated. Results of the originalDUGKS (DUGKS-T2S2CD), where meso-flux evaluation is implemented by central scheme, indicate the same conclusion. It is worth mentioning that the results obtained by DVM with $C = 1.0$ (see Table~\ref{tab:TD:LBMDVML2Phi}) are identical to the results achieved via LBM. The reason lies in the reconstruction process of DVM. To update the distribution function at cell center $\bm{x_c}$, particles would trace precisely back to the center of its neighbor cell $\bm{x_n}$ in the condition of $C = 1.0$. Therefore, there is no need to perform the interpolation step depicted by Eq.~(\ref{Eq:DVMDerivatives}) and thus no spatial dissipation is introduced. Particles will stream from one center point to another center point and then collide with each other, again and again. This is just the evolution process of LBM. Since the exact form of A-C equation can be obtained by applying Chapmann-Enskog analysis to the streaming and collision DVM, it is reasonable that both DVM and LBM present identical results.
What confuses us most is the second-order convergence rate of DUGKS employed with high-order spatial interpolation techniques, among which the DUGKS-T3S3 really provides bizarre results. Generally, it is expected that nearly third-order convergence rate should be observed if a scheme utilized in the simulation preserves third-order accuracy in both time and space. Two reasons should account for this phenomenon. The first one is the interface itself has gone through an unsteady transformation through its whole period in this case. Here ``unsteady'' means that even if we place a circular interface in a stationary flow, the $L_2$-norm error of index parameter would still keep increasing. The numerical scheme utilized in the simulation would lead to results different from analytical ones. It has been roughly estimated that the $L_2$-norm error of $\phi$ obtained by LBM rises to $10^{-3}$ after an iteration of 1000 steps in a stationary flow. The numerical dissipation of the scheme would always cause slight deviation between the analytical results and numerical results. This fact, somehow, affects the convergence rate of the methods being tested. The second reason, also the primary reason, is that the time accuracy cannot preserve third-order due to the precision loss of source information on the interface. Although the truncation error of trapezoidal rule in Eq.~(\ref{Eq:Characteristic-Integral-Solution}) is $O(h^3)$, the source term $S_i(\bm{x}_b,t+h)$ on cell interface is estimated by the average value of its two neighbor cells at time $t$. That means the estimation error of $S_i(\bm{x}_b,t+h)$ has a magnitude of $O(h)$ and the overall error is $O(h^2)$ when it is integrated over the half time step, which causes a precision loss in Eq.~(\ref{Eq:Characteristic-Integral-Solution}). Since the key feature of Allen-Cahn equation is that the interface is driven by the local curvature and the curvature information is contained in the source term, there is no wonder that the precision of source term has a crucial effect on the overall accuracy of all the methods being tested. This is the main cause why all of the schemes being tested have presented a second-order convergence rate in accuracy for the interface diagonal translation test.
\\
Next let us take a look at the impacts of CFL number $C$, i.e., the time step. Note that in Eq.~(\ref{Eq:CFL}), both $\Delta{x}$ and $\sqrt{3RT}$ have a constant value of 1, hence $C$ is equal to $\Delta{t}$. They will be used interchangeably in the following part. Fig.~\ref{FIG:TD:LBMDVML2Phi} illustrates the convergence rate of LBM as well as DVM with two different kinds of CFL number. The $L_2$-norm error obtained by DVM with $C = 0.8$ is larger than the results got with LBM and DVM, whose CFL number keeps a fixed value of 1.0. The rationality of this set of results depends on Eq.~(\ref{Eq:DVMDerivatives}). When the CFL number is not equal to unity, the interpolation procedure depicted by the above equation is performed, which in turn introduce numerical dissipation with an magnitude of $O(\Delta{x}^3)$. When the CFL number equals unity, just as we have explained above, the interpolation procedure is skipped and the whole dissipation of this method is reduced. Fig.~\ref{FIG:TD:DUGKST2S2CDL2Phi} illustrates the convergence rate of DUGKS whose meso-flux evaluation is implemented by central scheme (DUGKS-T2S2CD). The results achieved at various CFL number do not show much difference. This is mainly cause by the method used in meso-flux construction. The truncation error when interpolating $\bar{f_i}(\bm{x}_b-\bm{\xi}_ih,t)$ in Eq.~(\ref{Eq:BarPlus-Central}) is composed of two parts. The majority is introduced in the process of evaluating the auxiliary distribution function $\bar{f}^{+}(\bm{x}_b,t)$, which is computed by the average of its neighbor cell values. Here the magnitude of the truncation error introduced is $O(\Delta{x}^2)$. The minority is the remainder on the right hand side of Eq.~(\ref{Eq:BarPlus-Central}) and its magnitude should be no larger than $\nabla^2\bar{f_i}^{+}(\bm{x}_b,t)\cdot({\bm{\xi}_ih})^2$. Compared to the error of $O(\Delta{x}^2)$, this is a high-order term. Hence, the results obtained by DUGKS-T2S2CD do not show much sensitivity to the CFL number. Nevertheless, it is worth pointing out that the total time steps do have some effects on the numerical dissipation in this test. As is mentioned above, the interface diagonal translation is an ``unsteady" process. The less time steps it takes, the weaker the effects of the error accumulation would be. Thus the numerical results would be more accurate when compared to the analytical ones. This is the reason why the results with $C = 0.8$ have shown a slight advantage over the others. Fig.~\ref{FIG:TD:DUGKST2S3L2Phi} illustrates the convergence rate of DUGKS employed with the third-order interpolation scheme (DUGKS-T2S3). Note that when the accuracy of spatial interpolation scheme is third-order, the upwind scheme would be utilized in the evaluation of meso-flux. So the trailing term ``-UW'' is dropped in the abbreviation. There is only one single kind of truncation error in the upwind-based meso-flux construction method, i.e., the remainder in Eq.~(\ref{Eq:BarPlus-Upwind}), which has an expression of $\nabla^3\bar{f_i}^{+}\cdot(\bm{x}_b-\bm{\xi}_ih-\bm{x_n})^3$. The upper bound of this term would be $O(\vert{\Delta{x}/2-\bm{\xi}_ih}\vert^3)$. Noting that $h$ is the half time step, it can be concluded the spatial dissipation of DUGKS employed with the upwind-based meso-flux construction method have a direct connection with time step. The physical figure behind the upwind-based meso-flux construction method is that the particles would migrate from the nearest neighbor cells to the interface. To guarantee this property, the maximum discretized particle velocity times the migration time $h$, i.e., the distance traveled by the fastest group of particles, should be no longer than the characteristic length of its neighbor cell. For this research, the particle velocity in each component is either unity or zero and the maximum CFL number is unity. Thus, ${\Delta{x}/2-\bm{\xi}_ih}$ would always be positive. What it means is that an increase in CFL number would result in a decrease in the truncation error, which in turn leads to a reduction in numerical dissipation. Taken the ``unsteady'' property of this test into consideration, there is no wonder that the DUGKS-T2S3 appears sensitive to the CFL number. As for DUGKS implemented by third-stage third-order discretization scheme (DUGKS-T3S3), the results are presented in Fig.~\ref{FIG:TD:DUGKST3S3L2Phi}. The results with $C = 0.8$ are missing due to the limitation of $\Delta{t} < 12\tau$~\cite{Wu2018}. Since there exist double stages in the evolution of DUGKS-T3S3, the time step for each stage is smaller than that used in DUGKS-T2S3. According to the reason explained above, its numerical dissipation would be larger than DUGKS-T2S3, which has been proven by the comparative results presented in Table~\ref{tab:TD:DUGKS-T2S3-L2Phi} and~\ref{tab:TD:DUGKS-T3S3-L2Phi}. The last set of results shown in Fig.~\ref{FIG:TD:DUGKST2S5L2Phi} are obtained by DUGKS implemented with the fifth-order interpolation scheme (DUGKS-T2S5). It is used to investigate how effective high-order reconstruction scheme is in simulating source driven flows. Different from the isotropic finite-difference scheme used above, the method utilized here does not guarantee isotropy due to the complicated interpolation stencil. Detailed information are presented in Table~\ref{tab:TD:DUGKS-T2S5-L2Phi}. Although the convergence rate of DUGKS-T2S5 keeps the same second-order accuracy, the numerical dissipation of this method has shown a significant decrease when compared to that of other tested schemes. The upper bound of the truncation error in this scheme is $O(\Delta{x}^5)$, which is so tiny that the influence of error accumulation is nearly negligible. Hence the numerical dissipation should be mainly attributed to the temporal discretization error. Since the scale of temporal error is $O(\Delta{t}^2)$ and the error of source terms located on interface has a magnitude of $O(\Delta{x}^2)$, it is reasonable to get a second-order accurate scheme over the whole. As is illustrated in Fig.~\ref{FIG:TD:DUGKST2S5L2Phi}, the scheme with $C = 0.25$ provides results with minimum dissipation. It can be verified that the ``unsteady'' property discussed above are mainly due to the spatial discretization error. The temporal error would play its role in this case as long as the spatial error has been controlled within a tiny scale. Small CFL number indicates less numerical dissipation in time. Hence, the smaller the time step is, the less the numerical dissipation would be. This is just what the results of DUGKS-T2S5 reveal. However, the absolute differences among the results obtained by DUGKS-T2S5 with various CFL number are nearly indistinguishable, which demonstrates that the temporal error has limited impacts on the performance of this method.
\\
In the following part, the effects of P\'{e}clet number and mobility coefficient are analyzed. Fig.~\ref{FIG:TD:Pe-L2Phi} illustrates the variation of $L_2$-norm error of $\phi$ obtained by various methods with the adjustment of $Pe$ and Table~\ref{tab:TD:Pe-L2Phi} supplements the corresponding data with detail. Here $L_0 =  256$, $M_\phi = 0.02$ and $Cn = 4/256$. It can be observed clearly that LBM and DVM-CFL1.0 always provide identical results, which are also the best ones among all of the results obtained by multiple methods. As long as the reconstruction procedure is operated in the evolution process, there would be an notable rise in the overall $L_2$-norm error. Both DVM-CFL0.8 and DUGKS-T2S3 with $C = 0.8$ have exemplified this trend. Compared to its performance at high Pe, DUGKS-T2S2CD method provides better results when the $Pe$ is relatively low. This is a rational phenomenon since central scheme would be unable to burden the cost of coping with flows dominated by convection. It is worth noting that even the best results ($Pe = 256$) obtained by DUGKS-T2S2CD are still inferior to the results obtained by DUGKS-T2S3 at the same condition. As for the DUGKS-T3S3 scheme, due to the limitation of ${\Delta}t<12\tau$, the time step that should offer its most valuable results would not work. The DUGKS-T2S5 method, which provides the most excellent results in the framework of DUGKS, do not show much advantage over the DUGKS-T2S3 method when the algorithm complexity and time consumption are taken into consideration. Besides all of the discrepancy analyzed above, the results obtained by methods involving reconstruction procedure do show a general trend in that increased P\'{e}clet number would always lead to an rise in $L_2$-norm error. Another point that needs to be emphasized is DUGKS implemented by upwind-based meso-flux construction scheme with third-order spatial accuracy performs poorly when the CFL number is small. As is analyzed before, small time step results in large spatial dissipation. One should try to avoid the small time step condition when utilizing this kind of method.
\\
Fig.~\ref{FIG:TD:M-L2Phi} illustrates the $L_2$-norm error of $\phi$ obtained by multiple methods with the variation of mobility coefficient. The detailed information are presented in Table~\ref{tab:TD:M-L2Phi}. Other parameters are $L_0 = 256$, $Pe = 256$ and $Cn = 4/256$. Of all the results, LBM and DVM offer the most excellent ones, which are similar to the phenomenon observed above. The DUGKS-T2S2CD method do perform better since the $Pe$ number is small. However, it cannot yet compare with the results obtained by DUGKS-T2S3 with a CFL number of 0.8, which has little difference from the results achieved by DVM at the same CFL number. The effects of CFL number on the DUGKS-T2S2CD method is not that obvious, which is because the majority part of its truncation error has no relationship with CFL number. However, in the simulations conducted by DUGKS-T2S3 and DUGKS-T3S3, it makes a very big difference. As aforementioned, the smaller the time step is, the greater the spatial dissipation will be. Hence the results obtained by those two methods with the smallest time step are always worse than others. The unobtainable results are replaced with hyphen due to the limitations of DUGKS-T3S3. Among all the results obtained by the methods involving reconstruction process, DUGKS-T2S5 still offers the most excellent ones. However, the discrepancy between the results obtained by DUGKS-T2S5 and DVM with $C = 0.8$ is almost indistinguishable. The same goes for DUGKS-T2S3 with a CFL number of 0.8. Except the results achieved by DUGKS-T2S3 and DUGKS-T3S3 with small time steps, the overall magnitude of $L_2$-norm error for different mobility coefficients is about $10^{-3}$, which demonstrates that the results are not that sensitive to $M_\phi$. It is worth pointing out that when $M_\phi$ is set to 0.01, only DUGKS-T2S3 and DUGKS-T2S5 are able to provide results that is comparable with that obtained by LBM and DVM with unit time step. Even DVM with a time step of 0.8 is failed to give satisfactory results. Taking the cost of DUGKS-T2S5 into consideration, DUGKS-T2S3 can be viewed as an alternative method of LBM or DVM.
\begin{table}
\caption
{
  \label{tab:TD:Pe-L2Phi}
  $L_2$-norm error of $\phi$ for diagonal translation obtained by multiple methods with various $Pe$, $Cn = 4/256$, $M_\phi = 0.02$.
}
	\begin{minipage}{1.0\textwidth}
	\begin{subtable}{1.0\textwidth}
	\caption{\label{tab:TD:LBMDVM-Pe-L2Phi}LBM, DVM}
		\begin{ruledtabular}
		\begin{tabular}{ccccccc}
		Pe&256&512&640&819.2&1024&1280\\
		\colrule
		LBM & $2.22\times10^{-3}$ & $2.21\times10^{-3}$ & $2.19\times10^{-3}$ & $2.17\times10^{-3}$ & $2.14\times10^{-3}$ & $2.21\times10^{-3}$\\
		DVM-CFL1.0 & $2.22\times10^{-3}$ & $2.21\times10^{-3}$ & $2.19\times10^{-3}$ & $2.17\times10^{-3}$ & $2.14\times10^{-3}$ & $2.21\times10^{-3}$\\
		DVM-CFL0.8 & $4.67\times10^{-3}$ & $5.83\times10^{-3}$ & $6.57\times10^{-3}$ & $7.76\times10^{-3}$ & $9.30\times10^{-3}$ & $1.15\times10^{-2}$\\
		\end{tabular}
		\end{ruledtabular}
	\end{subtable}%
	\end{minipage}
	\par\medskip
	\begin{minipage}{1.0\textwidth}
	\begin{subtable}{1.0\textwidth}
	\caption{\label{tab:TD:DUGKS-T2S2CD-Pe-L2Phi}DUGKS-T2S2CD}
		\begin{ruledtabular}
		\begin{tabular}{ccccccc}
		Pe&256&512&640&819.2&1024&1280\\
		\colrule
		CFL0.25 & $6.34\times10^{-3}$ & $1.38\times10^{-2}$ & $2.14\times10^{-2}$ & $3.34\times10^{-2}$ & $4.60\times10^{-2}$ & $5.90\times10^{-2}$\\
		CFL0.5 & $6.80\times10^{-3}$ & $1.69\times10^{-2}$ & $2.36\times10^{-2}$ & $3.11\times10^{-2}$ & $4.29\times10^{-2}$ & $5.36\times10^{-2}$\\
		CFL0.8 & $6.69\times10^{-3}$ & $1.46\times10^{-2}$ & $1.95\times10^{-2}$ & $2.64\times10^{-2}$ & $3.40\times10^{-2}$ & $4.25\times10^{-2}$\\
		\end{tabular}
		\end{ruledtabular}
	\end{subtable}%
	\end{minipage}
	\par\medskip
	\begin{minipage}{1.0\textwidth}
	\begin{subtable}{1.0\textwidth}
	\caption{\label{tab:TD:DUGKS-T2S3-Pe-L2Phi}DUGKS-T2S3}
		\begin{ruledtabular}
		\begin{tabular}{ccccccc}
		Pe&256&512&640&819.2&1024&1280\\
		\colrule
		CFL0.25 & $2.73\times10^{-2}$ & $2.03\times10^{-2}$ & $1.93\times10^{-2}$ & $1.87\times10^{-2}$ & $1.86\times10^{-2}$ & $1.92\times10^{-2}$\\
		CFL0.5 & $8.53\times10^{-3}$ & $9.04\times10^{-3}$ & $9.52\times10^{-3}$ & $1.04\times10^{-2}$ & $1.16\times10^{-2}$ & $1.34\times10^{-2}$\\
		CFL0.8 & $4.37\times10^{-3}$ & $5.44\times10^{-3}$ & $6.11\times10^{-3}$ & $7.15\times10^{-3}$ & $8.50\times10^{-3}$ & $1.04\times10^{-2}$\\
		\end{tabular}
		\end{ruledtabular}
	\end{subtable}%
	\end{minipage}
	\par\medskip
	\begin{minipage}{1.0\textwidth}
	\begin{subtable}{1.0\textwidth}
	\caption{\label{tab:TD:DUGKS-T3S3-Pe-L2Phi}DUGKS-T3S3}
		\begin{ruledtabular}
		\begin{tabular}{ccccccc}
		Pe&256&512&640&819.2&1024&1280\\
		\colrule
		CFL0.25 & $4.43\times10^{-2}$ & $2.87\times10^{-2}$ & $2.60\times10^{-2}$ & $2.41\times10^{-2}$ & $2.29\times10^{-2}$ & $2.25\times10^{-2}$\\
		CFL0.5 & $2.42\times10^{-2}$ & $1.84\times10^{-2}$ & $1.76\times10^{-2}$ & $1.71\times10^{-2}$ & $1.70\times10^{-2}$ & $1.78\times10^{-2}$
		\end{tabular}
		\end{ruledtabular}
	\end{subtable}%
	\end{minipage}
	\par\medskip
	\begin{minipage}{1.0\textwidth}
	\begin{subtable}{1.0\textwidth}
	\caption{\label{tab:TD:DUGKS-T2S5-Pe-L2Phi}DUGKS-T2S5}
		\begin{ruledtabular}
		\begin{tabular}{ccccccc}
		Pe&256&512&640&819.2&1024&1280\\
		\colrule
		CFL0.25 & $2.71\times10^{-3}$ & $3.82\times10^{-3}$ & $4.46\times10^{-3}$ & $5.40\times10^{-3}$ & $6.51\times10^{-3}$ & $7.82\times10^{-3}$\\
		CFL0.5 & $2.98\times10^{-3}$ & $4.11\times10^{-3}$ & $4.77\times10^{-3}$ & $5.74\times10^{-3}$ & $6.89\times10^{-3}$ & $8.29\times10^{-3}$\\
		CFL0.8 & $3.32\times10^{-3}$ & $4.49\times10^{-3}$ & $5.18\times10^{-3}$ & $6.19\times10^{-3}$ & $7.40\times10^{-3}$ & $9.05\times10^{-3}$\\
		\end{tabular}
		\end{ruledtabular}
	\end{subtable}%
	\end{minipage}
\end{table}
\\
\begin{table}
\caption
{
  \label{tab:TD:M-L2Phi}
  $L_2$-norm error of $\phi$ for diagonal translation obtained by multiple methods with various $M_\phi$, $Cn = 4/256$, $Pe = 256$.
}
	\begin{minipage}{1.0\textwidth}
	\begin{subtable}{1.0\textwidth}
	\caption{\label{tab:TD:LBMDVM-M-L2Phi}LBM, DVM}
		\begin{ruledtabular}
		\begin{tabular}{cccccccc}
		$M_\phi$&0.01&0.02&0.04&0.05&0.064&0.08&0.1\\
		\colrule
		LBM & $2.23\times10^{-3}$ & $2.22\times10^{-3}$ & $2.20\times10^{-3}$ & $2.21\times10^{-3}$ & $2.26\times10^{-3}$ & $2.42\times10^{-3}$ & $2.79\times10^{-3}$\\
		DVM-CFL1.0 & $2.23\times10^{-3}$ & $2.22\times10^{-3}$ & $2.20\times10^{-3}$ & $2.21\times10^{-3}$ & $2.26\times10^{-3}$ & $2.42\times10^{-3}$ & $2.79\times10^{-3}$\\
		DVM-CFL0.8 & $8.36\times10^{-3}$ & $4.67\times10^{-3}$ & $3.62\times10^{-3}$ & $3.42\times10^{-3}$ & $3.25\times10^{-3}$ & $3.16\times10^{-3}$ & $3.23\times10^{-3}$\\
		\end{tabular}
		\end{ruledtabular}
	\end{subtable}%
	\end{minipage}
	\par\medskip
	\begin{minipage}{1.0\textwidth}
	\begin{subtable}{1.0\textwidth}
	\caption{\label{tab:TD:DUGKS-T2S2CD-M-L2Phi}DUGKS-T2S2CD}
		\begin{ruledtabular}
		\begin{tabular}{cccccccc}
		$M_\phi$&0.01&0.02&0.04&0.05&0.064&0.08&0.1\\
		\colrule
		CFL0.25 & $7.61\times10^{-3}$ & $6.34\times10^{-3}$ & $6.36\times10^{-3}$ & $6.49\times10^{-3}$ & $6.75\times10^{-3}$ & $7.12\times10^{-3}$ & $7.66\times10^{-3}$
		\\
		CFL0.5 & $1.04\times10^{-2}$ & $6.80\times10^{-3}$ & $5.93\times10^{-3}$ & $5.94\times10^{-3}$ & $6.05\times10^{-3}$ & $6.31\times10^{-3}$ & $6.79\times10^{-3}$
		\\
		CFL0.8 & $9.51\times10^{-3}$ & $6.69\times10^{-3}$ & $5.25\times10^{-3}$ & $5.13\times10^{-3}$ & $5.11\times10^{-3}$ & $5.24\times10^{-3}$ & $5.59\times10^{-3}$
		\\
		\end{tabular}
		\end{ruledtabular}
	\end{subtable}%
	\end{minipage}
	\par\medskip
	\begin{minipage}{1.0\textwidth}
	\begin{subtable}{1.0\textwidth}
	\caption{\label{tab:TD:DUGKS-T2S3-M-L2Phi}DUGKS-T2S3}
		\begin{ruledtabular}
		\begin{tabular}{cccccccc}
		$M_\phi$&0.01&0.02&0.04&0.05&0.064&0.08&0.1\\
		\colrule
		CFL0.25 & $4.57\times10^{-2}$ & $2.73\times10^{-2}$ & $1.15\times10^{-2}$ & $8.17\times10^{-3}$ & $5.90\times10^{-3}$ & $4.91\times10^{-3}$ & $4.73\times10^{-3}$\\
		CFL0.5 & $1.29\times10^{-2}$ & $8.53\times10^{-3}$ & $5.22\times10^{-3}$ & $4.64\times10^{-3}$ & $4.25\times10^{-3}$ & $4.12\times10^{-3}$ &$4.31\times10^{-3}$\\
		CFL0.8 & $4.82\times10^{-3}$ & $4.37\times10^{-3}$ & $3.85\times10^{-3}$ & $3.72\times10^{-3}$ & $3.68\times10^{-3}$ & $3.79\times10^{-3}$ & $4.17\times10^{-3}$\\
		\end{tabular}
		\end{ruledtabular}
	\end{subtable}%
	\end{minipage}
	\par\medskip
	\begin{minipage}{1.0\textwidth}
	\begin{subtable}{1.0\textwidth}
	\caption{\label{tab:TD:DUGKS-T3S3-M-L2Phi}DUGKS-T3S3}
		\begin{ruledtabular}
		\begin{tabular}{cccccccc}
		$M_\phi$&0.01&0.02&0.04&0.05&0.064&0.08&0.1\\
		\colrule
		CFL0.25 & $7.84\times10^{-2}$ & $4.43\times10^{-2}$ & $1.68\times10^{-2}$ & $1.09\times10^{-3}$ & $6.69\times10^{-3}$ & $4.97\times10^{-3}$ & $5.06\times10^{-3}$\\
		CFL0.5 & - & $2.42\times10^{-2}$ & $9.52\times10^{-3}$ & $6.67\times10^{-3}$ & $4.91\times10^{-3}$ & $4.49\times10^{-3}$ &$4.96\times10^{-3}$\\
		CFL0.8 & - & - & $5.54\times10^{-3}$ & $4.66\times10^{-3}$ & $4.17\times10^{-3}$ & $4.13\times10^{-3}$ & $4.77\times10^{-3}$\\
		\end{tabular}
		\end{ruledtabular}
	\end{subtable}%
	\end{minipage}
	\par\medskip
	\begin{minipage}{1.0\textwidth}
	\begin{subtable}{1.0\textwidth}
	\caption{\label{tab:TD:DUGKS-T2S5-M-L2Phi}DUGKS-T2S5}
		\begin{ruledtabular}
		\begin{tabular}{cccccccc}
		$M_\phi$&0.01&0.02&0.04&0.05&0.064&0.08&0.1\\
		\colrule
		CFL0.25 & $3.36\times10^{-3}$ & $2.71\times10^{-3}$ & $2.25\times10^{-3}$ & $2.23\times10^{-3}$ & $2.35\times10^{-3}$ & $2.66\times10^{-3}$ & $3.24\times10^{-3}$\\
		CFL0.5 & $3.39\times10^{-3}$ & $2.98\times10^{-3}$ & $2.61\times10^{-3}$ & $2.58\times10^{-3}$ & $2.67\times10^{-3}$ & $2.93\times10^{-3}$ & $3.46\times10^{-3}$\\
		CFL0.8 & $3.59\times10^{-3}$ & $3.32\times10^{-3}$ & $3.07\times10^{-3}$ & $3.05\times10^{-3}$ & $3.14\times10^{-3}$ & $3.38\times10^{-3}$ & $3.86\times10^{-3}$\\
		\end{tabular}
		\end{ruledtabular}
	\end{subtable}%
	\end{minipage}
\end{table}
\\
By far we have presented and analyzed the comparative results obtained by multiple methods. It can be concluded that LBM and DVM with unit time step would always present identical results, which are insensitive to P\'{e}clet number and mobility coefficient. The most cost-effective methods involving reconstruction process are DUGKS-T2S3 and DVM at the condition of large CFL number. DUGKS-T2S2CD does not provide any result better than DUGKS-T2S3 with a CFL number of 0.8 in this test. DUGKS-T3S3 is restricted by the native limitation of ${\Delta}t < 12\tau$, hence it could not exploit the advantages of DUGKS to the full. DUGKS-T2S5 offers the results with minimum dissipation, but the expensive cost spent on computation time and resources makes it less convenient.
\\
Aside from the comparative analysis given above, we here would like to deviate a little from our main subject to emphasize the importance of isotropic finite-difference scheme. To evaluate the unit vector normal to the interface in Eq.~(\ref{Eq:AC:Normal}), general finite-difference scheme needs to be utilized to calculate the value of $\nabla{\phi}$. Fig.~\ref{FIG:TD:Isotropy} illustrates the interface shapes obtained by LBM and DUGKS-T2S3 implemented with isotropy and anisotropy finite-difference schemes, respectively. It can be observed that there are apparent discrepancies between those shapes. The interface shown in Fig.~\ref{FIG:TD:LBMDVM-Isotropy} and~\ref{FIG:TD:DUGKST2S3-Isotropy}, obtained by LBM and DUGKS which take the property of isotropy into consideration, are in good agreement with the initial shape. The other results, achieved by methods ignoring the isotropy property, present shapes of diamond. It is self-evident that the isotropy property of the finite-difference scheme is crucial to physical symmetry of interfaces.
\\
\\
\subsection{\label{sec:sec3.B}Zalesak's disk rotation}
\begin{figure}[htbp]
    \centering
    \begin{subfigure}[b]{0.400\textwidth}
      \centering
      \includegraphics[width=1\linewidth]{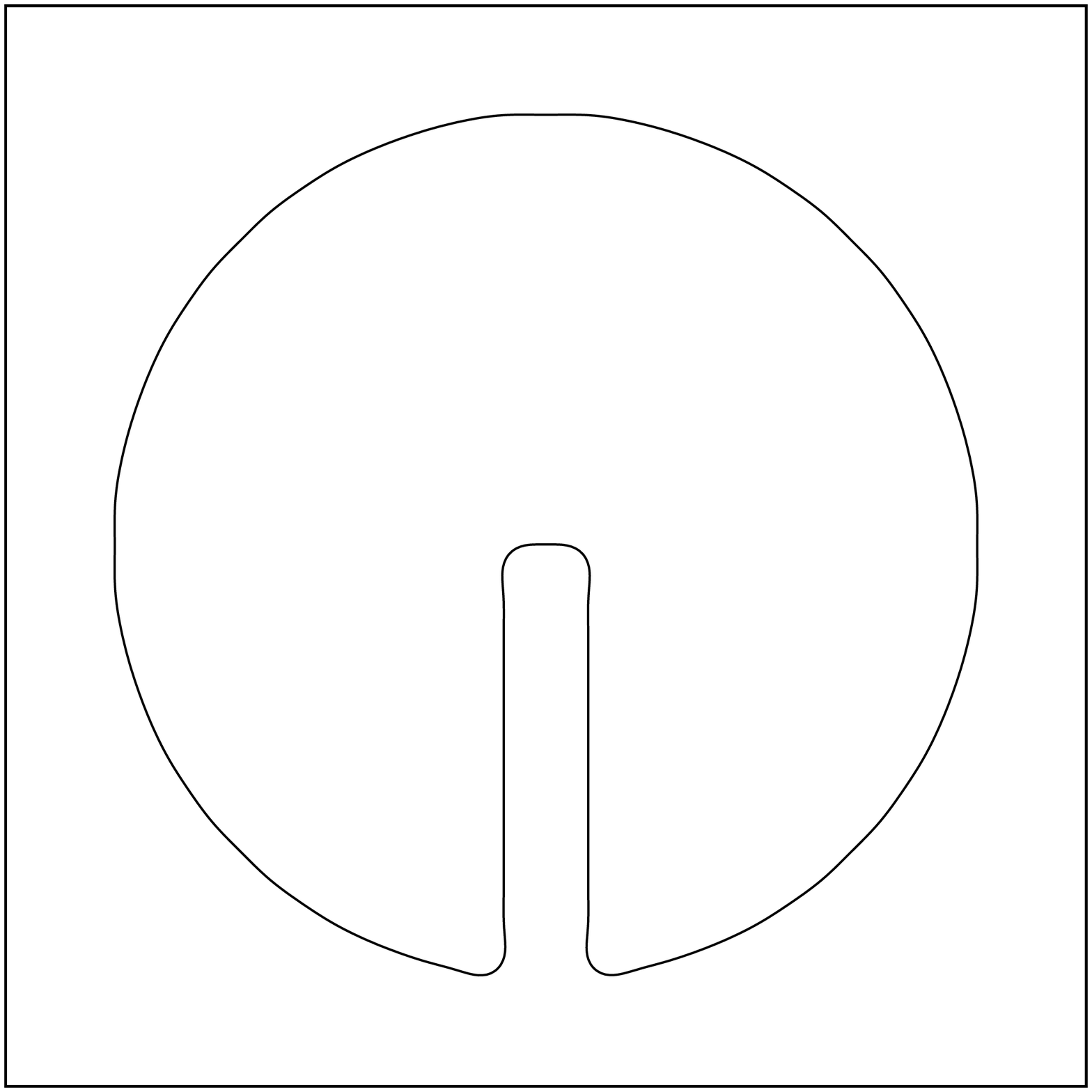}
      \caption[DVM,CFL=1.0]{{\scriptsize DVM, $C = 1.0$}}
      \label{FIG:ZD:DVM-Pe1024M0.02CFL1.0}
    \end{subfigure}
    \begin{subfigure}[b]{0.400\textwidth}
      \centering
      \includegraphics[width=1\linewidth]{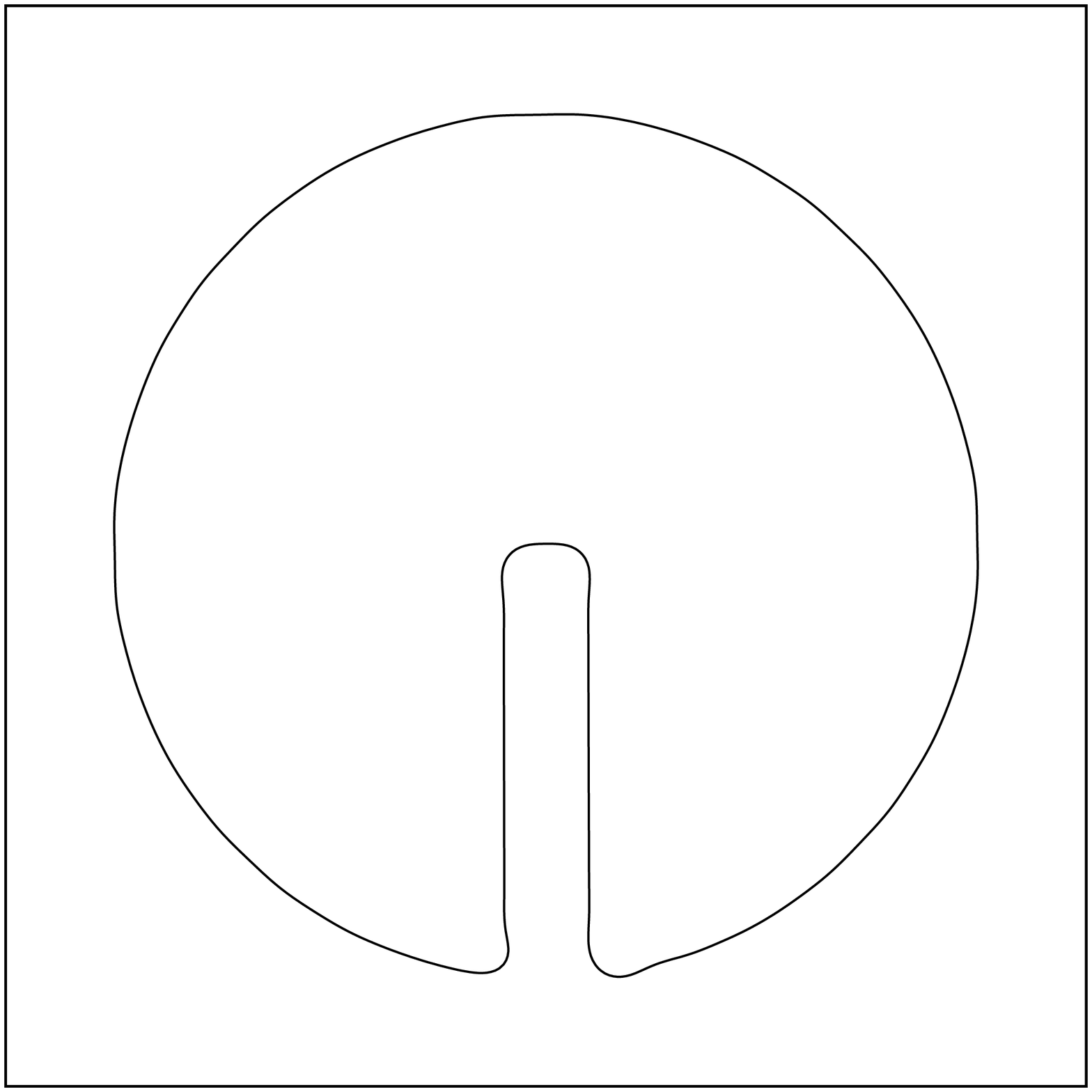}
      \caption[]{DVM, $C = 0.8$}
      \label{FIG:ZD:DVM-Pe1024M0.02CFL0.8}
    \end{subfigure}
    \hfill
    \begin{subfigure}{0.400 \columnwidth}
      \centering
      \includegraphics[width=1\linewidth]{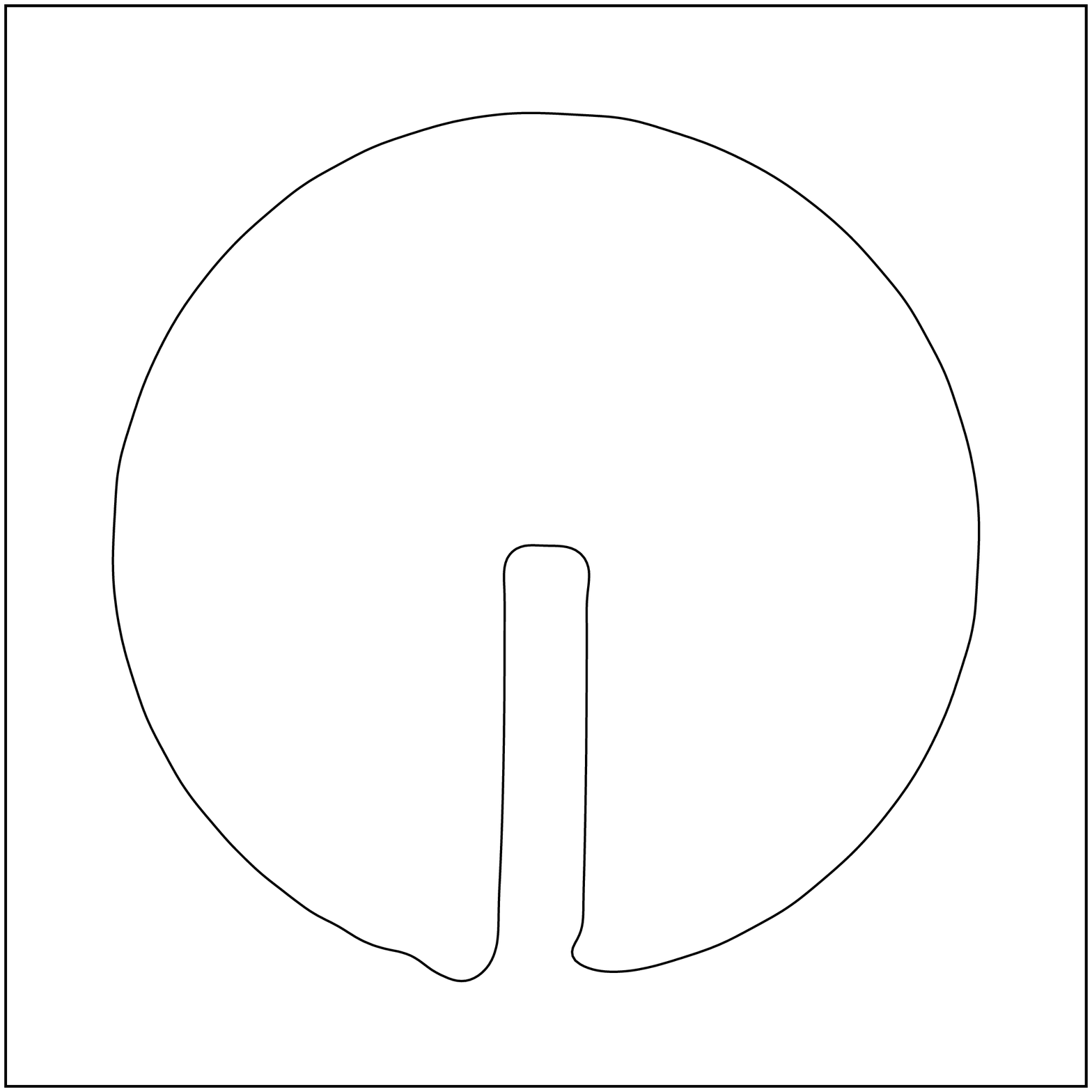}
      \caption[]{DUGKS-T2S2CD, $C = 0.25$}
      \label{FIG:ZD:DUGKST2S2CD-Pe1024M0.02CFL0.25}
    \end{subfigure}
    \begin{subfigure}{0.400 \columnwidth}
      \centering
      \includegraphics[width=1\linewidth]{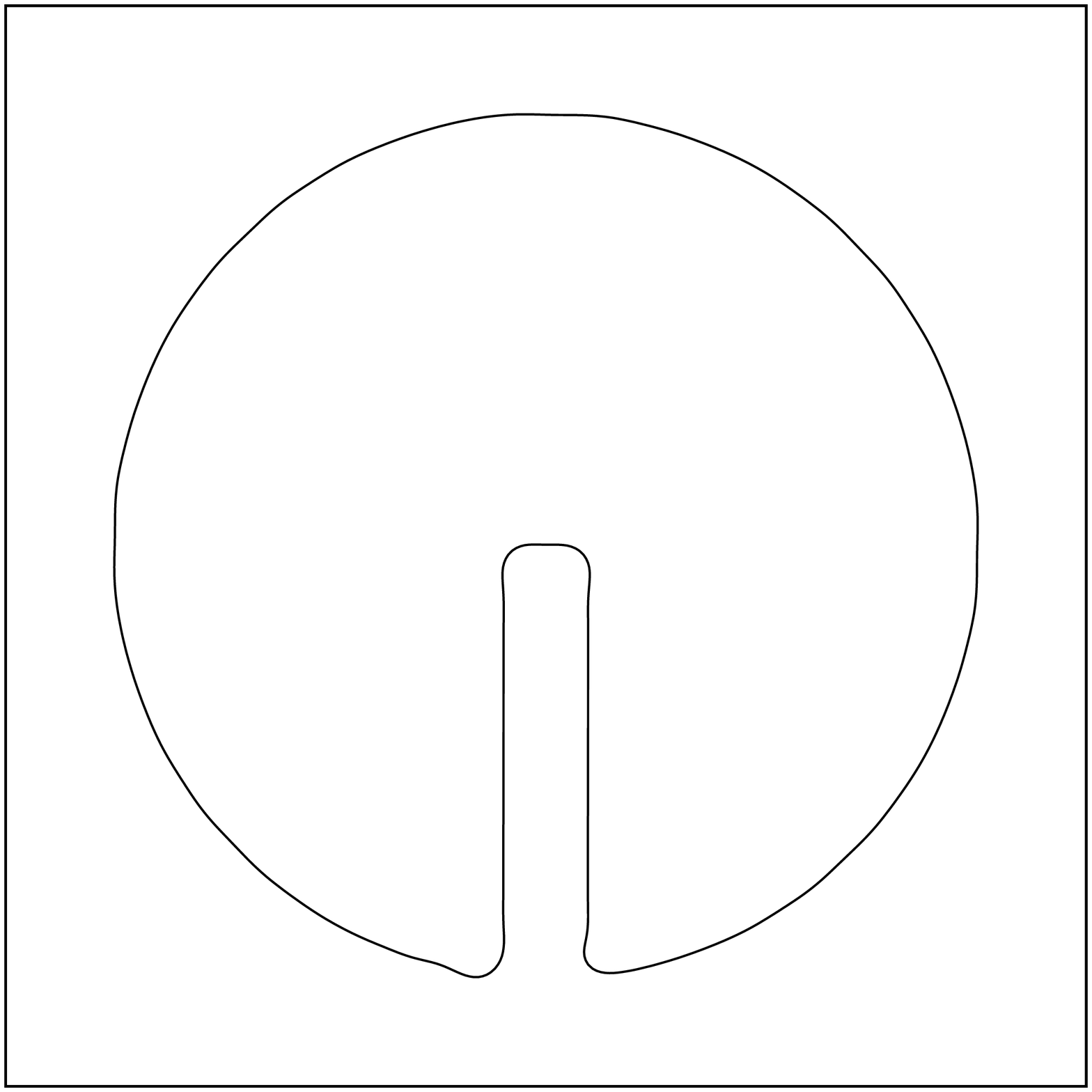}
      \caption[]{DUGKS-T2S5, $C = 0.25$}
      \label{FIG:ZD:DUGKST2S5-Pe1024M0.02CFL0.25}
    \end{subfigure}
    \hfill
    \begin{subfigure}{0.400 \columnwidth}
      \centering
      \includegraphics[width=1\linewidth]{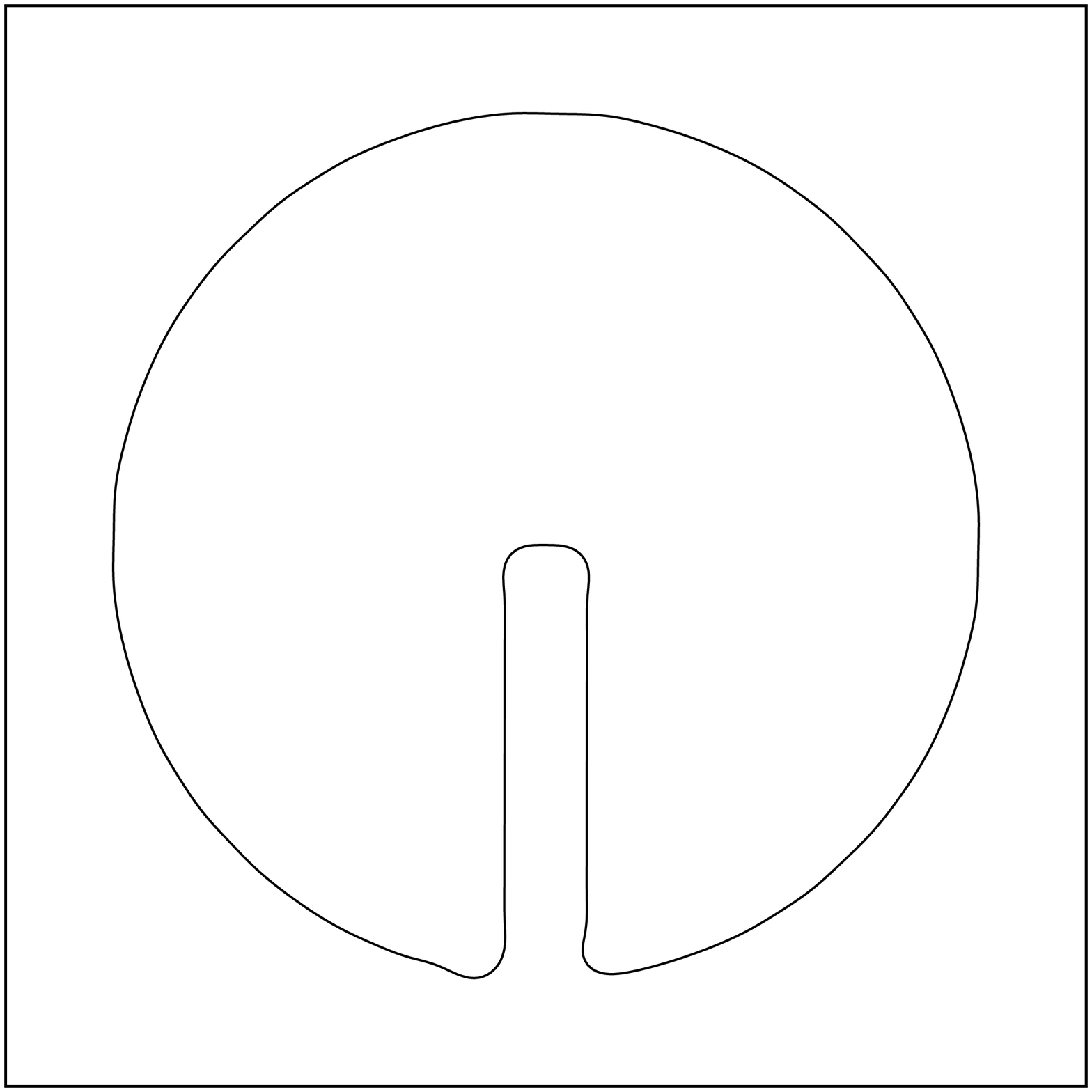}
      \caption[]{DUGKS-T2S3, $C = 0.8$}
      \label{FIG:ZD:DUGKST2S3-Pe1024M0.02CFL0.8}
    \end{subfigure}
    \begin{subfigure}{0.400 \columnwidth}
      \centering
      \includegraphics[width=1\linewidth]{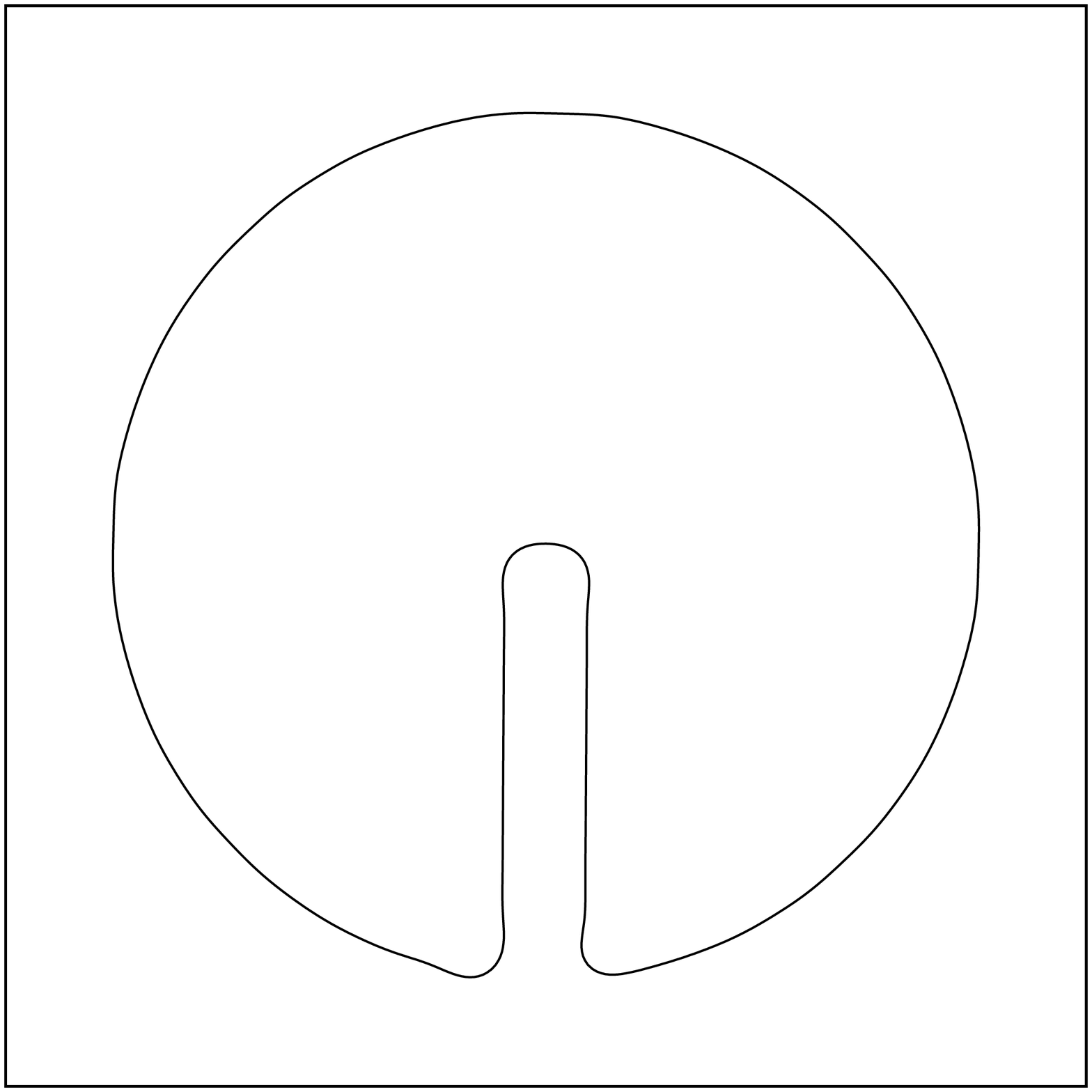}
      \caption{DUGKS-T3S3, $C = 0.5$}
      \label{FIG:ZD:DUGKST3S3-Pe1024M0.02CFL0.5}
    \end{subfigure}
\caption{Interface shape of Zalesak's disk obtained by multiple methods at $Pe = 1024$, $Cn = 4/256$, $M_\phi = 0.02$.}
\label{FIG:ZD:InterfaceShapePe1024M0.02}
\end{figure}

\begin{figure}[htbp]
    \centering
    \begin{minipage}[b]{0.9 \columnwidth}
    \begin{subfigure}{0.45 \columnwidth}
      \centering
      \includegraphics[width=1\linewidth]{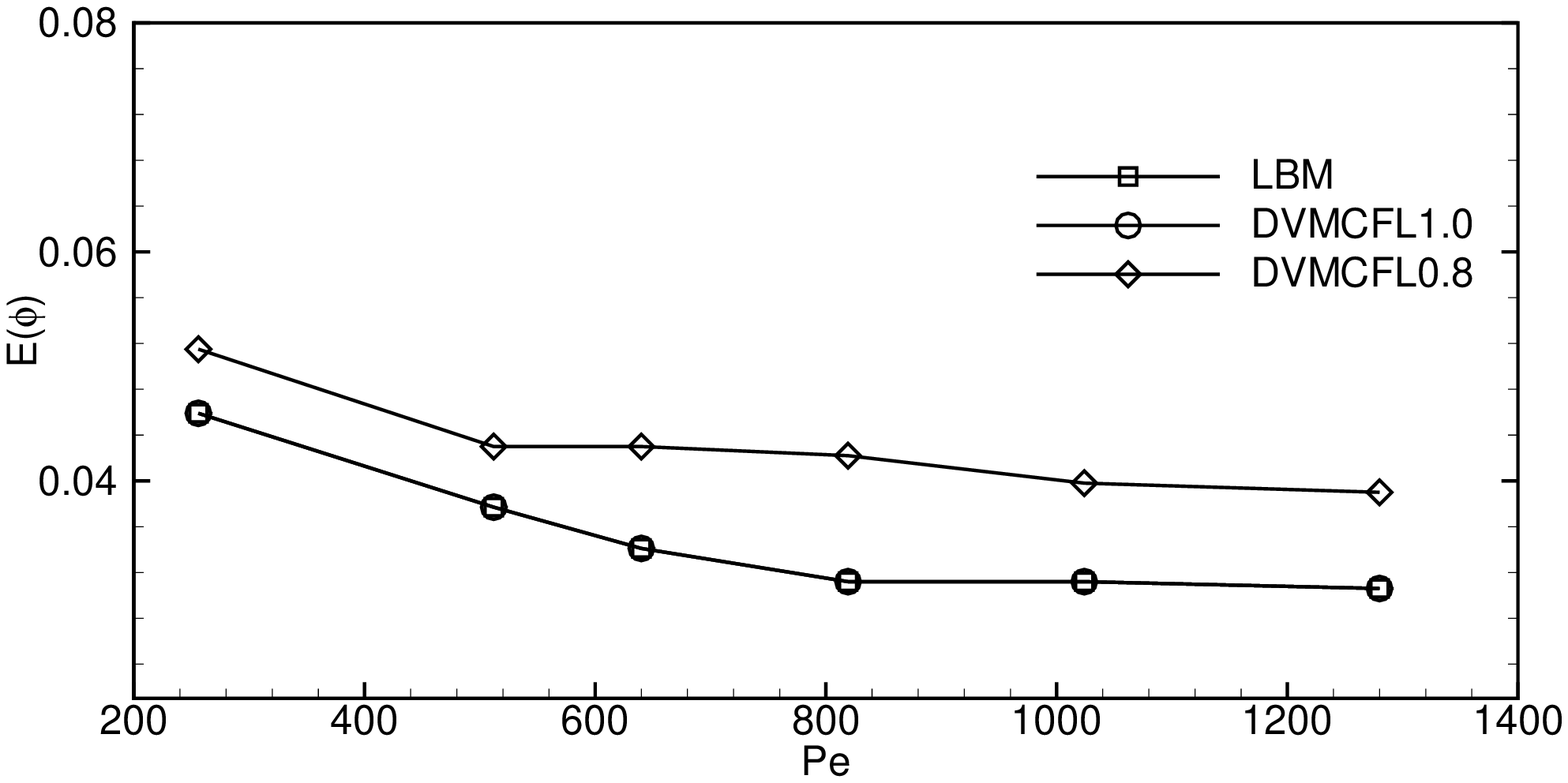}
      \caption{LBM, DVM}
      \label{FIG:ZD:LBMDVM-Pe-L2Phi}
    \end{subfigure}
    \end{minipage}
    \begin{minipage}[b]{0.9 \columnwidth}
    \begin{subfigure}{0.45 \columnwidth}
      \centering
      \includegraphics[width=1\linewidth]{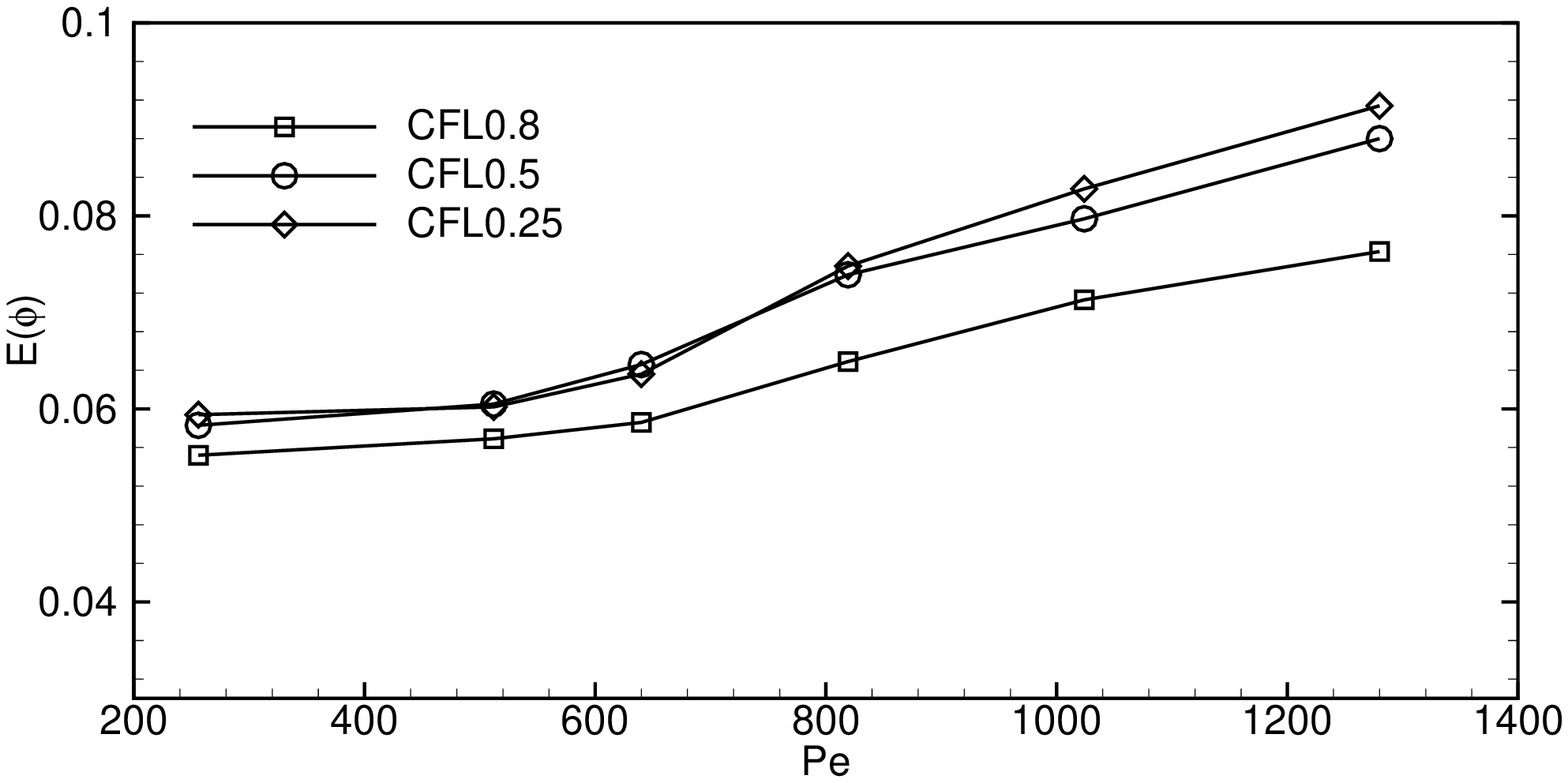}
      \caption{DUGKS-T2S2CD}
      \label{FIG:ZD:DUGKST2S2CD-Pe-L2Phi}
    \end{subfigure}
    \end{minipage}
    \begin{minipage}[b]{0.9 \columnwidth}
    \begin{subfigure}{0.45 \columnwidth}
      \centering
      \includegraphics[width=1\linewidth]{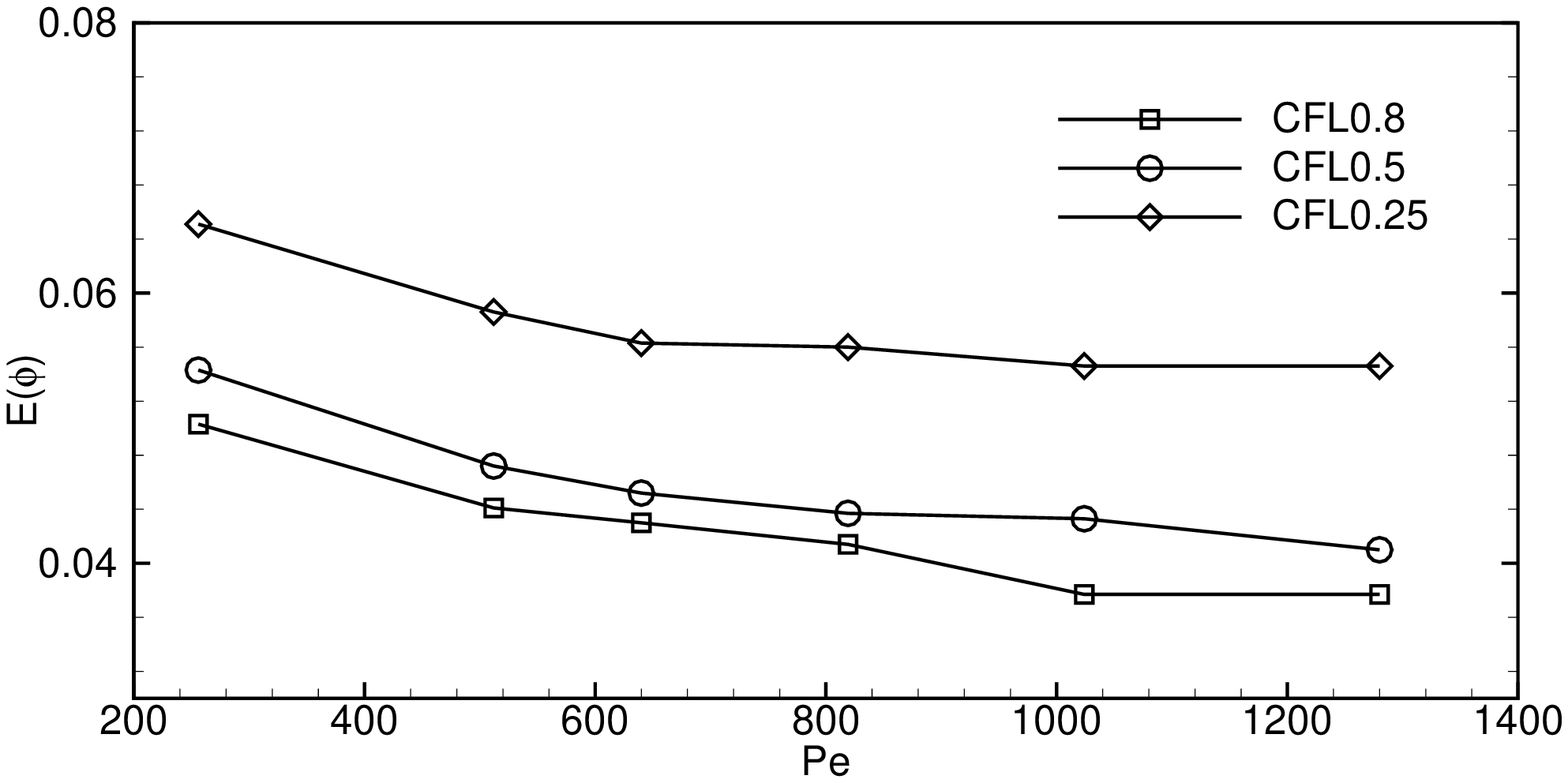}
      \caption{DUGKS-T2S3}
      \label{FIG:ZD:DUGKST2S3-Pe-L2Phi}
    \end{subfigure}
    \end{minipage}
    \begin{minipage}[b]{0.9 \columnwidth}
    \begin{subfigure}{0.45 \columnwidth}
      \centering
      \includegraphics[width=1\linewidth]{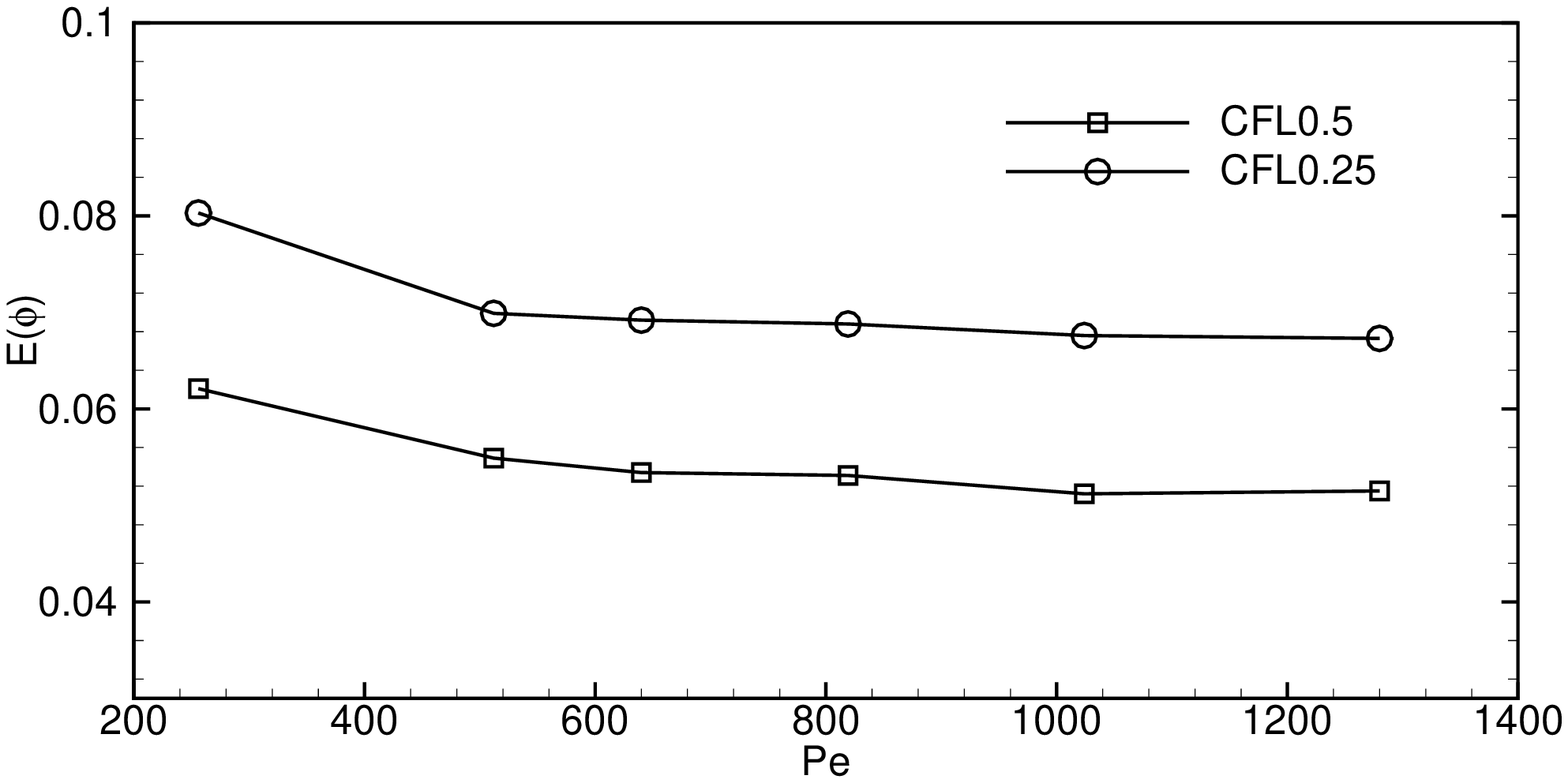}
      \caption{DUGKS-T3S3}
      \label{FIG:ZD:DUGKST3S3-Pe-L2Phi}
    \end{subfigure}
    \end{minipage}
    \begin{minipage}[b]{0.9 \columnwidth}
    \begin{subfigure}{0.45 \columnwidth}
      \centering
      \includegraphics[width=1\linewidth]{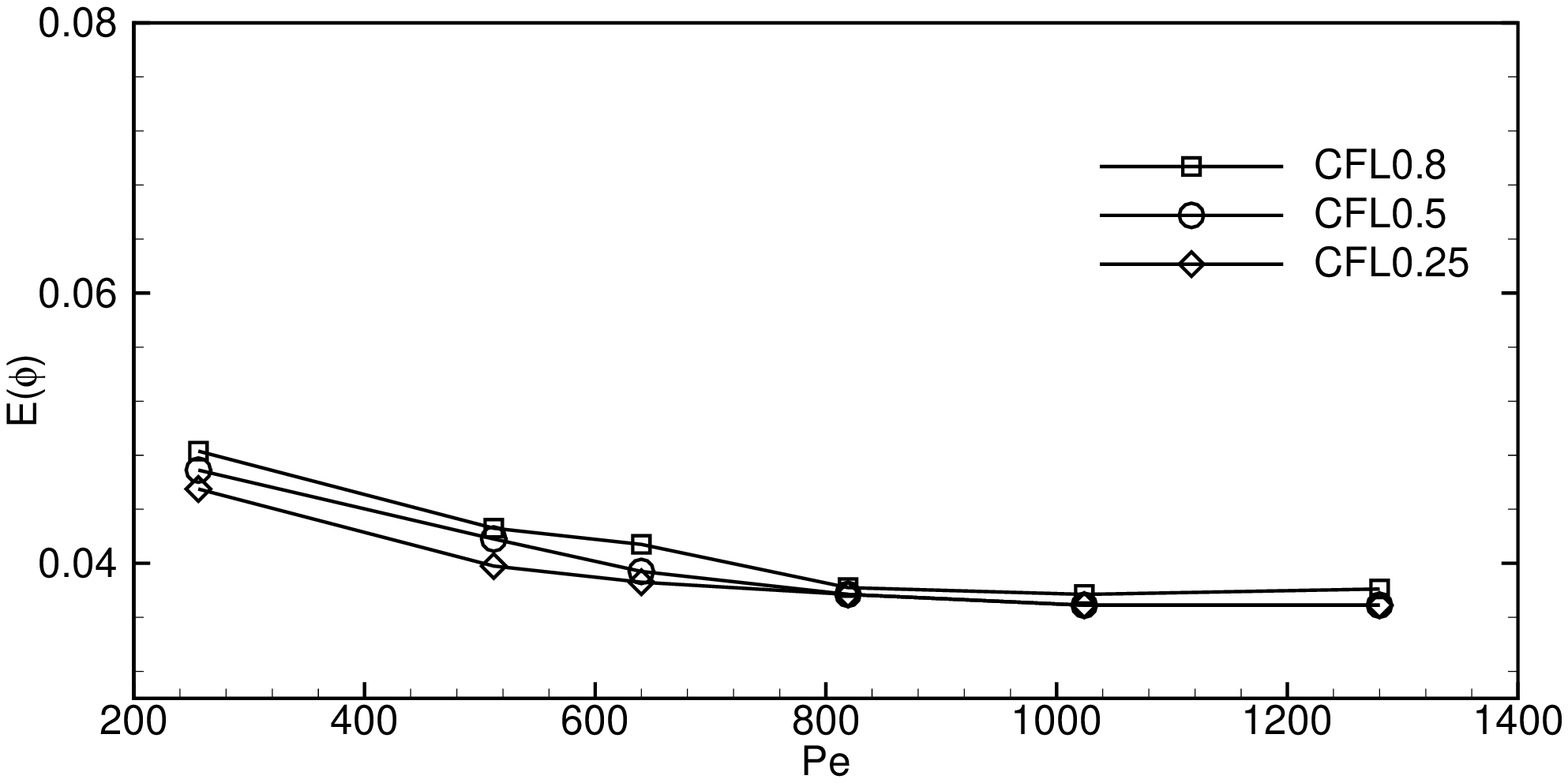}
      \caption{DUGKS-T2S5}
      \label{FIG:ZD:DUGKST2S5-Pe-L2Phi}
    \end{subfigure}
    \end{minipage}
\caption{$L_2$-norm error of $\phi$ for Zalesak's disk obtained by multiple methods with various $Pe$, $Cn = 4/256$, $M_\phi = 0.02$.}
\label{FIG:ZD:Pe-L2Phi}
\end{figure}

\begin{figure}[htbp]
    \centering
    \begin{minipage}[b]{0.9 \columnwidth}
    \begin{subfigure}{0.45 \columnwidth}
      \centering
      \includegraphics[width=1\linewidth]{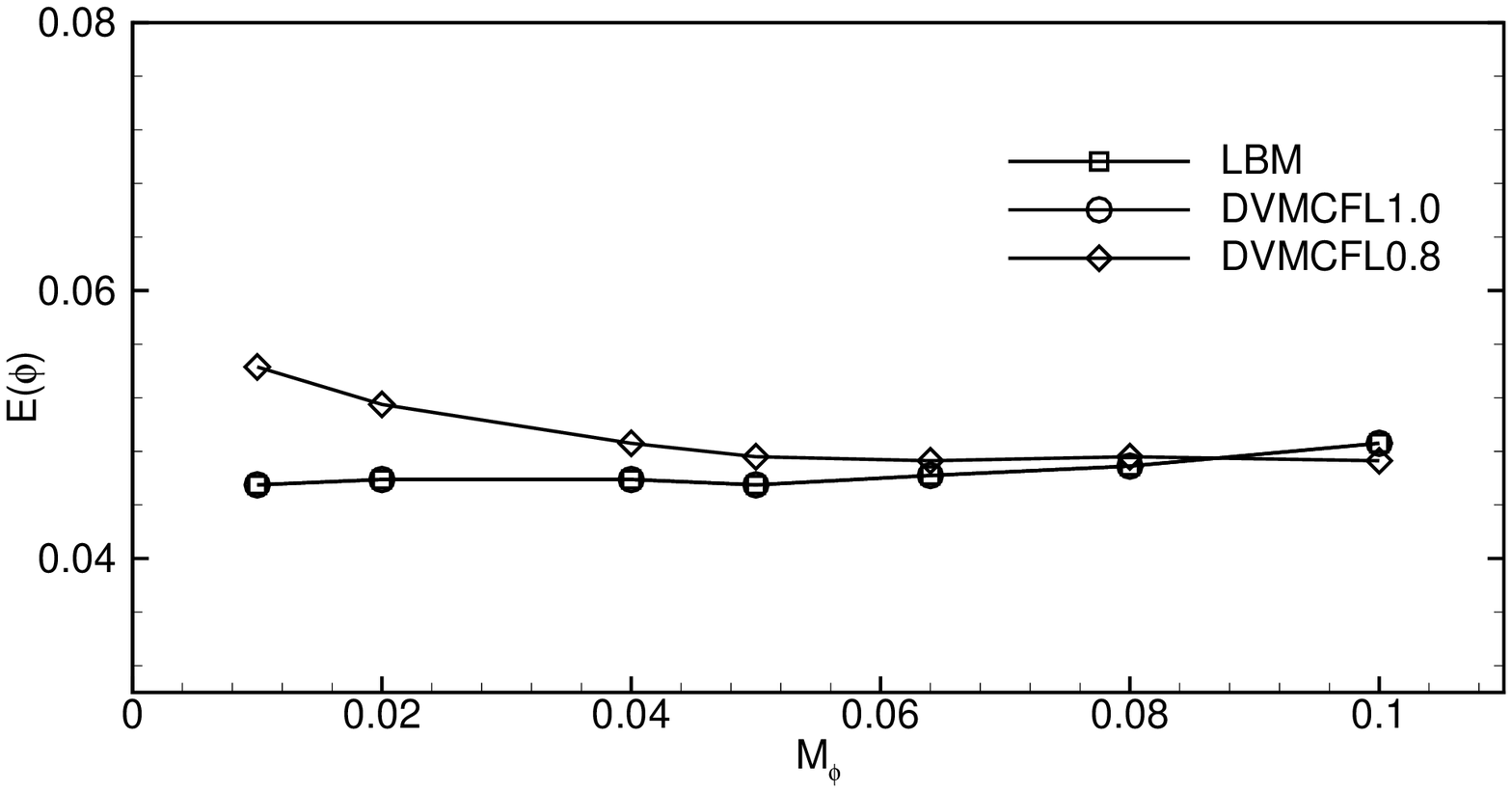}
      \caption{LBM, DVM}
      \label{FIG:ZD:LBMDVM-M-L2Phi}
    \end{subfigure}
    \end{minipage}
    \begin{minipage}[b]{0.9 \columnwidth}
    \begin{subfigure}{0.45 \columnwidth}
    \centering
      \centering
      \includegraphics[width=1\linewidth]{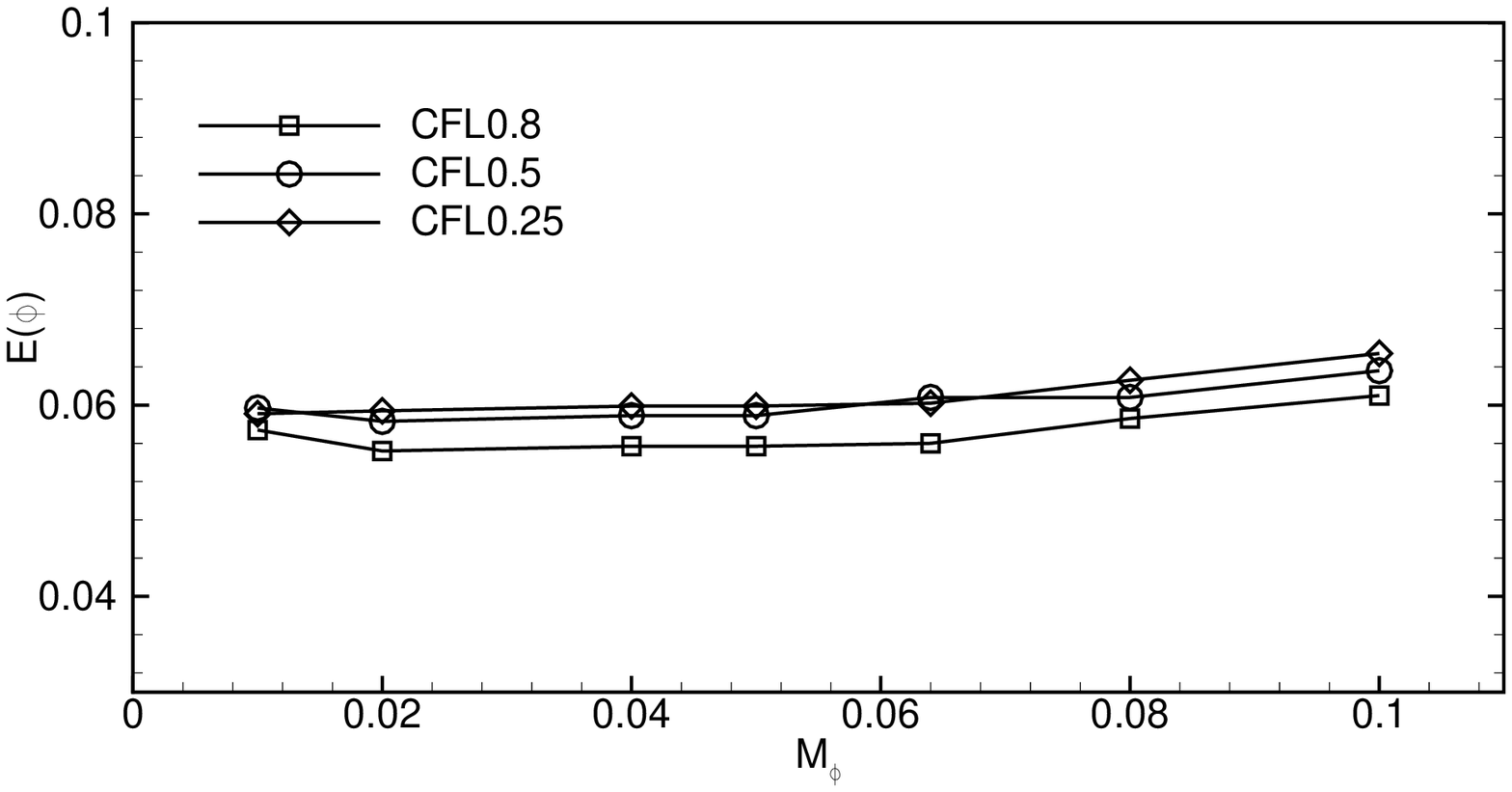}
      \caption{DUGKS-T2S2CD}
      \label{FIG:ZD:DUGKST2S2CD-M-L2Phi}
    \end{subfigure}
    \end{minipage}
    \begin{minipage}[b]{0.9 \columnwidth}
    \begin{subfigure}{0.45 \columnwidth}
      \centering
      \includegraphics[width=1\linewidth]{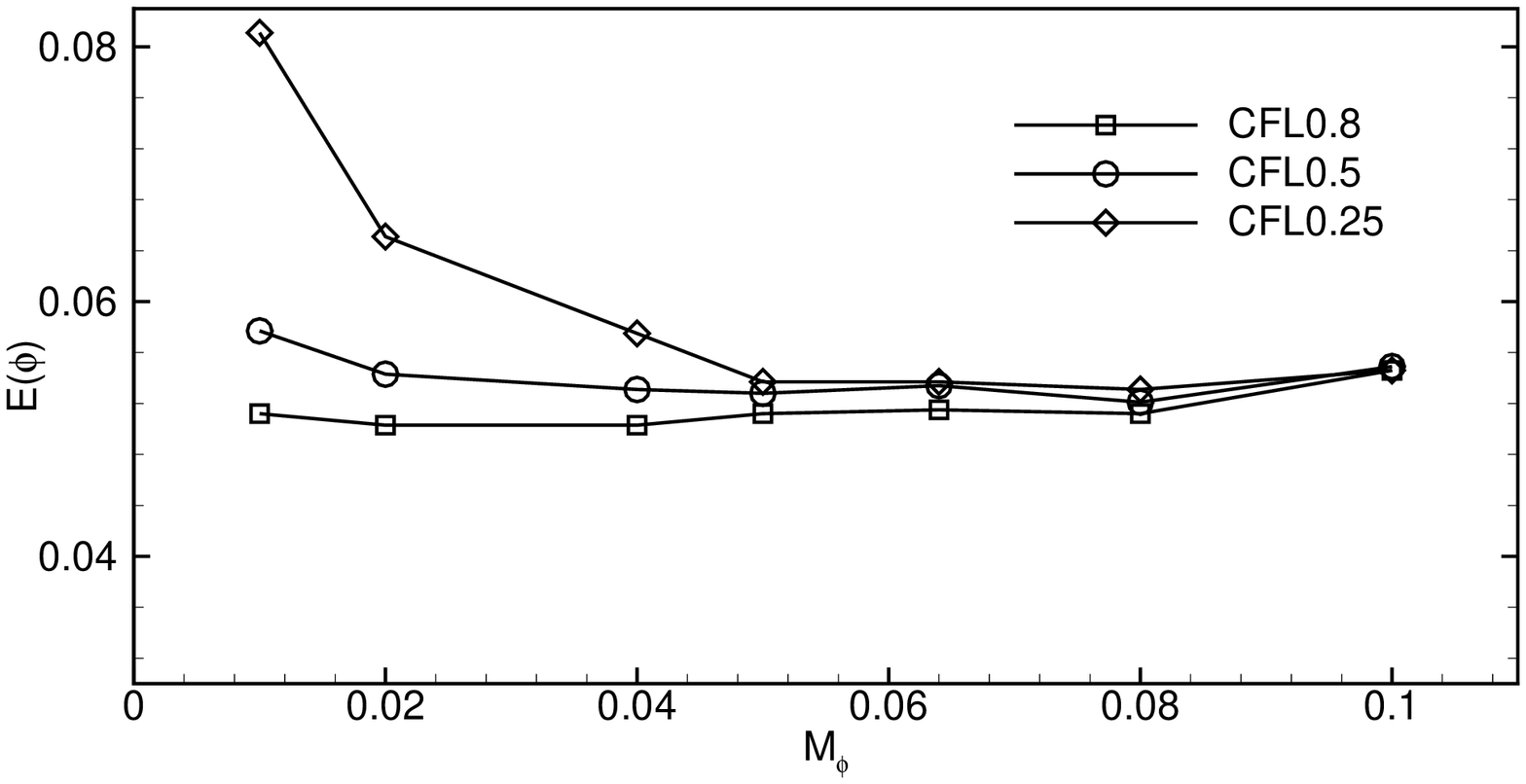}
      \caption{DUGKS-T2S3}
      \label{FIG:ZD:DUGKST2S3-M-L2Phi}
    \end{subfigure}
    \end{minipage}
    \begin{minipage}[b]{0.9 \columnwidth}
    \begin{subfigure}{0.45 \columnwidth}
      \centering
      \includegraphics[width=1\linewidth]{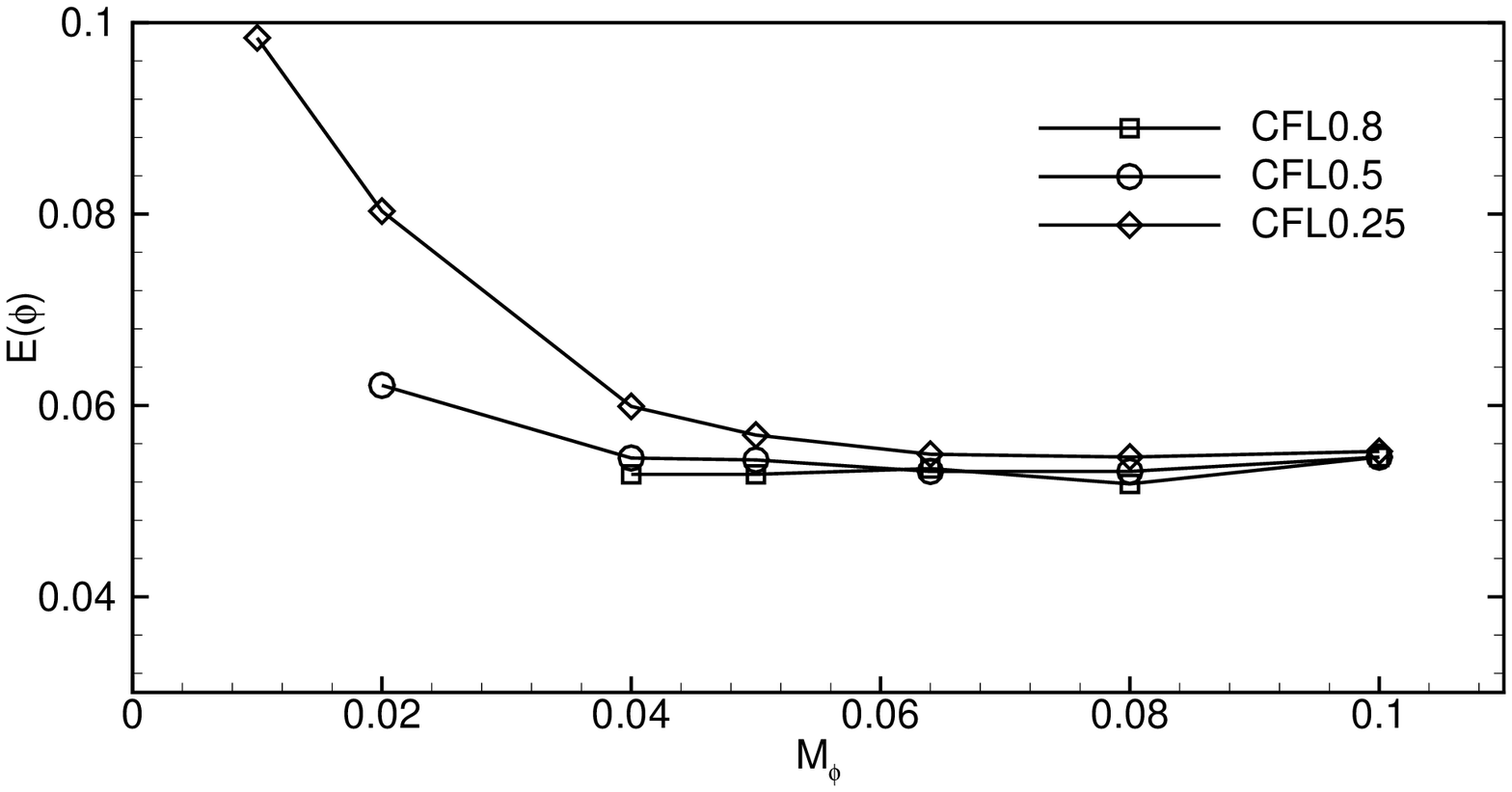}
      \caption{DUGKS-T3S3}
      \label{FIG:ZD:DUGKST3S3-M-L2Phi}
    \end{subfigure}
    \end{minipage}
    \begin{minipage}[b]{0.9 \columnwidth}
    \begin{subfigure}{0.45 \columnwidth}
      \centering
      \includegraphics[width=1\linewidth]{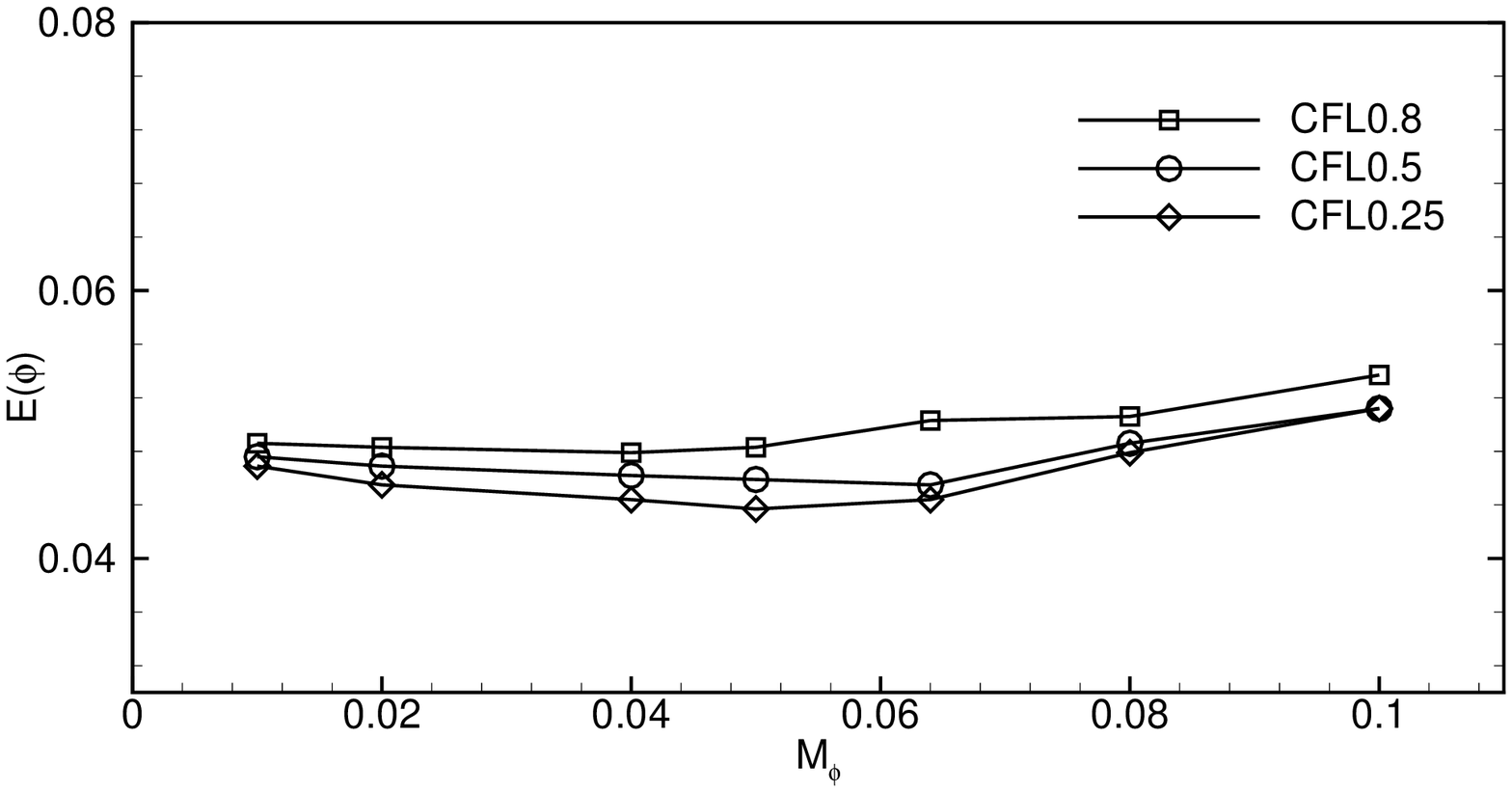}
      \caption{DUGKS-T2S5}
      \label{FIG:ZD:DUGKST2S5-M-L2Phi}
    \end{subfigure}
    \end{minipage}
\caption{$L_2$-norm error of $\phi$ for Zalesak's disk obtained by multiple methods with various $M_\phi$, $Cn = 4/256$, $Pe = 256$.}
\label{FIG:ZD:M-L2Phi}
\end{figure}

In this subsection, the rotating Zalesak's disk \cite{Liang2014PRE} is used to examine the performance of multiple methods on capturing sharp interfaces. The sharpness level of the interface is able to evaluate the dissipation of the method. A disk with a slot is initially placed at the center of the computational domain with $L_0{\times}L_0$ cells.The radius of the disk is 0.4$L_0$ and the width of the slot, orienting downward, is 0.08$L_0$. The disk is driven by a velocity field of constant vorticity,
\begin{equation}
        u(x,y)=-\frac{U_0\pi}{L_0}(y-0.5L_0),
        v(x,y)=\frac{U_0\pi}{L_0}(x-0.5L_0).
\end{equation}
Theoretically, the disk will return to its initial position after $T = 2T_f$ time. A well-designed scheme would maintain the distribution of $\phi$ close to its initial state as far as possible. In this test, $\phi$ is initialized as $\phi_A$ in the cells surrounded by the disk and $\phi_B$ outside the disk. Hence the thickness of the interface at initial time is zero. To make rational comparison between the final results and initial distribution of $\phi$, a simple step function is utilized to redistribute the index parameter at the final moment. Thus, the effects of the evolved interface with a certain width will be eliminated. The dimensionless parameters are set as $L_0 = 256$ and $Cn = 4/256$.
\\
Various interface shapes of Zalesak' disk evolved after a single period are illustrated in Fig.~\ref{FIG:ZD:InterfaceShapePe1024M0.02}. Those shapes hold the similar pattern except the sharp corners located around the slot. The physical mechanism behind this phenomenon is that the curvature of interface far away from the slot is pretty close to the equilibrium value while the interface near the slot is considerably different from its equilibrium state. Hence, sharp corners around the slot are the touchstone for the performance of various methods. Since LBM and DVM-CFL1.0 provide almost identical results, we here would not show the interface shape achieved by LBM. Due to the low dissipation of DVM-CFL1.0, the interface shape shown in Fig.~\ref{FIG:ZD:DVM-Pe1024M0.02CFL1.0} is almost identical to its initial state. When it comes to Fig.~\ref{FIG:ZD:DVM-Pe1024M0.02CFL0.8}, the result obtained by DVM-CFL0.8, it can be observed that sharp corner at the tip of the slot is skewed to a slight extent. This is caused by the error introduced in the process of interpolation, which has been explained in the former subsection. The result of DUGKS-T2S2CD method, shown in Fig.~\ref{FIG:ZD:DUGKST2S2CD-Pe1024M0.02CFL0.25}, is different from the result shown in Fig.~\ref{FIG:ZD:DVM-Pe1024M0.02CFL1.0} apparently. The twisted sharp corners at the tip of the slot have exemplified the weakness capability of this method in the condition of large P\'{e}clet number. DUGKS-T2S5 offers a results that is little different from the result presented in Fig.~\ref{FIG:ZD:DVM-Pe1024M0.02CFL1.0}. Taking a close look the sharp corners at the tip of the slot, we can indeed observe a bit of discrepancies between those two shapes. Nevertheless, it is too indistinguishable to be paid attention to. Both DUGKS-T2S3 and DUGKS-T3S3 provide results comparable to that obtained by DVM-CFL1.0. The CFL number of DUGKS-T3S3 is 0.5 due to the restriction of $\Delta{t} < 12\tau$, which reminds us of the inherent limitation of this method.
\\
\\
Next let us take a look at the effects of P\'{e}clet number and CFL number. Fig.~\ref{FIG:ZD:Pe-L2Phi} illustrates the results obtained by multiple methods varying in CFL number and reconstruction schemes. The detailed information corresponding to this figure is presented in Table~\ref{tab:ZD:Pe-L2Phi}. LBM and DVM-CFL1.0 take the lead in the performance of capturing sharp interface, followed by DUGKS-T2S5, DUGKS-T2S3 and DUGKS-T2S2CD. The differences between results obtained by DUGKS-T2S5 ($C = 0.25$) and results provided by DUGKS-T2S3 ($C = 0.8$) are so tiny that it is practically negligible. With an increase in P\'{e}lect number, the $L_2$-norm error tend to decrease in the results presented by DUGKS-T2S5 and DUGKS-T2S3, which can be attributed to the upwind scheme utilized in the evaluation of meso-flux. DUGKS-T3S3 still fails to exploit its advantage because of the restriction on time step. As the P\'{e}lect number increases, DUGKS-T2S2CD offers the worst results, which is rational since central scheme are poor at coping with flows dominated by advection. As for the effects of CFL number, it can be generalized that higher CFL number would result in smaller $L_2$-norm error, except for DUGKS-T2S5. As is mentioned above, an increase in time step would reduce the spatial dissipation of DUGKS implemented with upwind-based meso-flux construction method. Hence, there is no doubt that DUGKS-T2S3 and DUGKS-T3S3 present such a tendency. The most confusing thing is why DUGKS-T2S2CD has shown the same tendency. The answer lies in the unsteady evolution process of Zalesak's disk. The initial shape of Zalesak's disk is far different from the equilibrium state. The curvature of the sharp corners around the slot has changed dramatically. To evolve toward the equilibrium state, these acute corners tend to become smooth and soft gradually. Since the terminal time of this simulation has been set, small CFL number indicates more time steps and the accumulated error caused by the spatial dissipation would grow as well. Hence, DUGKS-T2S2CD with a larger CFL number behaves better than that with smaller ones.\\
When the numerical dissipation in space has been reduced to a lower level, the temporal discretization error comes into play. Small CFL number means short time step, which in turn leads to a reduce in temporal dissipation. This is exactly what the results of DUGKS-T2S5 shown in Fig.~\ref{FIG:ZD:DUGKST2S5-Pe-L2Phi} and Table~\ref{tab:ZD:DUGKS-T2S5-Pe-L2Phi} tells. The results obtained by DVM with a CFL number of 0.8 (DVM-CFL0.8) have shown higher dissipation than that produced by DVM-CFL1.0, which is rational since the process of interpolation and reconstruction introduces new spatial dissipation. Another peculiar phenomenon that needs explanation is the duplicate figures shown constantly in Table~\ref{tab:ZD:Pe-L2Phi}. This is due to the step function applied to the index parameter in the final moment. There $\phi$ is reset to be either $\phi_A$ or $\phi_B$ dependent on the disparity between the value of $\phi$ and $(\phi_A + \phi_B)/2$. In this way, the probability of obtaining same index distribution at different conditions tends to rise up. This can also account for the similar phenomenon shown in Table~\ref{tab:ZD:M-L2Phi}.
\\
Fig.~\ref{FIG:ZD:M-L2Phi} illustrates the $L_2$-norm error of $\phi$ for Zalesak's disk obtained by multiple methods with various mobility coefficients. The detailed information is provided by Table~\ref{tab:ZD:M-L2Phi}. It can be observed that the results provide by DUGKS-T2S5 with a CFL number of 0.25 have advantages over that of LBM when the simulation is conducted at moderate $M_\phi$ (0.02 - 0.064). And the difference between them is about $1.0\times10^{-3}$. When the mobility coefficient is pretty small (0.01) or extremely large (0.1), LBM and DVM-CFL1.0 regain the dominance. The results of DUGKS-T2S3 ($C = 0.8$) are a little worse than those obtained by DUGKS-T2S5 by an average value of $5.0\times10^{-3}$. DUGKS-T3S3 ($C = 0.8$) offers results comparable with that obtained by DUGKS-T2S3 ($C = 0.8$) and the absolute difference is no more than $2.0\times10^{-3}$. However, due to the limitation of DUGKS-T3S3, results at small mobility coefficient are missing. Among the results produced by DUGKS-T2S2CD, even the best ones ($C = 0.8$) are inferior to the results obtained by DUGKS-T2S3 ($C = 0.8$), which testifies that the capability of DUGKS-T2S2CD could not compare with that of DUGKS-T2S3 even at the condition of low P\'{e}clet number. It should be noticed that the majority of the lines shown in Fig.~{\ref{FIG:ZD:M-L2Phi}} are flat, which indicates that the value of $L_2$-norm error is insensitive to the variation of mobility coefficient. As for the effects of CFL number, it is easy to reach the same conclusion mentioned above. Both DUGKS-T2S3 and DUGKS-T2S3 prefer a large CFL number, which could reduce the accumulated error caused by the spatial dissipation. DUGKS-T2S5 performs better when the CFL number is small, which is due to the combined action of tiny spatial dissipation of $O({\Delta}x^5)$ and temporal dissipation of $O({\Delta}t^2)$. The reason why DUGKS-T2S2CD tends to provide better results when the CFL number is larger has been explained in the former paragraph. DVM-CFL0.8, behaving as expected, provides worse results than that obtained by LBM and DVM-CFL1.0 because of the numerical dissipation introduced in the process of reconstruction.
\\
\begin{table}[htbp]
\caption
{
  \label{tab:ZD:Pe-L2Phi}
  $L_2$-norm error of $\phi$ for Zalesak's disk obtained by multiple methods with various $Pe$, $Cn = 4/256$, $M_\phi = 0.02$.
}
	\begin{minipage}{1.0\textwidth}
	\begin{subtable}{1.0\textwidth}
	\caption{\label{tab:ZD:LBMDVM-Pe-L2Phi}LBM, DVM}
		\begin{ruledtabular}
		\begin{tabular}{ccccccc}
		Pe&256&512&640&819.2&1024&1280\\
		\colrule
		LBM & $4.59\times10^{-2}$ & $3.77\times10^{-2}$ & $3.41\times10^{-2}$ & $3.12\times10^{-2}$ & $3.12\times10^{-2}$ & $3.06\times10^{-2}$\\
		DVM-CFL1.0 & $4.59\times10^{-2}$ & $3.77\times10^{-2}$ & $3.41\times10^{-2}$ & $3.12\times10^{-2}$ & $3.12\times10^{-2}$ & $3.06\times10^{-2}$\\
		DVM-CFL0.8 & $5.15\times10^{-2}$ & $4.30\times10^{-2}$ & $4.30\times10^{-2}$ & $4.22\times10^{-2}$ & $3.98\times10^{-2}$ & $3.90\times10^{-2}$\\
		\end{tabular}
		\end{ruledtabular}
	\end{subtable}%
	\end{minipage}
	\par\medskip
	\begin{minipage}{1.0\textwidth}
	\begin{subtable}{1.0\textwidth}
	\caption{\label{tab:ZD:DUGKS-T2S2CD-Pe-L2Phi}DUGKS-T2S2CD}
		\begin{ruledtabular}
		\begin{tabular}{ccccccc}
		Pe&256&512&640&819.2&1024&1280\\
		\colrule		
		CFL0.25 & $5.94\times10^{-2}$ & $6.02\times10^{-2}$ & $6.36\times10^{-2}$ & $7.48\times10^{-2}$ & $8.28\times10^{-2}$ & $9.14\times10^{-2}$\\
		CFL0.5 & $5.83\times10^{-2}$ & $6.05\times10^{-2}$ & $6.46\times10^{-2}$ & $7.39\times10^{-2}$ & $7.97\times10^{-2}$ & $8.80\times10^{-2}$\\
		CFL0.8 & $5.52\times10^{-2}$ & $5.69\times10^{-2}$ & $5.86\times10^{-2}$ & $6.49\times10^{-2}$ & $7.13\times10^{-2}$ & $7.63\times10^{-2}$\\
		\end{tabular}
		\end{ruledtabular}
	\end{subtable}%
	\end{minipage}
	\par\medskip
	\begin{minipage}{1.0\textwidth}
	\begin{subtable}{1.0\textwidth}
	\caption{\label{tab:ZD:DUGKS-T2S3-Pe-L2Phi}DUGKS-T2S3}
		\begin{ruledtabular}
		\begin{tabular}{ccccccc}
		Pe&256&512&640&819.2&1024&1280\\
		\colrule
		CFL0.25 & $6.51\times10^{-2}$ & $5.86\times10^{-2}$ & $5.63\times10^{-2}$ & $5.60\times10^{-2}$ & $5.46\times10^{-2}$ & $5.46\times10^{-2}$\\
		CFL0.5 & $5.43\times10^{-2}$ & $4.72\times10^{-2}$ & $4.52\times10^{-2}$ & $4.37\times10^{-2}$ & $4.33\times10^{-2}$ & $4.10\times10^{-2}$\\
		CFL0.8 & $5.03\times10^{-2}$ & $4.41\times10^{-2}$ & $4.30\times10^{-2}$ & $4.14\times10^{-2}$ & $3.77\times10^{-2}$ & $3.77\times10^{-2}$\\
		\end{tabular}
		\end{ruledtabular}
	\end{subtable}%
	\end{minipage}
	\par\medskip
	\begin{minipage}{1.0\textwidth}
	\begin{subtable}{1.0\textwidth}
	\caption{\label{tab:ZD:DUGKS-T3S3-Pe-L2Phi}DUGKS-T3S3}
		\begin{ruledtabular}
		\begin{tabular}{ccccccc}
		Pe&256&512&640&819.2&1024&1280\\
		\colrule
		CFL0.25 & $8.03\times10^{-2}$ & $6.99\times10^{-2}$ & $6.92\times10^{-2}$ & $6.88\times10^{-2}$ & $6.76\times10^{-2}$ & $6.73\times10^{-2}$\\
		CFL0.5 & $6.21\times10^{-2}$ & $5.49\times10^{-2}$ & $5.34\times10^{-2}$ & $5.31\times10^{-2}$ & $5.12\times10^{-2}$ & $5.15\times10^{-2}$
		\end{tabular}
		\end{ruledtabular}
	\end{subtable}%
	\end{minipage}
	\par\medskip
	\begin{minipage}{1.0\textwidth}
	\begin{subtable}{1.0\textwidth}
	\caption{\label{tab:ZD:DUGKS-T2S5-Pe-L2Phi}DUGKS-T2S5}
		\begin{ruledtabular}
		\begin{tabular}{ccccccc}
		Pe&256&512&640&819.2&1024&1280\\
		\colrule
		CFL0.25 & $4.55\times10^{-2}$ & $3.98\times10^{-2}$ & $3.86\times10^{-2}$ & $3.77\times10^{-2}$ & $3.69\times10^{-2}$ & $3.69\times10^{-2}$\\
		CFL0.5 & $4.69\times10^{-2}$ & $4.18\times10^{-2}$ & $3.94\times10^{-2}$ & $3.77\times10^{-2}$ & $3.69\times10^{-2}$ & $3.69\times10^{-2}$\\
		CFL0.8 & $4.83\times10^{-2}$ & $4.26\times10^{-2}$ & $4.14\times10^{-2}$ & $3.82\times10^{-2}$ & $3.77\times10^{-2}$ & $3.81\times10^{-2}$\\
		\end{tabular}
		\end{ruledtabular}
	\end{subtable}%
	\end{minipage}
\end{table}
\begin{table}[htbp]
\caption
{
  \label{tab:ZD:M-L2Phi}
  $L_2$-norm error of $\phi$ for Zalesak's disk obtained by multiple methods with various $M_\phi$, $Cn = 4/256$, $Pe = 256$.
}
	\begin{minipage}{1.0\textwidth}
	\begin{subtable}{1.0\textwidth}
	\caption{\label{tab:ZD:LBMDVM-M-L2Phi}LBM, DVM}
		\begin{ruledtabular}
		\begin{tabular}{cccccccc}
		$M_\phi$&0.01&0.02&0.04&0.05&0.064&0.08&0.1\\
		\colrule
		LBM & $4.55\times10^{-2}$ & $4.59\times10^{-2}$ & $4.59\times10^{-2}$ & $4.55\times10^{-2}$ & $4.62\times10^{-2}$ & $4.69\times10^{-2}$ & $4.86\times10^{-2}$\\
		DVM-CFL1.0 & $4.55\times10^{-2}$ & $4.59\times10^{-2}$ & $4.59\times10^{-2}$ & $4.55\times10^{-2}$ & $4.62\times10^{-2}$ & $4.69\times10^{-2}$ & $4.86\times10^{-2}$\\
		DVM-CFL0.8 & $5.43\times10^{-2}$ & $5.15\times10^{-2}$ & $4.86\times10^{-2}$ & $4.76\times10^{-2}$ & $4.73\times10^{-2}$ & $4.76\times10^{-2}$ & $4.73\times10^{-2}$\\
		\end{tabular}
		\end{ruledtabular}
	\end{subtable}%
	\end{minipage}
	\par\medskip
	\begin{minipage}{1.0\textwidth}
	\begin{subtable}{1.0\textwidth}
	\caption{\label{tab:ZD:DUGKS-T2S2CD-M-L2Phi}DUGKS-T2S2CD}
		\begin{ruledtabular}
		\begin{tabular}{cccccccc}
		$M_\phi$&0.01&0.02&0.04&0.05&0.064&0.08&0.1\\
		\colrule
		CFL0.25 & $5.91\times10^{-2}$ & $5.94\times10^{-2}$ & $5.99\times10^{-2}$ & $5.99\times10^{-2}$ & $6.02\times10^{-2}$ & $6.26\times10^{-2}$ & $6.54\times10^{-2}$
		\\
		CFL0.5 & $5.97\times10^{-2}$ & $5.83\times10^{-2}$ & $5.89\times10^{-2}$ & $5.89\times10^{-2}$ & $6.08\times10^{-2}$ & $6.08\times10^{-2}$ & $6.36\times10^{-2}$
		\\
		CFL0.8 & $5.74\times10^{-2}$ & $5.52\times10^{-2}$ & $5.57\times10^{-2}$ & $5.57\times10^{-2}$ & $5.60\times10^{-2}$ & $5.86\times10^{-2}$ & $6.10\times10^{-2}$
		\\
		\end{tabular}
		\end{ruledtabular}
	\end{subtable}%
	\end{minipage}
	\par\medskip
	\begin{minipage}{1.0\textwidth}
	\begin{subtable}{1.0\textwidth}
	\caption{\label{tab:ZD:DUGKS-T2S3-M-L2Phi}DUGKS-T2S3}
		\begin{ruledtabular}
		\begin{tabular}{cccccccc}
		$M_\phi$&0.01&0.02&0.04&0.05&0.064&0.08&0.1\\
		\colrule
		CFL0.25 & $8.11\times10^{-2}$ & $6.51\times10^{-2}$ & $5.75\times10^{-2}$ & $5.37\times10^{-2}$ & $5.37\times10^{-2}$ & $5.31\times10^{-2}$ & $5.46\times10^{-2}$\\
		CFL0.5 & $5.77\times10^{-2}$ & $5.43\times10^{-2}$ & $5.31\times10^{-2}$ & $5.28\times10^{-2}$ & $5.34\times10^{-2}$ & $5.21\times10^{-2}$ &$5.49\times10^{-2}$\\
		CFL0.8 & $5.12\times10^{-2}$ & $5.03\times10^{-2}$ & $5.03\times10^{-2}$ & $5.12\times10^{-2}$ & $5.15\times10^{-2}$ & $5.12\times10^{-2}$ & $5.46\times10^{-2}$\\
		\end{tabular}
		\end{ruledtabular}
	\end{subtable}%
	\end{minipage}
	\par\medskip
	\begin{minipage}{1.0\textwidth}
	\begin{subtable}{1.0\textwidth}
	\caption{\label{tab:ZD:DUGKS-T3S3-M-L2Phi}DUGKS-T3S3}
		\begin{ruledtabular}
		\begin{tabular}{cccccccc}
		$M_\phi$&0.01&0.02&0.04&0.05&0.064&0.08&0.1\\
		\colrule
		CFL0.25 & $9.84\times10^{-2}$ & $8.03\times10^{-2}$ & $5.99\times10^{-2}$ & $5.69\times10^{-2}$ & $5.49\times10^{-2}$ & $5.46\times10^{-2}$ & $5.52\times10^{-2}$\\
		CFL0.5 & - & $6.21\times10^{-2}$ & $5.45\times10^{-2}$ & $5.43\times10^{-2}$ & $5.31\times10^{-2}$ & $5.31\times10^{-2}$ &$5.46\times10^{-2}$\\
		CFL0.8 & - & - & $5.28\times10^{-2}$ & $5.28\times10^{-2}$ & $5.34\times10^{-2}$ & $5.18\times10^{-2}$ & $5.46\times10^{-2}$\\
		\end{tabular}
		\end{ruledtabular}
	\end{subtable}%
	\end{minipage}
	\par\medskip
	\begin{minipage}{1.0\textwidth}
	\begin{subtable}{1.0\textwidth}
	\caption{\label{tab:ZD:DUGKS-T2S5-M-L2Phi}DUGKS-T2S5}
		\begin{ruledtabular}
		\begin{tabular}{cccccccc}
		$M_\phi$&0.01&0.02&0.04&0.05&0.064&0.08&0.1\\
		\colrule
		CFL0.25 & $4.69\times10^{-2}$ & $4.55\times10^{-2}$ & $4.44\times10^{-2}$ & $4.37\times10^{-2}$ & $4.44\times10^{-2}$ & $4.79\times10^{-2}$ & $5.12\times10^{-2}$\\
		CFL0.5 & $4.76\times10^{-2}$ & $4.69\times10^{-2}$ & $4.62\times10^{-2}$ & $4.59\times10^{-2}$ & $4.55\times10^{-2}$ & $4.86\times10^{-2}$ & $5.12\times10^{-2}$\\
		CFL0.8 & $4.86\times10^{-2}$ & $4.83\times10^{-2}$ & $4.79\times10^{-2}$ & $4.83\times10^{-2}$ & $5.03\times10^{-2}$ & $5.06\times10^{-2}$ & $5.37\times10^{-2}$\\
		\end{tabular}
		\end{ruledtabular}
	\end{subtable}%
	\end{minipage}
\end{table}
\\
Another point that needs to be paid attention to is although we try to compare those results quantitatively, the step function applied at the terminal moment somehow contaminates our results. However, this operation is indispensable because of the zero-thickness interface at initial state. To some extent, the comparisons made among those results are not that quantitative.
\subsection{\label{sec:sec3.C}Interface extension}
\begin{figure}[htbp]
    \centering
    \begin{subfigure}{0.400 \columnwidth}
      \centering
      \includegraphics[width=1\linewidth]{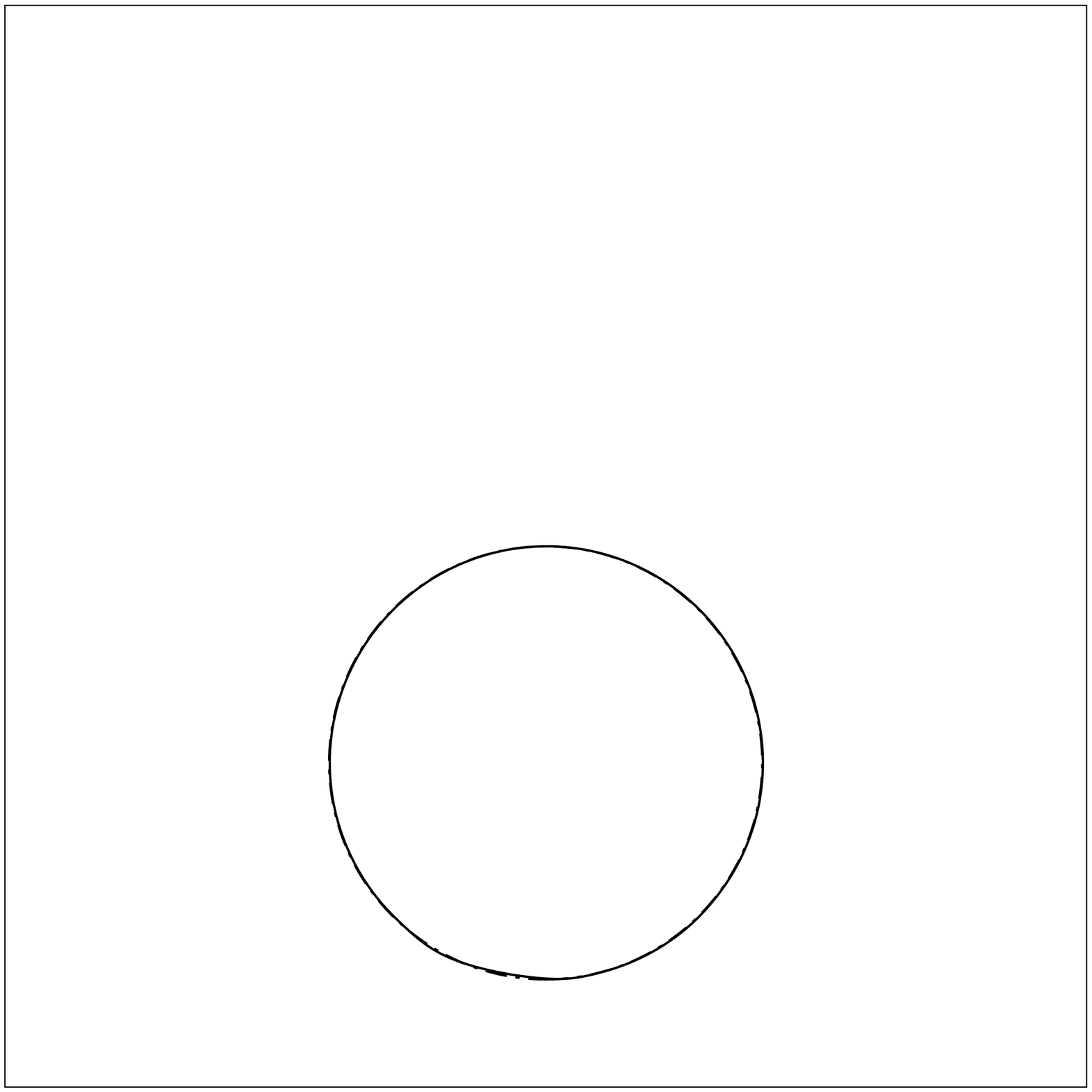}
      \caption[]{DVM, $C = 1.0$}
      \label{FIG:IE:DVMCFL1.0-4-4T}
    \end{subfigure}
    \begin{subfigure}{0.400 \columnwidth}
      \centering
      \includegraphics[width=1\linewidth]{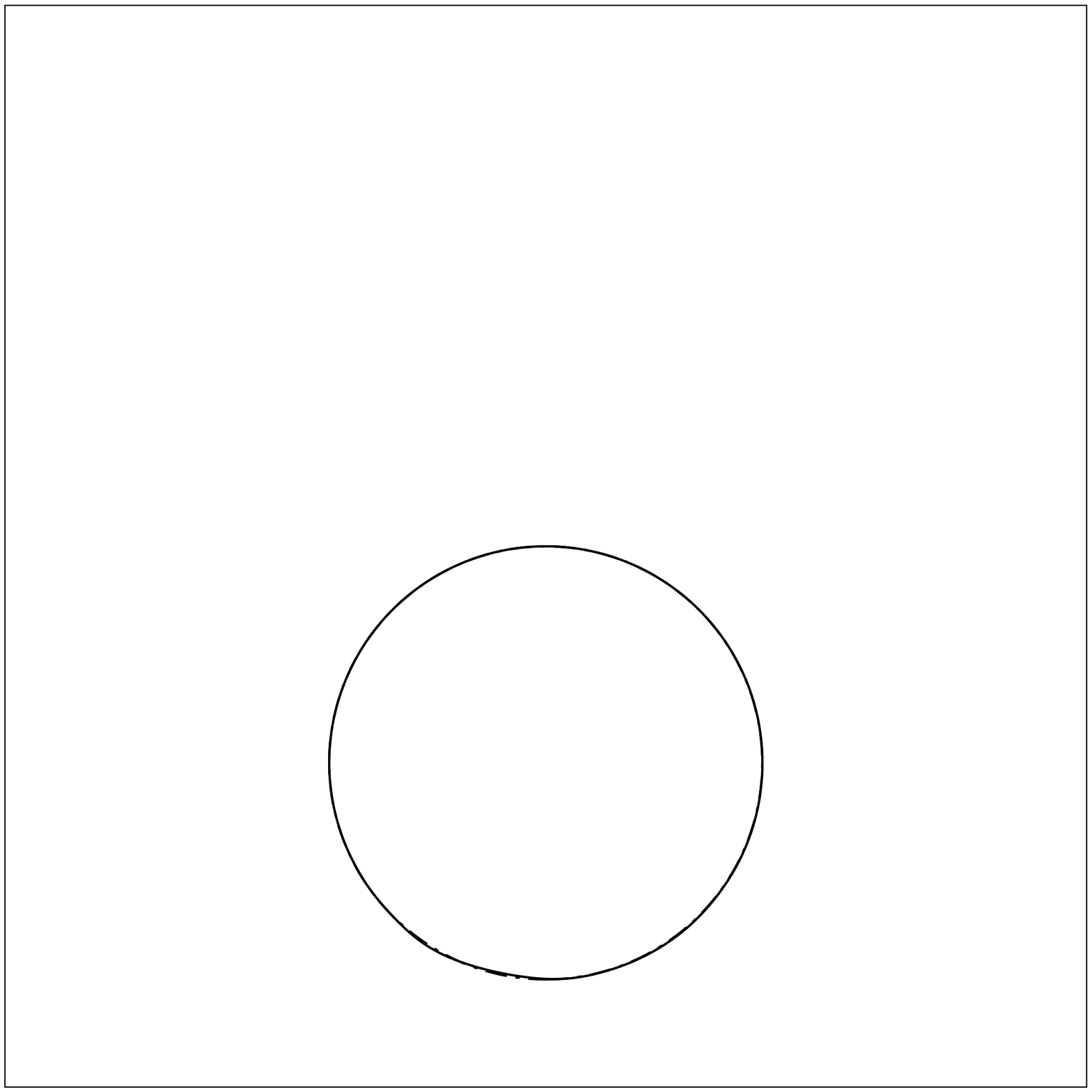}
      \caption{DVM, $C = 0.8$}
      \label{FIG:IE:DVMCFL0.8-4-4T}
    \end{subfigure}
    \hfill
    \begin{subfigure}{0.400 \columnwidth}
      \centering
      \includegraphics[width=1\linewidth]{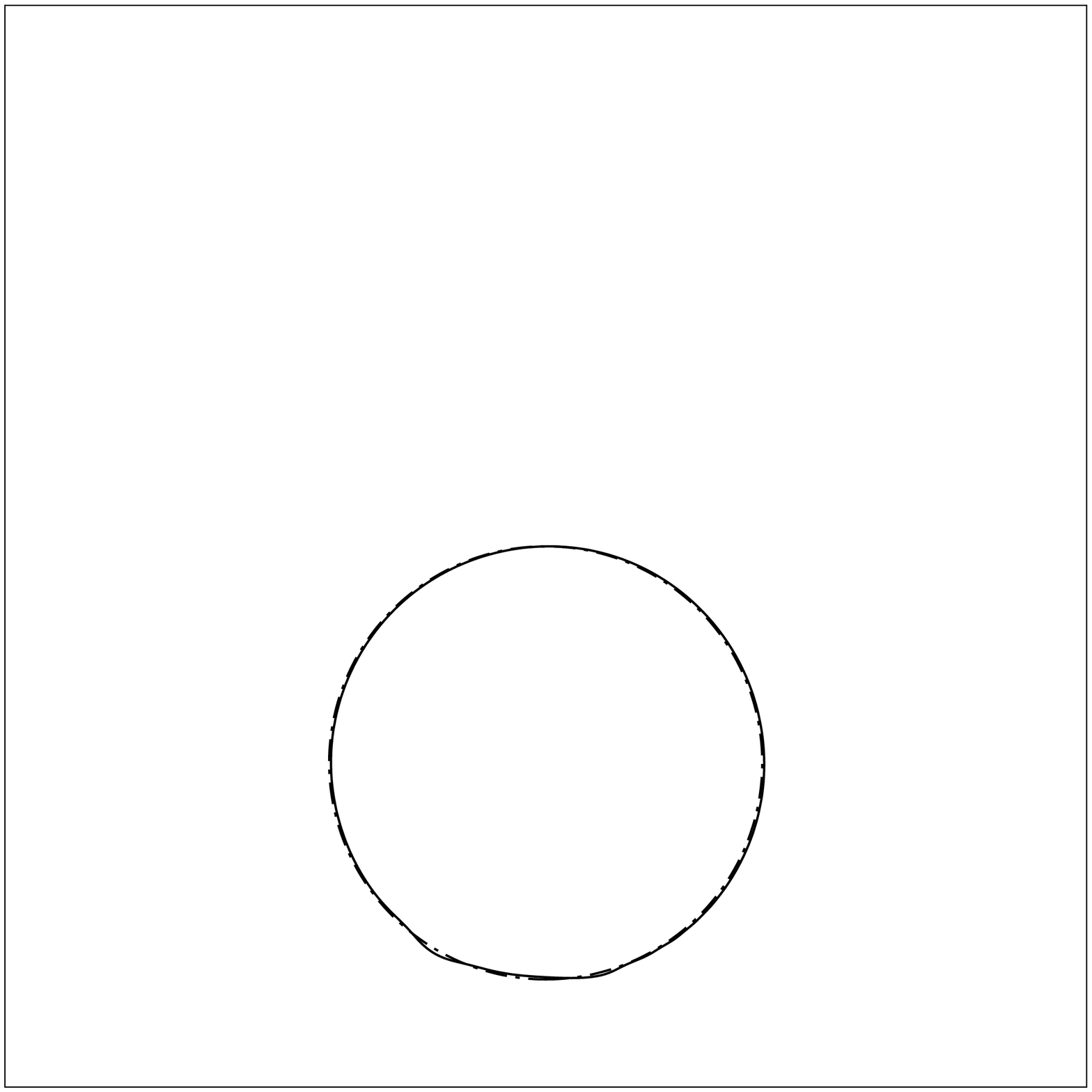}
      \caption[]{DUGKS-T2S2CD, $C = 0.8$}
      \label{FIG:IE:DUGKS-T2S2CD-4-4T}
    \end{subfigure}
    \begin{subfigure}{0.400 \columnwidth}
      \centering
      \includegraphics[width=1\linewidth]{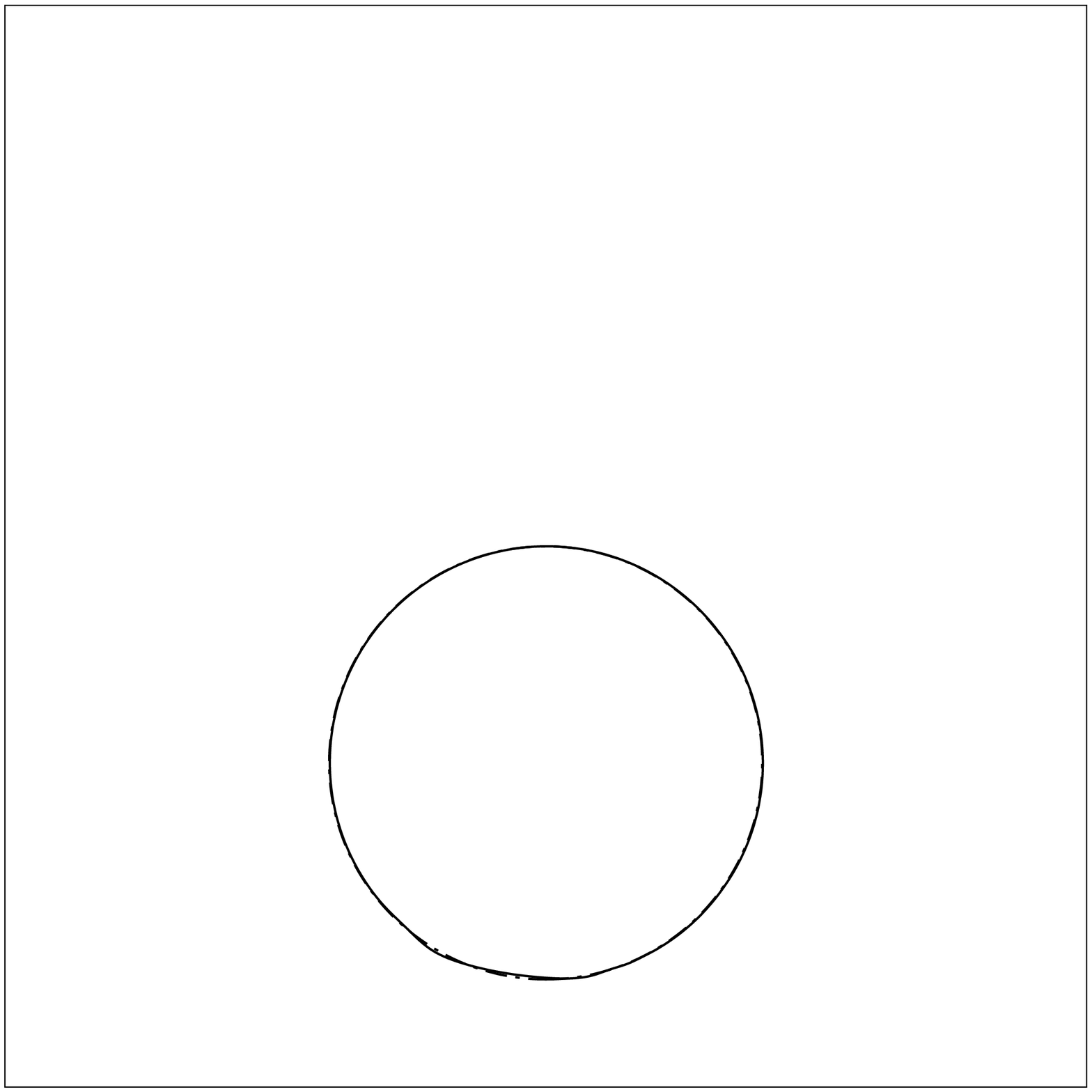}
      \caption[]{DUGKS-T2S5, $C = 0.25$}
      \label{FIG:IE:DUGKS-T2S5-4-4T}
    \end{subfigure}
    \hfill
    \begin{subfigure}[b]{0.400\textwidth}
      \centering
      \includegraphics[width=1\linewidth]{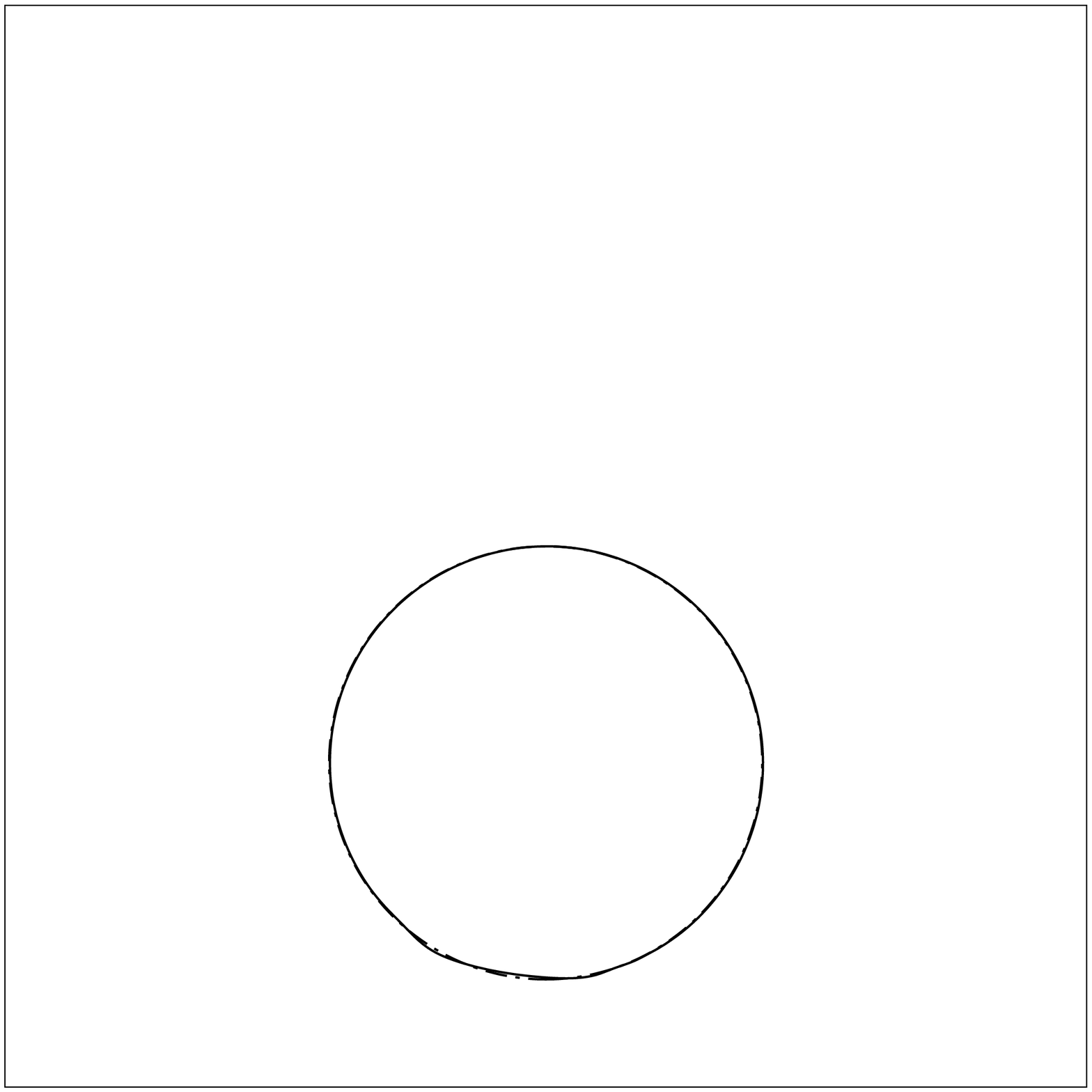}
      \caption[]{DUGKS-T2S3, $C = 0.8$}
      \label{FIG:IE:DUGKS-T2S3-4-4T}
    \end{subfigure}
    \begin{subfigure}[b]{0.400\textwidth}
      \centering
      \includegraphics[width=1\linewidth]{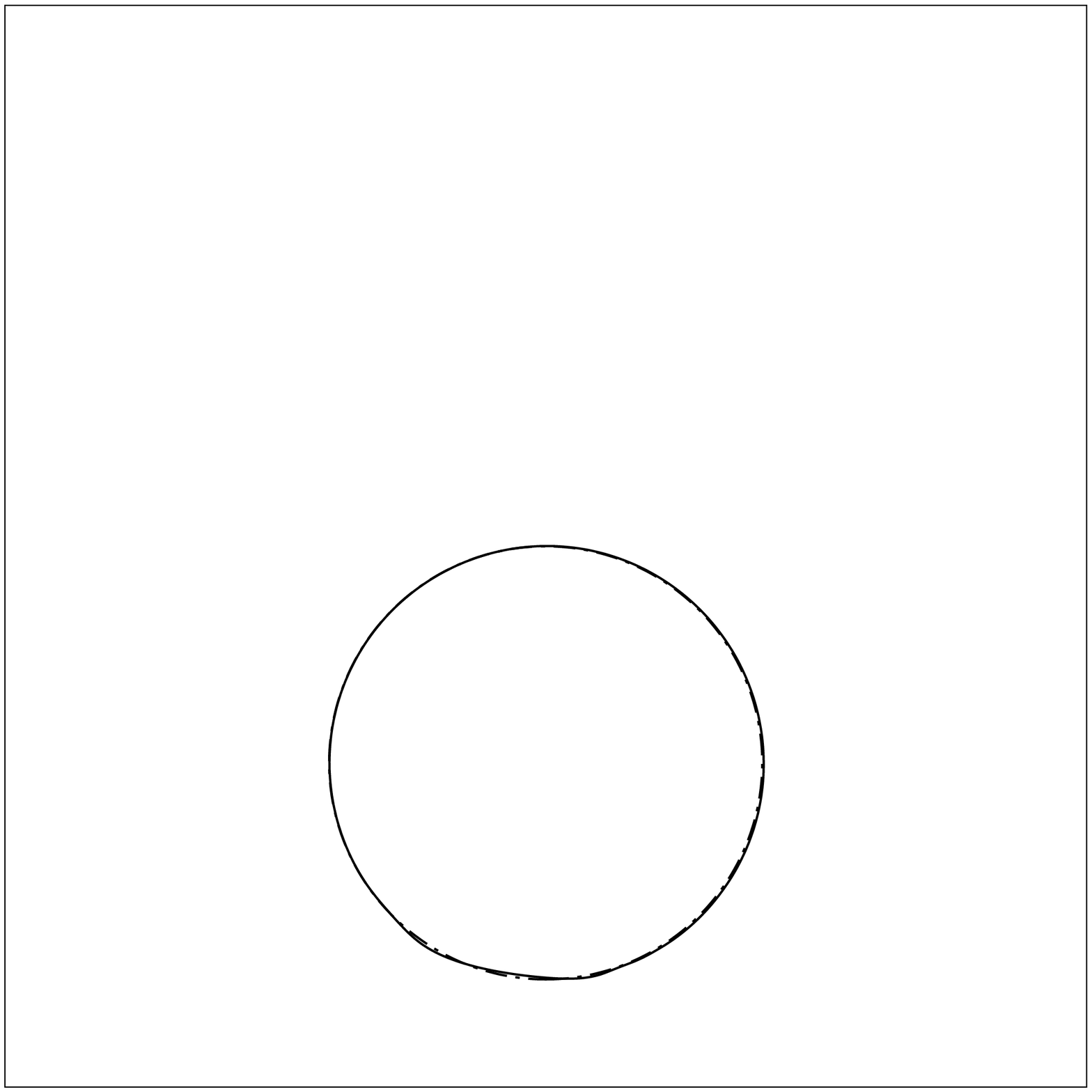}
      \caption[]{DUGKS-T3S3, $C = 0.5$}
      \label{FIG:IE:DUGKS-T3S3-4-4T}
    \end{subfigure}
\caption{Interface extension in a shear flow with $Pe = 1024$, $Cn = 4/256$, $M_\phi = 0.02$.}
\label{FIG:IE:Scheme-4-4T}
\end{figure}

\begin{figure}[htbp]
    \centering
    \begin{minipage}[b]{0.9 \columnwidth}
    \begin{subfigure}{0.45 \columnwidth}
      \centering
      \includegraphics[width=1\linewidth]{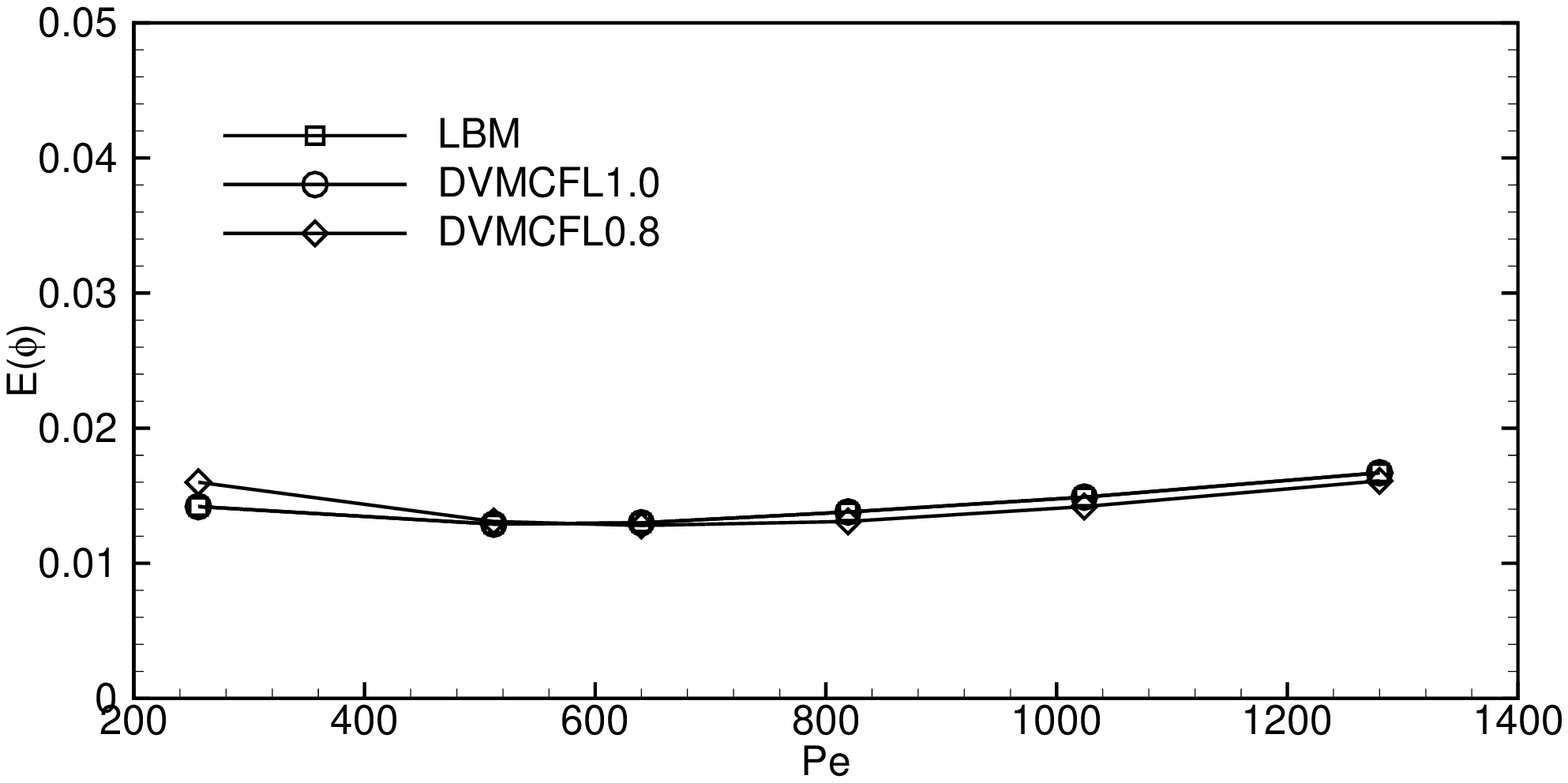}
      \caption{LBM, DVM}
      \label{FIG:IE:LBMDVM-Pe-L2Phi}
    \end{subfigure}
    \end{minipage}
    \begin{minipage}[b]{0.9 \columnwidth}
    \begin{subfigure}{0.45 \columnwidth}
      \centering
      \includegraphics[width=1\linewidth]{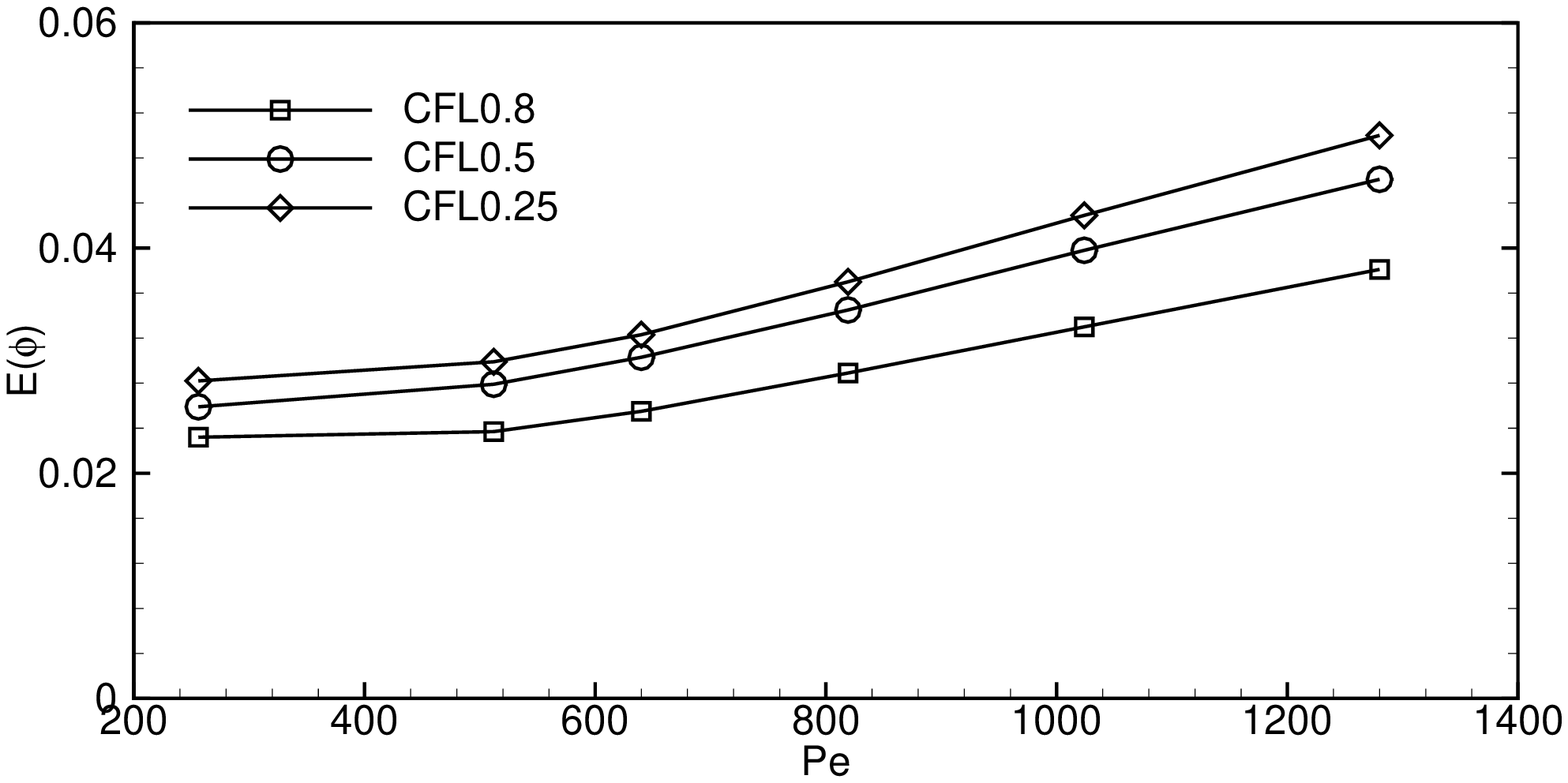}
      \caption{DUGKS-T2S2CD}
      \label{FIG:IE:DUGKST2S2CD-Pe-L2Phi}
    \end{subfigure}
    \end{minipage}
    \begin{minipage}[b]{0.9 \columnwidth}
    \begin{subfigure}{0.45 \columnwidth}
      \centering
      \includegraphics[width=1\linewidth]{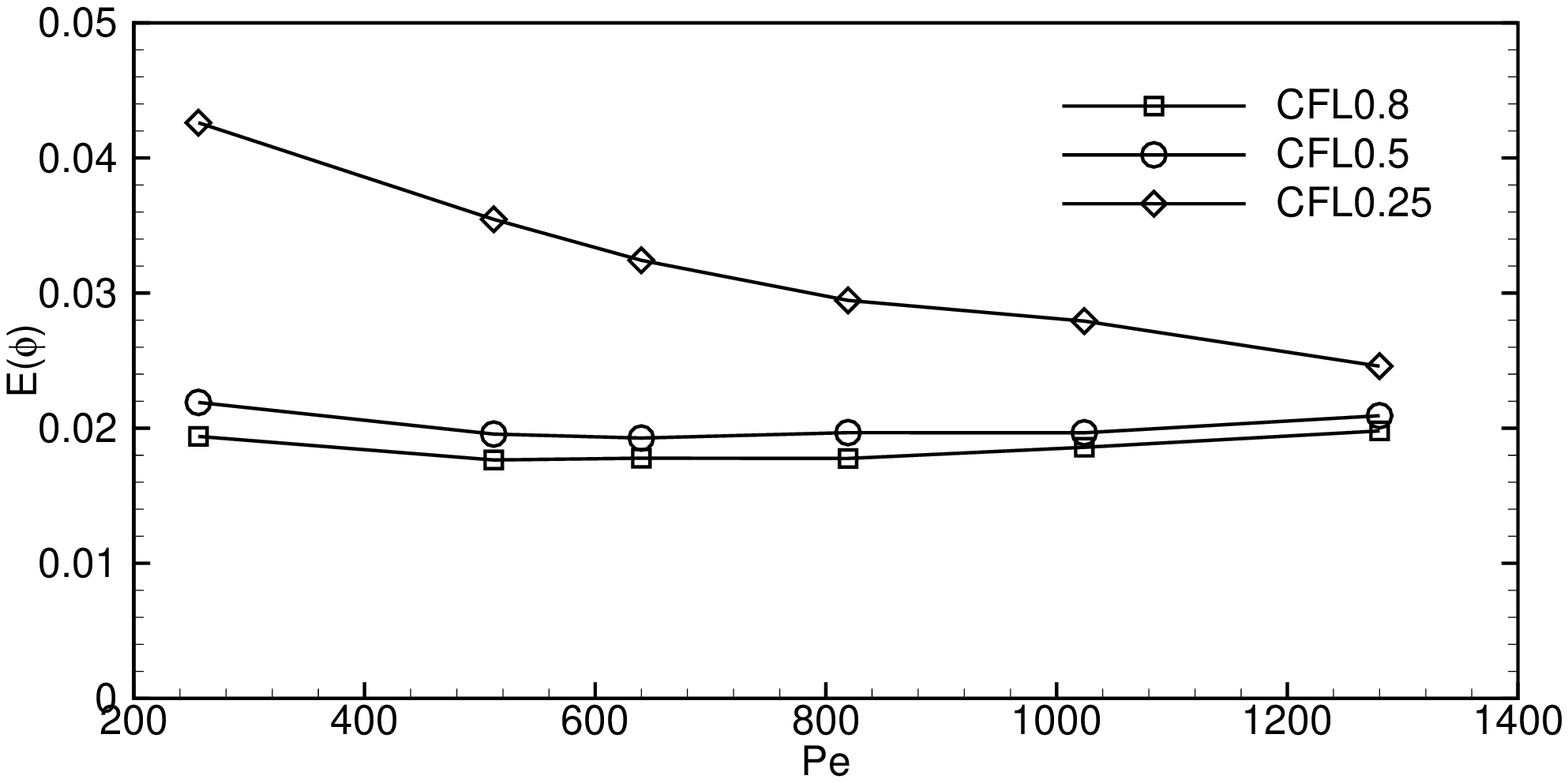}
      \caption{DUGKS-T2S3}
      \label{FIG:IE:DUGKST2S3-Pe-L2Phi}
    \end{subfigure}
    \end{minipage}
    \begin{minipage}[b]{0.9 \columnwidth}
    \begin{subfigure}{0.45 \columnwidth}
      \centering
      \includegraphics[width=1\linewidth]{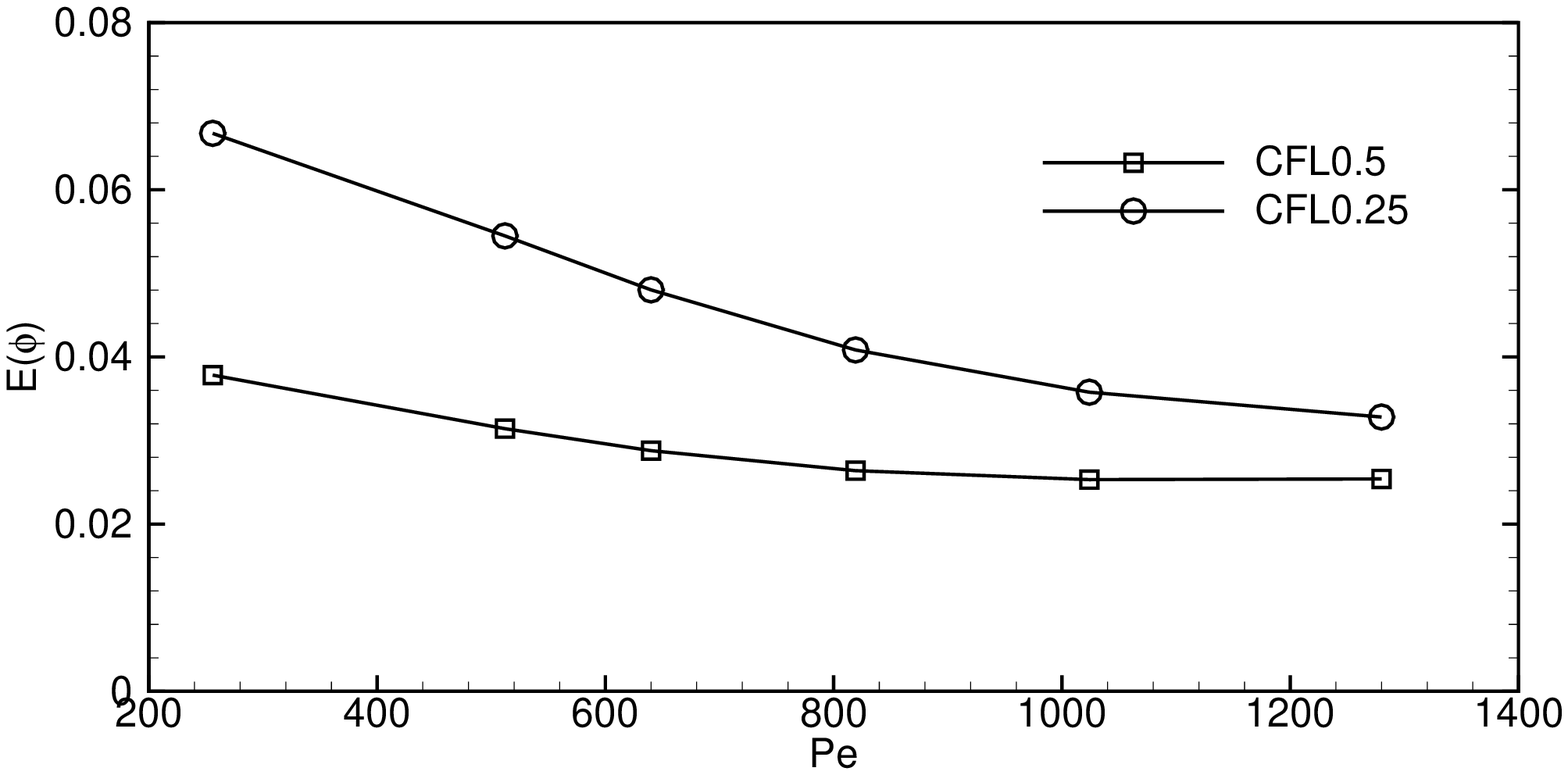}
      \caption{DUGKS-T3S3}
      \label{FIG:IE:DUGKST3S3-Pe-L2Phi}
    \end{subfigure}
    \end{minipage}
    \begin{minipage}[b]{0.9 \columnwidth}
    \begin{subfigure}{0.45 \columnwidth}
      \centering
      \includegraphics[width=1\linewidth]{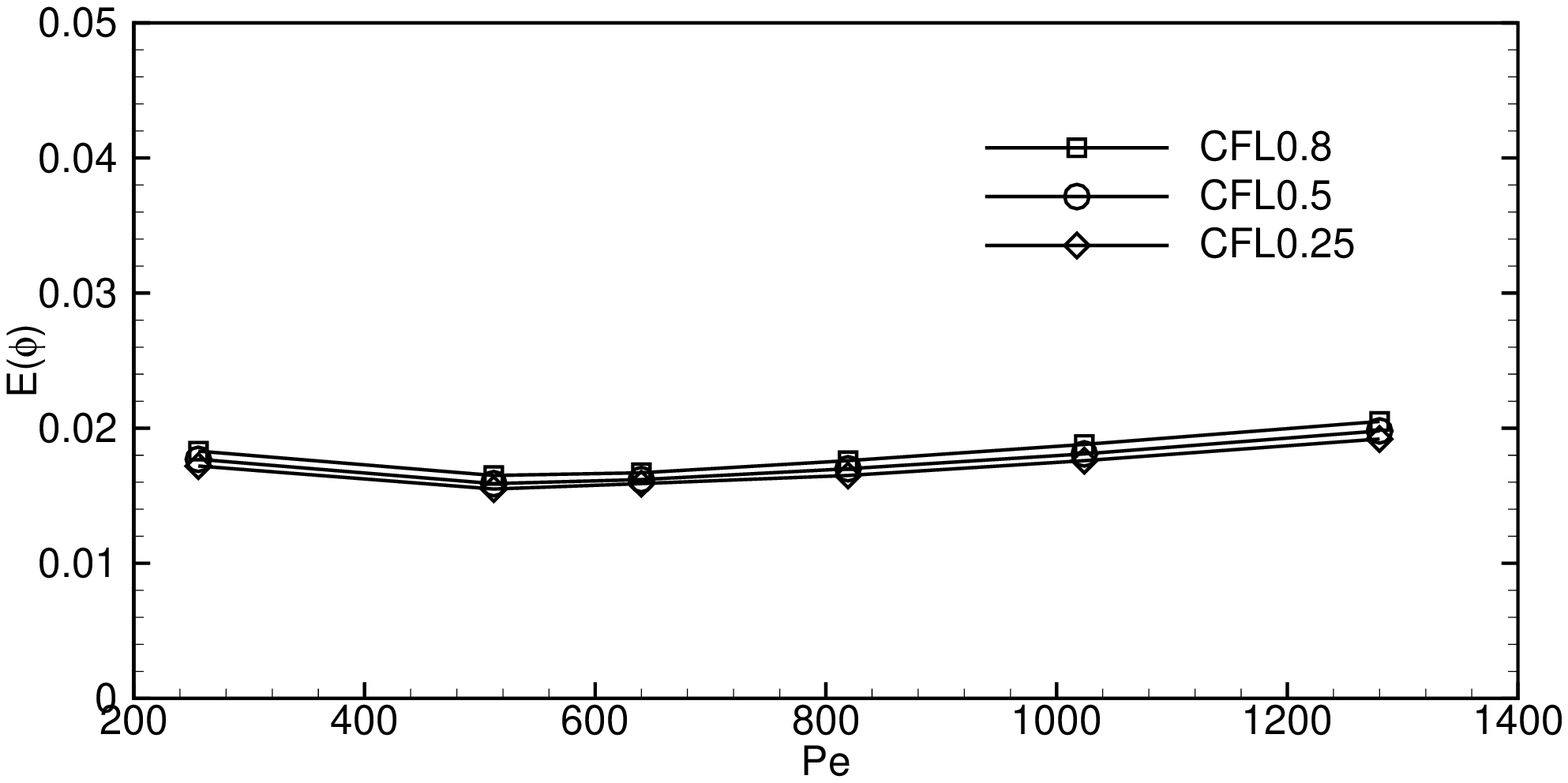}
      \caption{DUGKS-T2S5}
      \label{FIG:IE:DUGKST2S5-Pe-L2Phi}
    \end{subfigure}
    \end{minipage}
\caption{$L_2$-norm error of $\phi$ for interface extension obtained by multiple methods with various $Pe$, $Cn = 4/256$, $M_\phi = 0.02$.}
\label{FIG:IE:Pe-L2Phi}
\end{figure}

\begin{figure}[htbp]
    \centering
    \begin{minipage}[b]{0.9 \columnwidth}
    \begin{subfigure}{0.45 \columnwidth}
      \centering
      \includegraphics[width=1\linewidth]{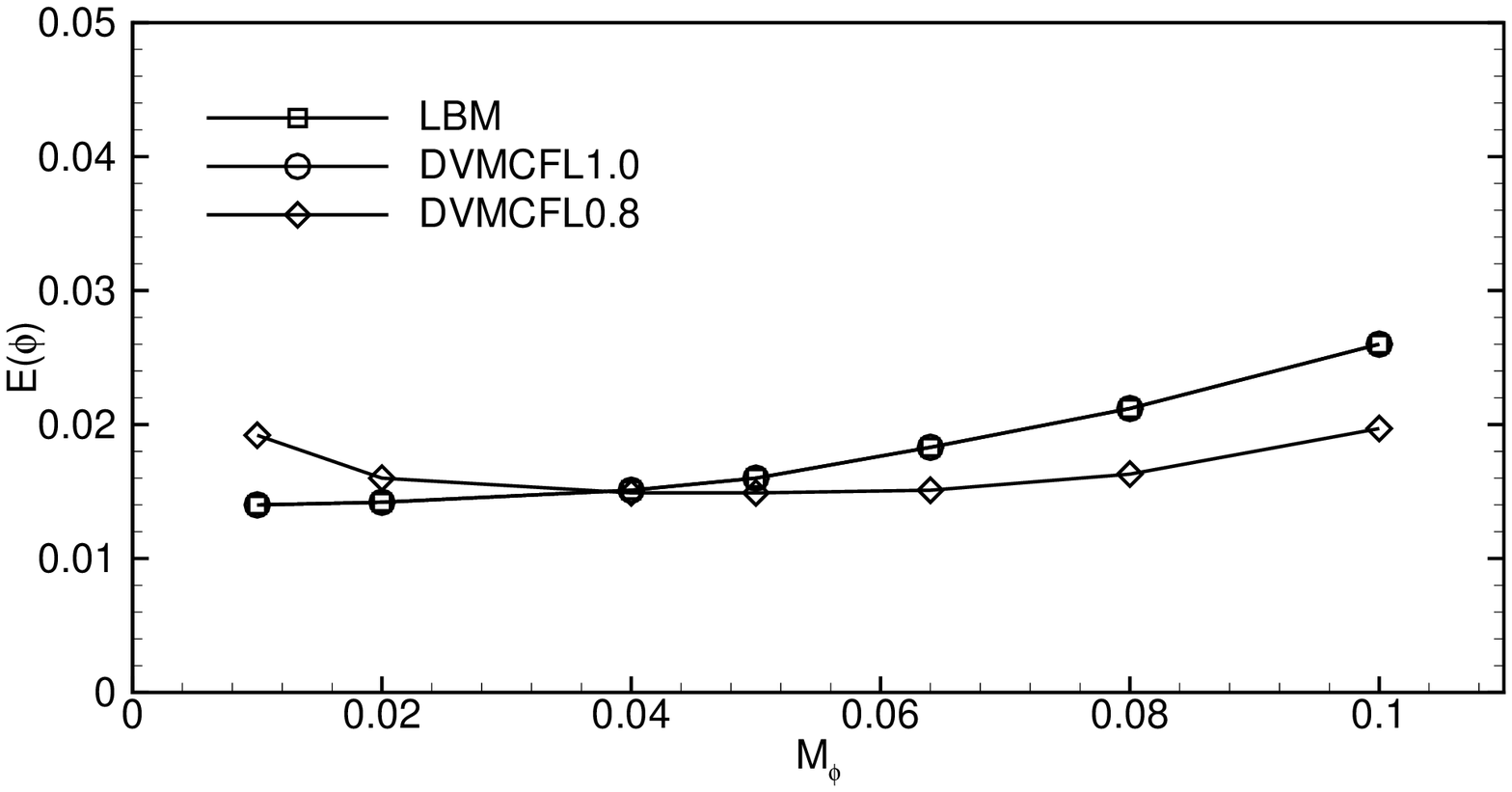}
      \caption{LBM, DVM}
      \label{FIG:IE:LBMDVM-M-L2Phi}
    \end{subfigure}
    \end{minipage}
    \begin{minipage}[b]{0.9 \columnwidth}
    \begin{subfigure}{0.45 \columnwidth}
    \centering
      \centering
      \includegraphics[width=1\linewidth]{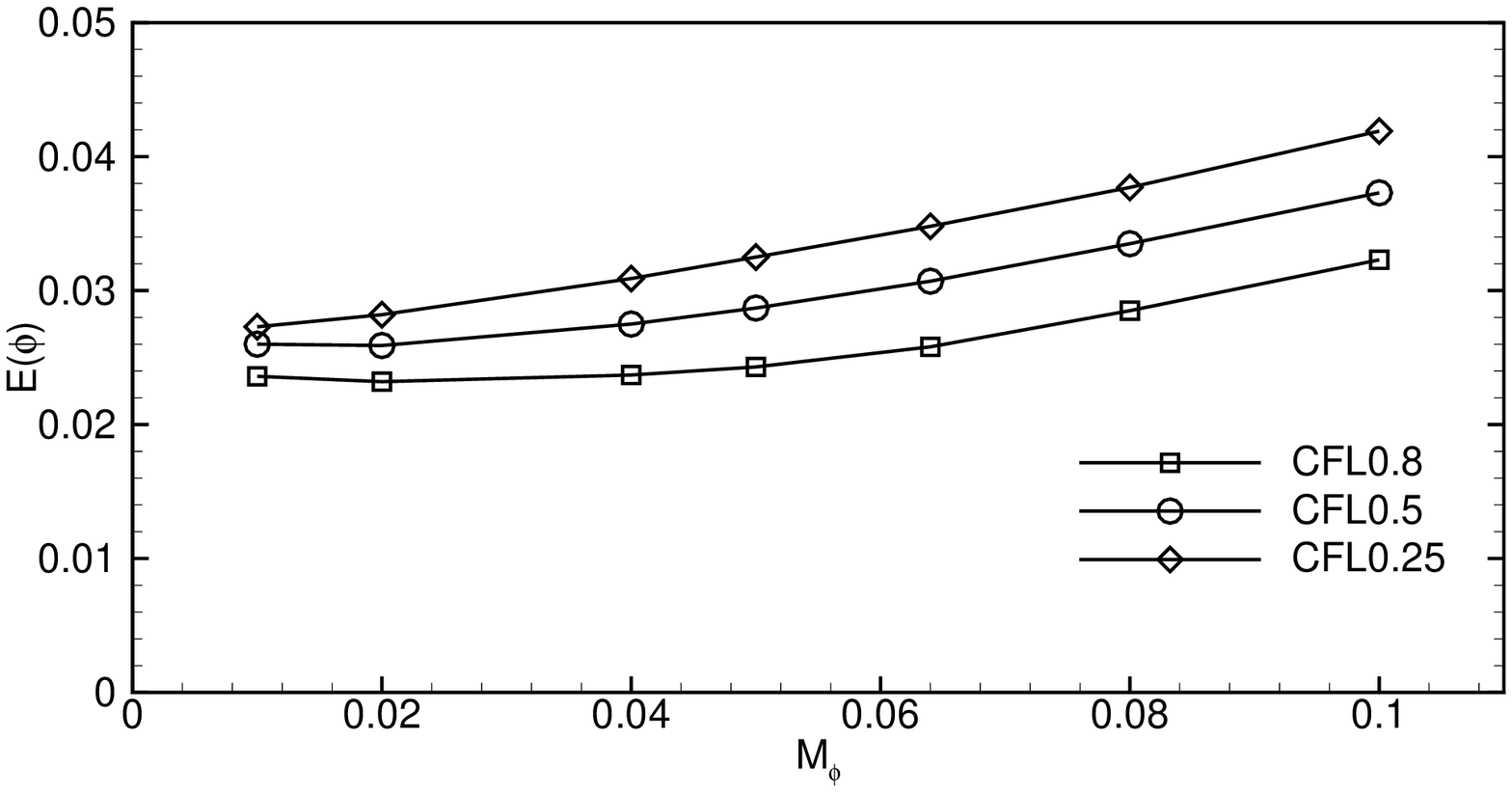}
      \caption{DUGKS-T2S2CD}
      \label{FIG:IE:DUGKST2S2CD-M-L2Phi}
    \end{subfigure}
    \end{minipage}
    \begin{minipage}[b]{0.9 \columnwidth}
    \begin{subfigure}{0.45 \columnwidth}
      \centering
      \includegraphics[width=1\linewidth]{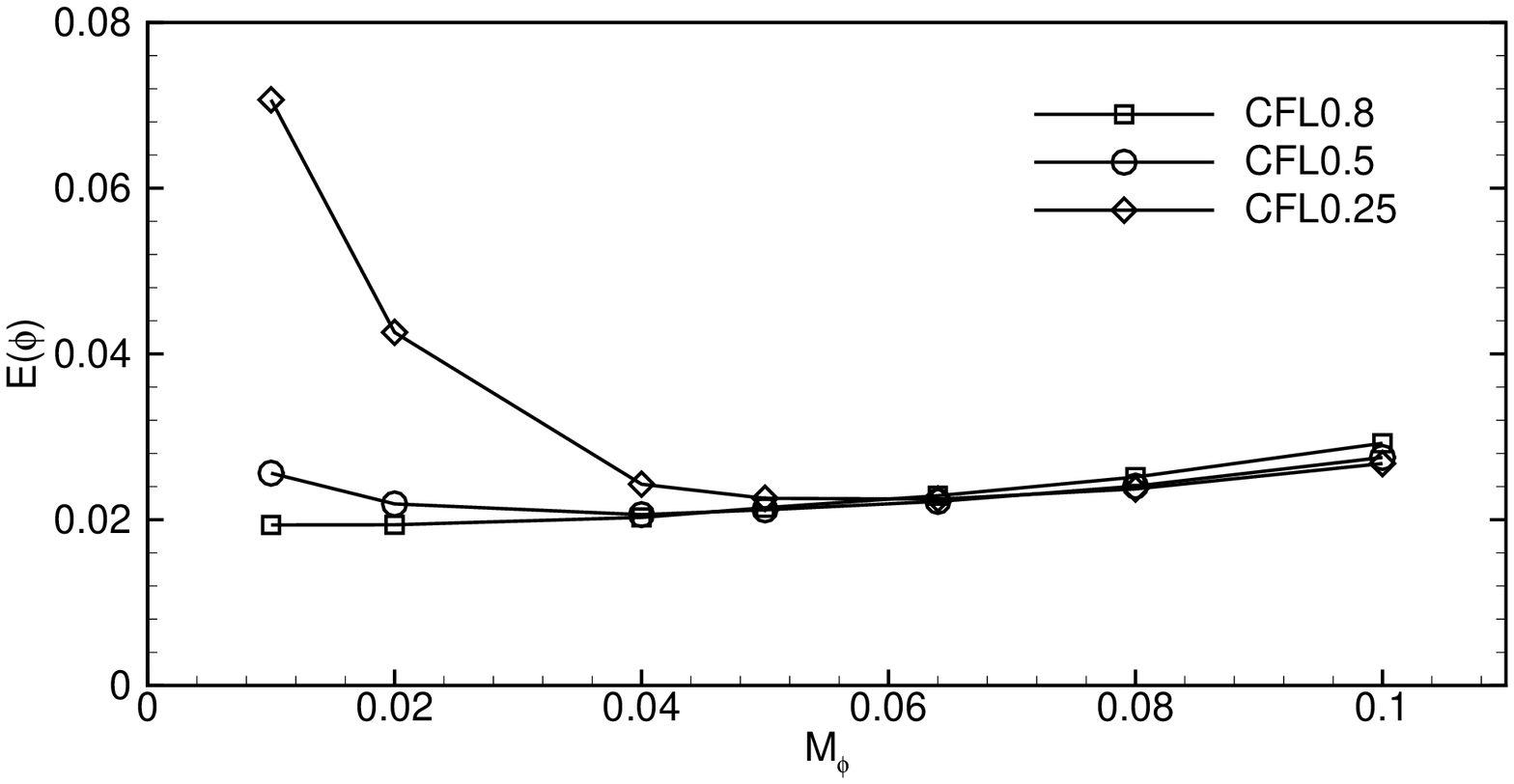}
      \caption{DUGKS-T2S3}
      \label{FIG:IE:DUGKST2S3-M-L2Phi}
    \end{subfigure}
    \end{minipage}
    \begin{minipage}[b]{0.9 \columnwidth}
    \begin{subfigure}{0.45 \columnwidth}
      \centering
      \includegraphics[width=1\linewidth]{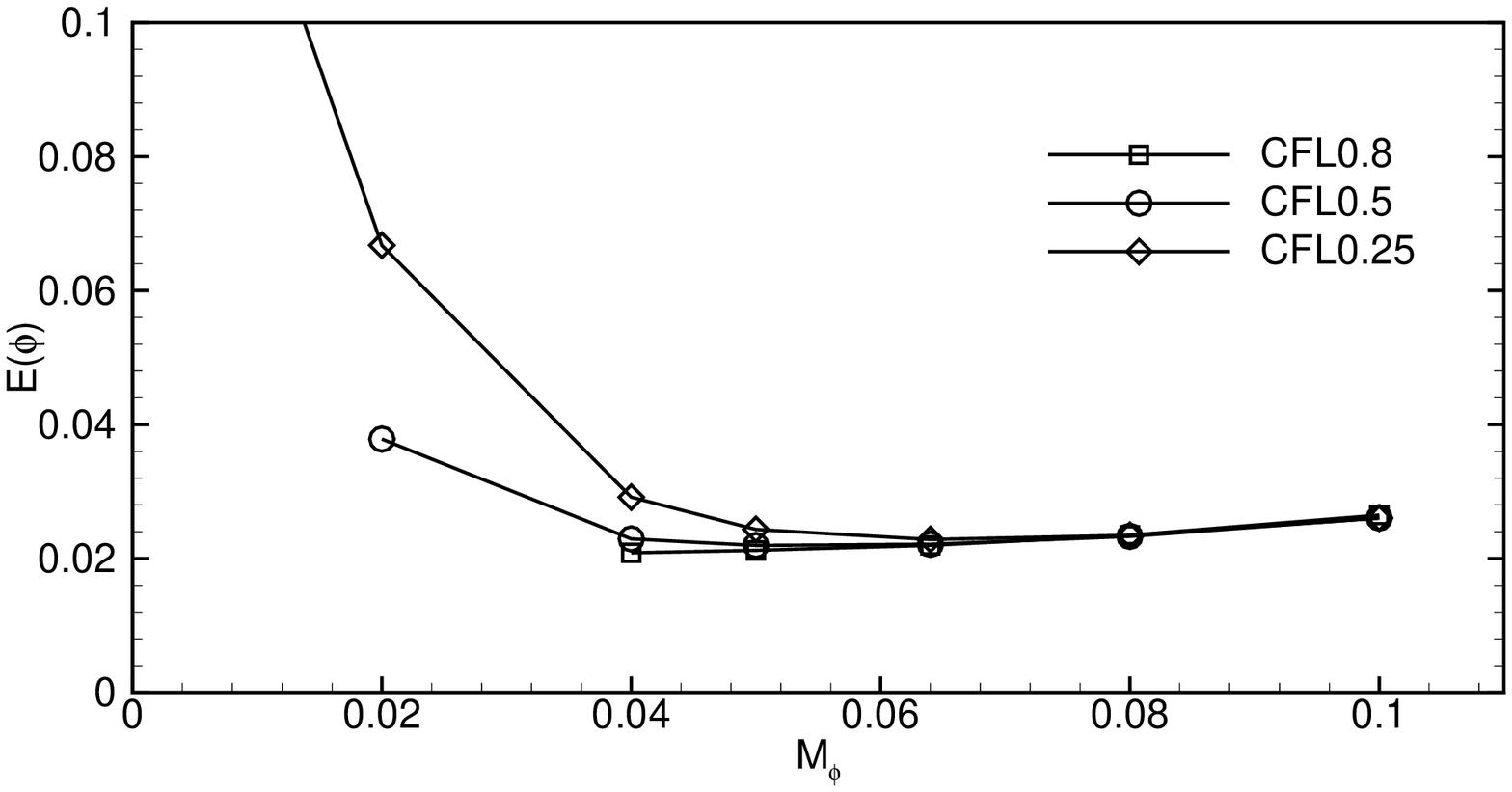}
      \caption{DUGKS-T3S3}
      \label{FIG:IE:DUGKST3S3-M-L2Phi}
    \end{subfigure}
    \end{minipage}
    \begin{minipage}[b]{0.9 \columnwidth}
    \begin{subfigure}{0.45 \columnwidth}
      \centering
      \includegraphics[width=1\linewidth]{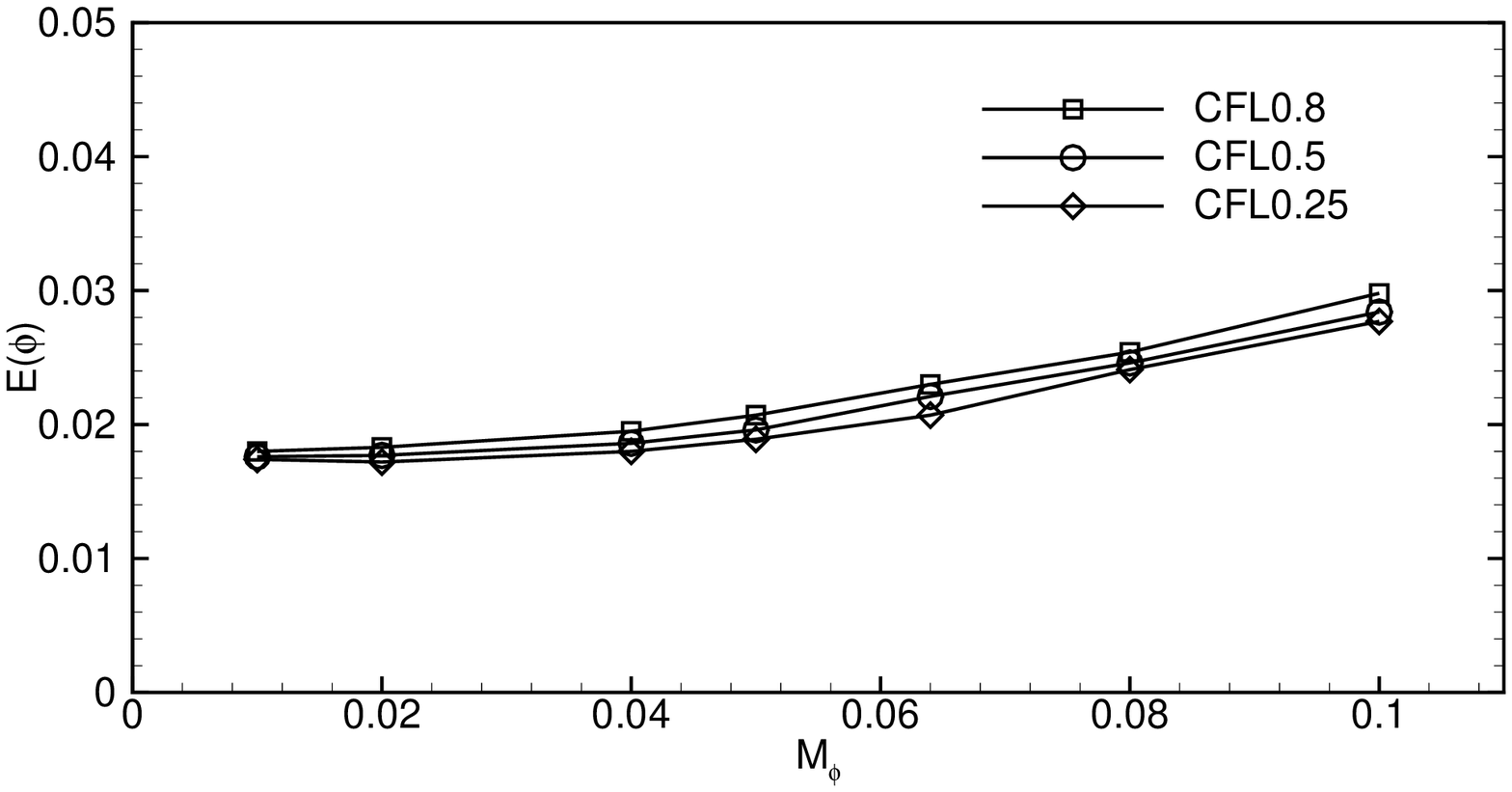}
      \caption{DUGKS-T2S5}
      \label{FIG:IE:DUGKST2S5-M-L2Phi}
    \end{subfigure}
    \end{minipage}
\caption{$L_2$-norm error of $\phi$ for interface extension obtained by multiple methods with various $M_\phi$, $Cn = 4/256$, $Pe = 256$.}
\label{FIG:IE:M-L2Phi}
\end{figure}

This benchmark test aims at evaluating the performance of multiple methods on capturing interface with large topological changes. A circular body with a radius of $R = L_0/5$ is placed in a square domain with $L_0{\times}L_0$ cells. The center of this body is located at $x = 0.5L_0$ and $y = 0.3L_0$. Here $L_0 = 256$ is the reference length. The velocity field is governed by
\begin{equation}
\begin{aligned}
&u(x,y)=U_0{\pi}\textrm{sin}\big(\frac{{\pi}x}{L_0}\big)\textrm{cos}{\big(\frac{{\pi}y}{L_0}\big)},
\\
&v(x,y)=-U_0{\pi}\textrm{cos}\big(\frac{{\pi}x}{L_0}\big)\textrm{sin}{\big(\frac{{\pi}y}{L_0}\big)}.
\end{aligned}
\end{equation}
After an elapsed period of time $T_f$, the velocity field would be reversed and the elongated interface starts to evolve toward its initial state. After another time of $T_f$, the interface would be restored to its initial shape, despite the slight differences caused by the numerical dissipation. The discrepancy between initial and final interface is rated as the evaluation criteria for the performance of various methods and fewer differences signify better performance of the method employed in this interface tracking test.
\\
Fig.~\ref{FIG:IE:Scheme-4-4T} illustrates the differences between the restored interface (solid line) and the initial interface (dash line) at $Pe = 1024,M_{\phi} = 0.02$ provided by various methods. Base on the data in Table~\ref{tab:IE:Pe-L2Phi}, it can be observed that DVM-CFL0.8 provides the best result due to the tiny difference between the interfaces at initial and final moments. The results obtained by DVM-CFL1.0, illustrated in Fig.~\ref{FIG:IE:DVMCFL1.0-4-4T} and presented in Table~\ref{tab:IE:LBMDVM-Pe-L2Phi}, is almost identical to those shown in Fig.~\ref{FIG:IE:DVMCFL0.8-4-4T}. Although the absolute difference between the results presented in Table~\ref{tab:IE:LBMDVM-Pe-L2Phi} is tiny, this outcome truly perplexes us because DVM-CFL0.8, which involves the reconstruction step where spatial dissipation is introduced, outperforms the pure streaming and collision methods, DVM-CFL1.0 and LBM. DUGKS-T2S5 with a CFL number of 0.25 offers the next-best result. And then comes DUGKS-T2S3 with a fixed CFL number of 0.8. Although DUGKS-T3S3 does not exploit its advantage to the full due to the inherent restriction, its performance is still superior to that of DUGKS-T2S2CD. As the P\'{e}clet number is large, it is reasonable that flux based on upwind scheme is more accurate than flux based on central scheme.
\\
Fig.~\ref{FIG:IE:Pe-L2Phi} illustrates the results of $L_2$ norm error obtained by multiple methods at various P\'{e}clet number for the interface extension test. And the detailed data are presented in Table~\ref{tab:IE:Pe-L2Phi}. As the P\'{e}clet number increases, the overall error offered by DUGKS-T2S2CD tends to rise up. This tendency has its rationality since flux based on central scheme performs worse when the flow is dominated by advection. The results obtained by DUGKS-T2S3 and DUGKS-T3S3 have shown an opposite trend since upwind scheme does well in such a situation. Not much correlation can be observed between the variation of P\'{e}clet number and the results of $L_2$ norm error offered by LBM, DVM and DUGKS-T2S5. For the DUGKS-T2S2CD scheme, an increase in CFL number would generate a decrease in the overall $L_2$ norm error. As has been explained earlier, the spatial dissipation dominates in this whole process and less time steps (large CFL number) would alleviate the error accumulation. This explanation can also account for the similar trends observed in the results offered by DUGKS-T2S3 and DUGKS-T3S3. when the spatial dissipation is reduced by employing higher-order interpolation algorithm, as is implemented in DUGKS-T2S5, the accumulated error would be suppressed. The $L_2$ norm error is mainly influenced by the temporal dissipation under such a circumstance. Increases in CFL number would enhance the temporal dissipation. Hence the best result obtained from DUGKS-T2S5 is the one with minimum CFL number. It is worth pointing out that the distinction amongst the results from DUGKS-T2S5 is nearly unnoticeable, which demonstrates that the temporal dissipation has limited effects on the performance of those methods. The results obtained by DVM with different CFL number are presented in Table~\ref{tab:IE:LBMDVM-Pe-L2Phi}. Although the difference between them is tiny, DVM-CFL0.8 has shown an advantage over DVM-CFL1.0 when the P\'{e}clet number is relatively large. As is mentioned, this outcome is really abnormal because the spatial dissipation of DVM-CFL0.8 is greater than that of DVM-CFL1.0. The reason may be attributed to the sudden reverse of velocity field in this test. The evolution of interface relies directly on the information calculated from the previous time step. The sudden reverse of velocity field would bring in nonphysical distortion in this process. Given a bigger time step, this distortion would play a greater role in such a situation. From another point of view, the interface extension test, which involves the sudden reverse of velocity field, is not a perfect benchmark test for validating performance of numerical methods.
\\
Fig.~{\ref{FIG:IE:M-L2Phi}} illustrates the results of $L_2$ norm error obtained by multiple methods with different mobility coefficients and the corresponding data are summarized in Table~\ref{tab:IE:M-L2Phi} exhaustively. Except for the results achieved at relatively small mobility coefficients, the $L_2$ norm error obtained by all of schemes tends to rise up as the mobility coefficient increases. The sudden reverse of velocity field should take some responsibility for this phenomenon since larger mobility coefficient means higher velocity in such a condition. Thus the distortion in velocity field would be more severe as the mobility coefficient increases. The similar phenomenon that DVM-CFL0.8 outperforms DVM-CFL1.0 at relatively large mobility coefficients can be observed and the underlying reason has been explained in the previous paragraph. For the method of DUGKS-T2S2CD, the best result is still achieved with a CFL number of 0.8, which is in accord with the explanation provided above. When the mobility coefficient is small (no more than 0.04), DUGKS-T2S3 with small CFL numbers have produced terrible results, which can be observed in Fig.~\ref{FIG:IE:DUGKST2S3-M-L2Phi}. The same condition can also be found in the results provided by DUGKS-T3S3 with small CFL numbers. However, when the CFL number is increased to 0.8, the improvements in the results are apparent. As we have described earlier, larger CFL number in DUGKS based on upwind flux would be able to reduce the spatial truncation error and in turn prevent the numerical dissipation. It is recommended to maximize the CFL number if DUGKS with upwind flux is selected to conduct a simulation. The differences among the results provided by DUGKS-T2S5 with various CFL numbers have proved again that temporal dissipation is almost negligible in this test. The minimum $L_2$ norm error is obtained when the CFL number is set to the smallest value. This tendency is in accordance with the one presented in Fig.~\ref{FIG:IE:DUGKST2S5-Pe-L2Phi} and the underlying reason has been stated above.
\\
\begin{table}[htbp]
\caption
{
  \label{tab:IE:Pe-L2Phi}
  $L_2$-norm error of $\phi$ for interface extension obtained by multiple methods with various $Pe$, $Cn = 4/256$, $M_\phi = 0.02$.
}
	\begin{minipage}{1.0\textwidth}
	\begin{subtable}{1.0\textwidth}
	\caption{\label{tab:IE:LBMDVM-Pe-L2Phi}LBM, DVM}
		\begin{ruledtabular}
		\begin{tabular}{ccccccc}
		Pe&256&512&640&819.2&1024&1280\\
		\colrule
		LBM & $1.42\times10^{-2}$ & $1.29\times10^{-2}$ & $1.30\times10^{-2}$ & $1.38\times10^{-2}$ & $1.49\times10^{-2}$ & $1.67\times10^{-2}$\\
		DVM-CFL1.0 & $1.42\times10^{-2}$ & $1.29\times10^{-2}$ & $1.30\times10^{-2}$ & $1.38\times10^{-2}$ & $1.49\times10^{-2}$ & $1.67\times10^{-2}$\\
		DVM-CFL0.8 & $1.60\times10^{-2}$ & $1.31\times10^{-2}$ & $1.28\times10^{-2}$ & $1.31\times10^{-2}$ & $1.42\times10^{-2}$ & $1.61\times10^{-2}$\\
		\end{tabular}
		\end{ruledtabular}
	\end{subtable}%
	\end{minipage}
	\par\medskip
	\begin{minipage}{1.0\textwidth}
	\begin{subtable}{1.0\textwidth}
	\caption{\label{tab:IE:DUGKS-T2S2CD-Pe-L2Phi}DUGKS-T2S2CD}
		\begin{ruledtabular}
		\begin{tabular}{ccccccc}
		Pe&256&512&640&819.2&1024&1280\\
		\colrule		
		CFL0.25 & $2.82\times10^{-2}$ & $2.99\times10^{-2}$ & $3.23\times10^{-2}$ & $3.70\times10^{-2}$ & $4.29\times10^{-2}$ & $5.00\times10^{-2}$\\
		CFL0.5 & $2.59\times10^{-2}$ & $2.79\times10^{-2}$ & $3.03\times10^{-2}$ & $3.45\times10^{-2}$ & $3.98\times10^{-2}$ & $4.61\times10^{-2}$\\
		CFL0.8 & $2.32\times10^{-2}$ & $2.37\times10^{-2}$ & $2.55\times10^{-2}$ & $2.89\times10^{-2}$ & $3.30\times10^{-2}$ & $3.81\times10^{-2}$\\
		\end{tabular}
		\end{ruledtabular}
	\end{subtable}%
	\end{minipage}
	\par\medskip
	\begin{minipage}{1.0\textwidth}
	\begin{subtable}{1.0\textwidth}
	\caption{\label{tab:IE:DUGKS-T2S3-Pe-L2Phi}DUGKS-T2S3}
		\begin{ruledtabular}
		\begin{tabular}{ccccccc}
		Pe&256&512&640&819.2&1024&1280\\
		\colrule
		CFL0.25 & $4.26\times10^{-2}$ & $3.54\times10^{-2}$ & $3.24\times10^{-2}$ & $2.94\times10^{-2}$ & $2.79\times10^{-2}$ & $2.45\times10^{-2}$\\
		CFL0.5 & $2.19\times10^{-2}$ & $1.95\times10^{-2}$ & $1.92\times10^{-2}$ & $1.96\times10^{-2}$ & $1.97\times10^{-2}$ & $2.09\times10^{-2}$\\
		CFL0.8 & $1.93\times10^{-2}$ & $1.76\times10^{-2}$ & $1.77\times10^{-2}$ & $1.78\times10^{-2}$ & $1.85\times10^{-2}$ & $1.98\times10^{-2}$\\
		\end{tabular}
		\end{ruledtabular}
	\end{subtable}%
	\end{minipage}
	\par\medskip
	\begin{minipage}{1.0\textwidth}
	\begin{subtable}{1.0\textwidth}
	\caption{\label{tab:IE:DUGKS-T3S3-Pe-L2Phi}DUGKS-T3S3}
		\begin{ruledtabular}
		\begin{tabular}{ccccccc}
		Pe&256&512&640&819.2&1024&1280\\
		\colrule
		CFL0.25 & $6.67\times10^{-2}$ & $5.45\times10^{-2}$ & $4.80\times10^{-2}$ & $4.08\times10^{-2}$ & $3.58\times10^{-2}$ & $3.28\times10^{-2}$\\
		CFL0.5 & $3.78\times10^{-2}$ & $3.14\times10^{-2}$ & $2.88\times10^{-2}$ & $2.64\times10^{-2}$ & $2.53\times10^{-2}$ & $2.54\times10^{-2}$
		\end{tabular}
		\end{ruledtabular}
	\end{subtable}%
	\end{minipage}
	\par\medskip
	\begin{minipage}{1.0\textwidth}
	\begin{subtable}{1.0\textwidth}
	\caption{\label{tab:IE:DUGKS-T2S5-Pe-L2Phi}DUGKS-T2S5}
		\begin{ruledtabular}
		\begin{tabular}{ccccccc}
		Pe&256&512&640&819.2&1024&1280\\
		\colrule
		CFL0.25 & $1.72\times10^{-2}$ & $1.55\times10^{-2}$ & $1.59\times10^{-2}$ & $1.65\times10^{-2}$ & $1.76\times10^{-2}$ & $1.92\times10^{-2}$\\
		CFL0.5 & $1.77\times10^{-2}$ & $1.59\times10^{-2}$ & $1.62\times10^{-2}$ & $1.70\times10^{-2}$ & $1.81\times10^{-2}$ & $1.98\times10^{-2}$\\
		CFL0.8 & $1.83\times10^{-2}$ & $1.65\times10^{-2}$ & $1.67\times10^{-2}$ & $1.76\times10^{-2}$ & $1.88\times10^{-2}$ & $2.05\times10^{-2}$\\
		\end{tabular}
		\end{ruledtabular}
	\end{subtable}%
	\end{minipage}
\end{table}
\begin{table}[htbp]
\caption
{
  \label{tab:IE:M-L2Phi}
  $L_2$-norm error of $\phi$ for interface extension obtained by multiple methods with various $M_\phi$, $Cn = 4/256$, $Pe = 256$.
}
	\begin{minipage}{1.0\textwidth}
	\begin{subtable}{1.0\textwidth}
	\caption{\label{tab:IE:LBMDVM-M-L2Phi}LBM, DVM}
		\begin{ruledtabular}
		\begin{tabular}{cccccccc}
		$M_\phi$&0.01&0.02&0.04&0.05&0.064&0.08&0.1\\
		\colrule
		LBM & $1.40\times10^{-2}$ & $1.42\times10^{-2}$ & $1.51\times10^{-2}$ & $1.60\times10^{-2}$ & $1.83\times10^{-2}$ & $2.12\times10^{-2}$ & $2.60\times10^{-2}$\\
		DVM-CFL1.0 & $1.40\times10^{-2}$ & $1.42\times10^{-2}$ & $1.51\times10^{-2}$ & $1.60\times10^{-2}$ & $1.83\times10^{-2}$ & $2.12\times10^{-2}$ & $2.60\times10^{-2}$\\
		DVM-CFL0.8 & $1.92\times10^{-2}$ & $1.60\times10^{-2}$ & $1.49\times10^{-2}$ & $1.49\times10^{-2}$ & $1.51\times10^{-2}$ & $1.63\times10^{-2}$ & $1.97\times10^{-2}$\\
		\end{tabular}
		\end{ruledtabular}
	\end{subtable}%
	\end{minipage}
	\par\medskip
	\begin{minipage}{1.0\textwidth}
	\begin{subtable}{1.0\textwidth}
	\caption{\label{tab:IE:DUGKS-T2S2CD-M-L2Phi}DUGKS-T2S2CD}
		\begin{ruledtabular}
		\begin{tabular}{cccccccc}
		$M_\phi$&0.01&0.02&0.04&0.05&0.064&0.08&0.1\\
		\colrule
		CFL0.25 & $2.73\times10^{-2}$ & $2.82\times10^{-2}$ & $3.09\times10^{-2}$ & $3.25\times10^{-2}$ & $3.48\times10^{-2}$ & $3.77\times10^{-2}$ & $4.19\times10^{-2}$
		\\
		CFL0.5 & $2.60\times10^{-2}$ & $2.59\times10^{-2}$ & $2.75\times10^{-2}$ & $2.87\times10^{-2}$ & $3.07\times10^{-2}$ & $3.35\times10^{-2}$ & $3.73\times10^{-2}$
		\\
		CFL0.8 & $2.36\times10^{-2}$ & $2.32\times10^{-2}$ & $2.37\times10^{-2}$ & $2.43\times10^{-2}$ & $2.58\times10^{-2}$ & $2.85\times10^{-2}$ & $3.23\times10^{-2}$
		\\
		\end{tabular}
		\end{ruledtabular}
	\end{subtable}%
	\end{minipage}
	\par\medskip
	\begin{minipage}{1.0\textwidth}
	\begin{subtable}{1.0\textwidth}
	\caption{\label{tab:IE:DUGKS-T2S3-M-L2Phi}DUGKS-T2S3}
		\begin{ruledtabular}
		\begin{tabular}{cccccccc}
		$M_\phi$&0.01&0.02&0.04&0.05&0.064&0.08&0.1\\
		\colrule
		CFL0.25 & $7.06\times10^{-2}$ & $4.26\times10^{-2}$ & $2.43\times10^{-2}$ & $2.26\times10^{-2}$ & $2.25\times10^{-2}$ & $2.37\times10^{-2}$ & $2.68\times10^{-2}$\\
		CFL0.5 & $2.56\times10^{-2}$ & $2.19\times10^{-2}$ & $2.06\times10^{-2}$ & $2.12\times10^{-2}$ & $2.22\times10^{-2}$ & $2.40\times10^{-2}$ &$2.75\times10^{-2}$\\
		CFL0.8 & $1.94\times10^{-2}$ & $1.96\times10^{-2}$ & $2.02\times10^{-2}$ & $2.14\times10^{-2}$ & $2.29\times10^{-2}$ & $2.52\times10^{-2}$ & $2.92\times10^{-2}$\\
		\end{tabular}
		\end{ruledtabular}
	\end{subtable}%
	\end{minipage}
	\par\medskip
	\begin{minipage}{1.0\textwidth}
	\begin{subtable}{1.0\textwidth}
	\caption{\label{tab:IE:DUGKS-T3S3-M-L2Phi}DUGKS-T3S3}
		\begin{ruledtabular}
		\begin{tabular}{cccccccc}
		$M_\phi$&0.01&0.02&0.04&0.05&0.064&0.08&0.1\\
		\colrule
		CFL0.25 & $1.20\times10^{-1}$ & $6.67\times10^{-2}$ & $2.92\times10^{-2}$ & $2.43\times10^{-2}$ & $2.28\times10^{-2}$ & $2.35\times10^{-2}$ & $2.60\times10^{-2}$\\
		CFL0.5 & - & $3.78\times10^{-2}$ & $2.29\times10^{-2}$ & $2.20\times10^{-2}$ & $2.21\times10^{-2}$ & $2.33\times10^{-2}$ &$2.61\times10^{-2}$\\
		CFL0.8 & - & - & $2.08\times10^{-2}$ & $2.12\times10^{-2}$ & $2.20\times10^{-2}$ & $2.35\times10^{-2}$ & $2.65\times10^{-2}$\\
		\end{tabular}
		\end{ruledtabular}
	\end{subtable}%
	\end{minipage}
	\par\medskip
	\begin{minipage}{1.0\textwidth}
	\begin{subtable}{1.0\textwidth}
	\caption{\label{tab:IE:DUGKS-T2S5-M-L2Phi}DUGKS-T2S5}
		\begin{ruledtabular}
		\begin{tabular}{cccccccc}
		$M_\phi$&0.01&0.02&0.04&0.05&0.064&0.08&0.1\\
		\colrule
		CFL0.25 & $1.74\times10^{-2}$ & $1.72\times10^{-2}$ & $1.80\times10^{-2}$ & $1.89\times10^{-2}$ & $2.07\times10^{-2}$ & $2.41\times10^{-2}$ & $2.77\times10^{-2}$\\
		CFL0.5 & $1.76\times10^{-2}$ & $1.77\times10^{-2}$ & $1.86\times10^{-2}$ & $1.96\times10^{-2}$ & $2.21\times10^{-2}$ & $2.46\times10^{-2}$ & $2.84\times10^{-2}$\\
		CFL0.8 & $1.80\times10^{-2}$ & $1.83\times10^{-2}$ & $1.95\times10^{-2}$ & $2.07\times10^{-2}$ & $2.30\times10^{-2}$ & $2.54\times10^{-2}$ & $2.98\times10^{-2}$\\
		\end{tabular}
		\end{ruledtabular}
	\end{subtable}%
	\end{minipage}
\end{table}
\subsection{\label{sec:sec3.D}Smoothed Deformation}
\begin{figure}[htbp]
    \centering
    \begin{subfigure}{0.400 \columnwidth}
      \centering
      \includegraphics[width=1\linewidth]{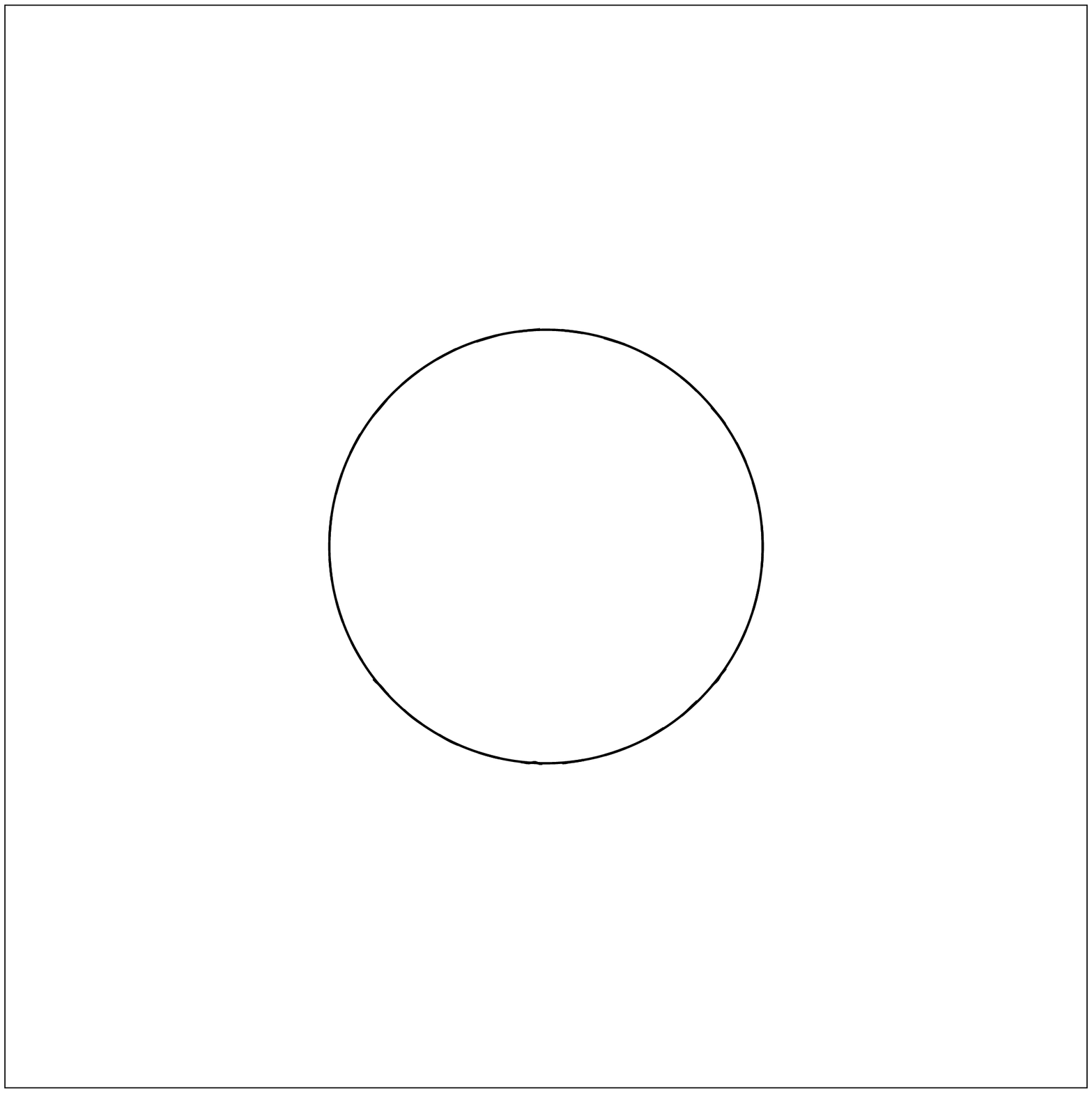}
      \caption[]{DVM, $C = 1.0$}
      \label{FIG:SD:DVMCFL1.0-4-4T}
    \end{subfigure}
    \begin{subfigure}{0.400 \columnwidth}
      \centering
      \includegraphics[width=1\linewidth]{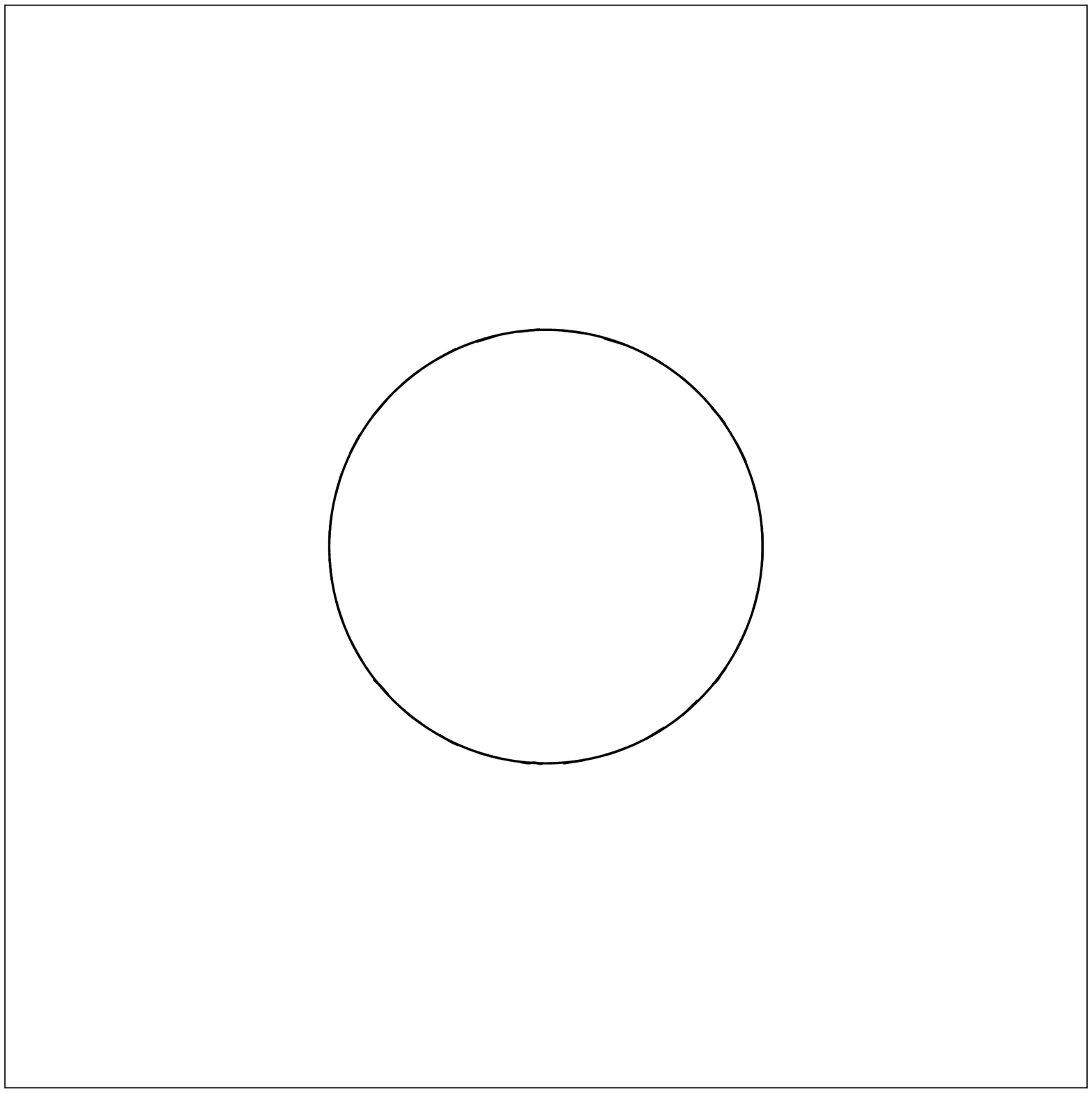}
      \caption{DVM, $C = 0.8$}
      \label{FIG:SD:DVMCFL0.8-4-4T}
    \end{subfigure}
    \hfill
    \begin{subfigure}{0.400 \columnwidth}
      \centering
      \includegraphics[width=1\linewidth]{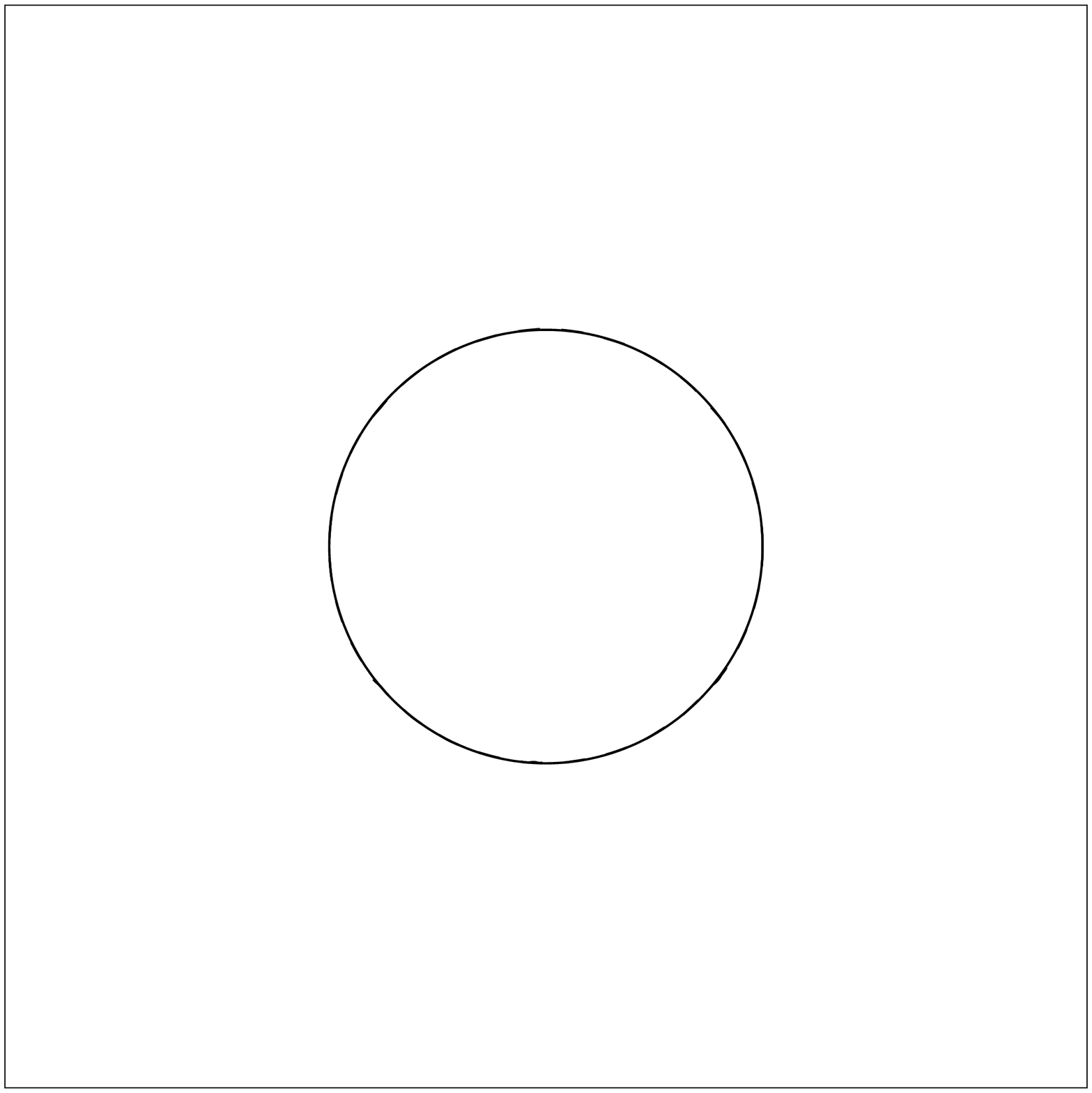}
      \caption[]{DUGKS-T2S2CD, $C = 0.8$}
      \label{FIG:SD:DUGKS-T2S2CD-4-4T}
    \end{subfigure}
    \begin{subfigure}{0.400 \columnwidth}
      \centering
      \includegraphics[width=1\linewidth]{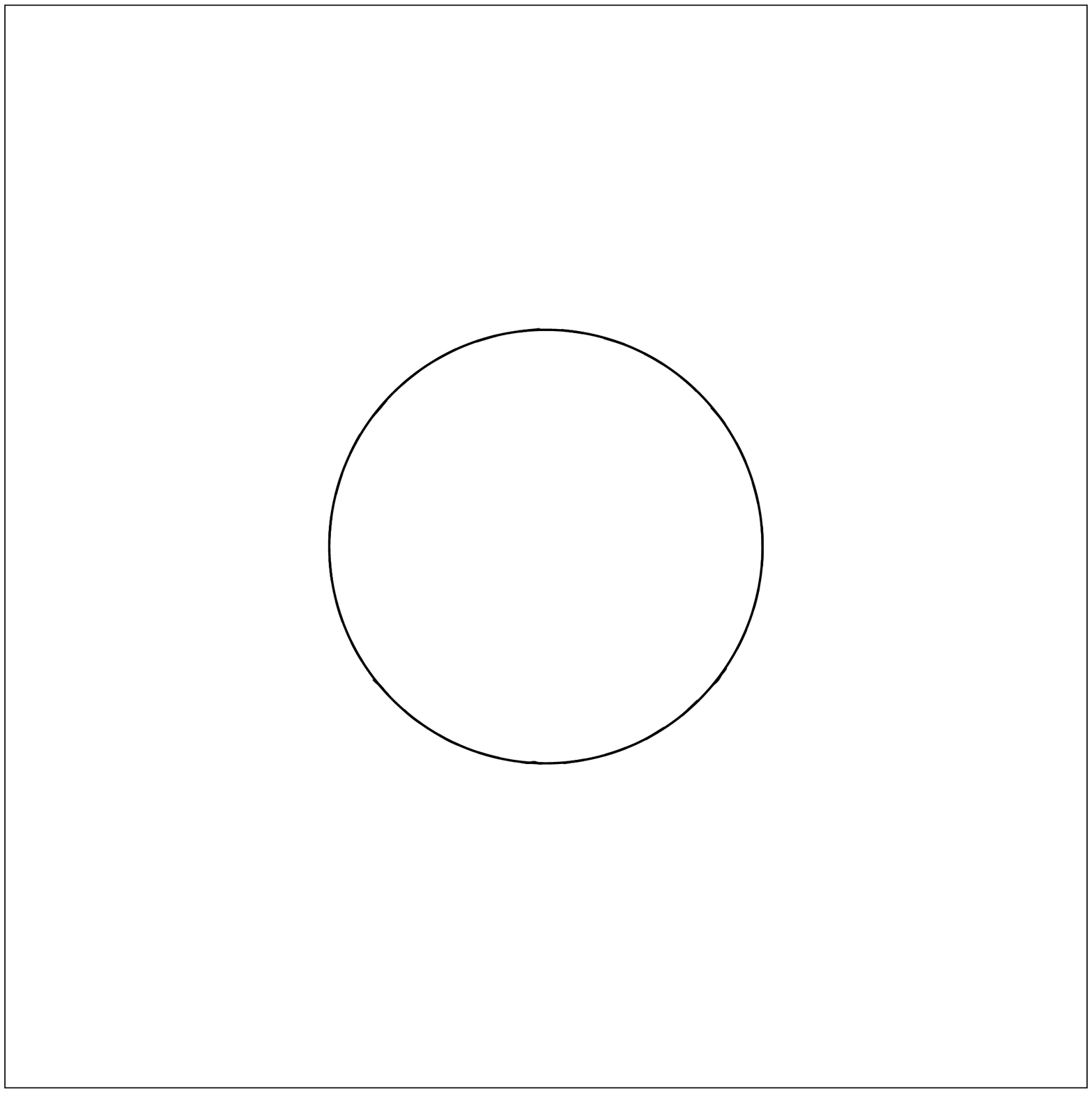}
      \caption[]{DUGKS-T2S5, $C = 0.25$}
      \label{FIG:SD:DUGKS-T2S5-4-4T}
    \end{subfigure}
    \hfill
    \begin{subfigure}[b]{0.400\textwidth}
      \centering
      \includegraphics[width=1\linewidth]{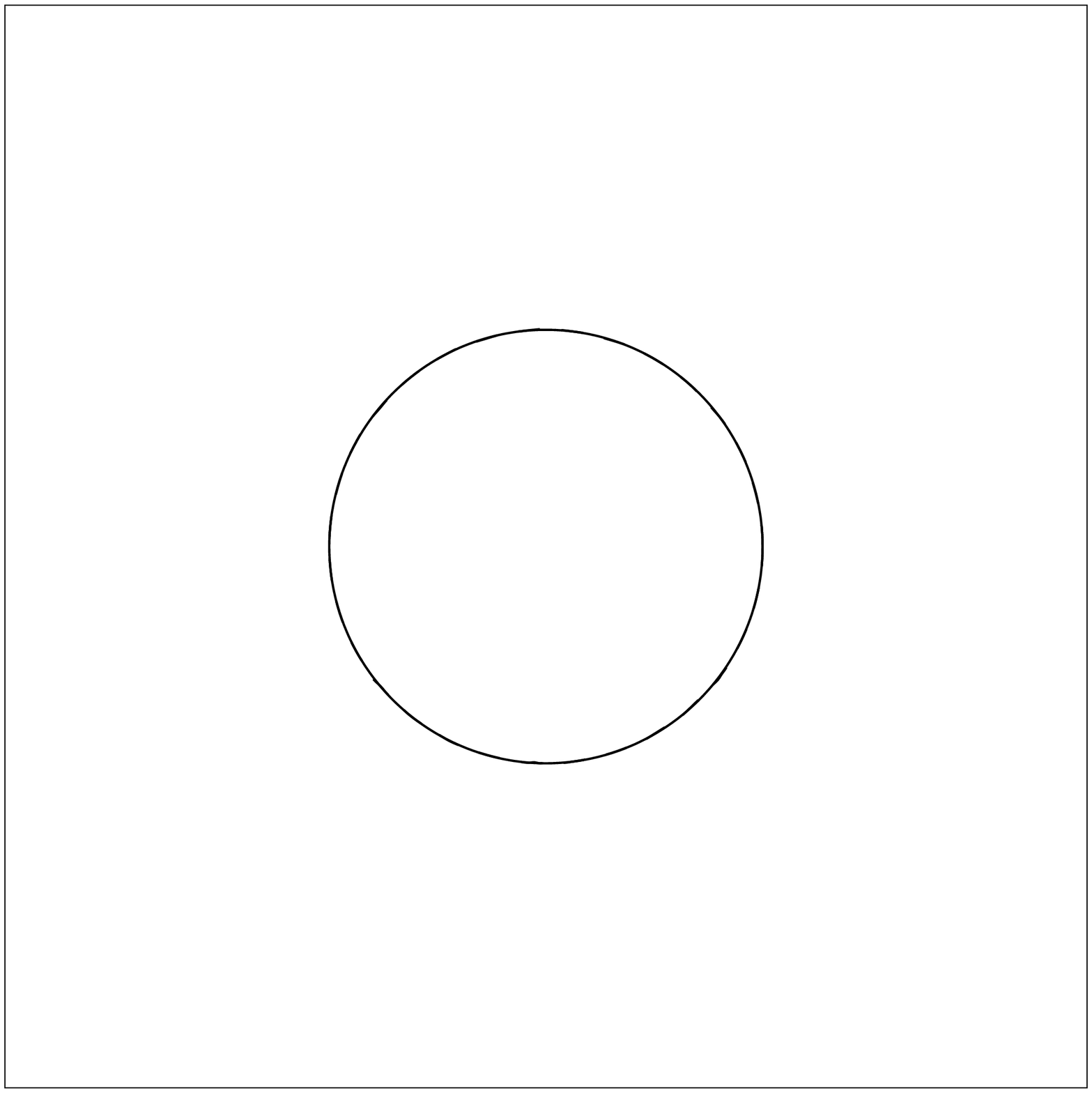}
      \caption[]{DUGKS-T2S3, $C = 0.8$}
      \label{FIG:SD:DUGKS-T2S3-4-4T}
    \end{subfigure}
    \begin{subfigure}[b]{0.400\textwidth}
      \centering
      \includegraphics[width=1\linewidth]{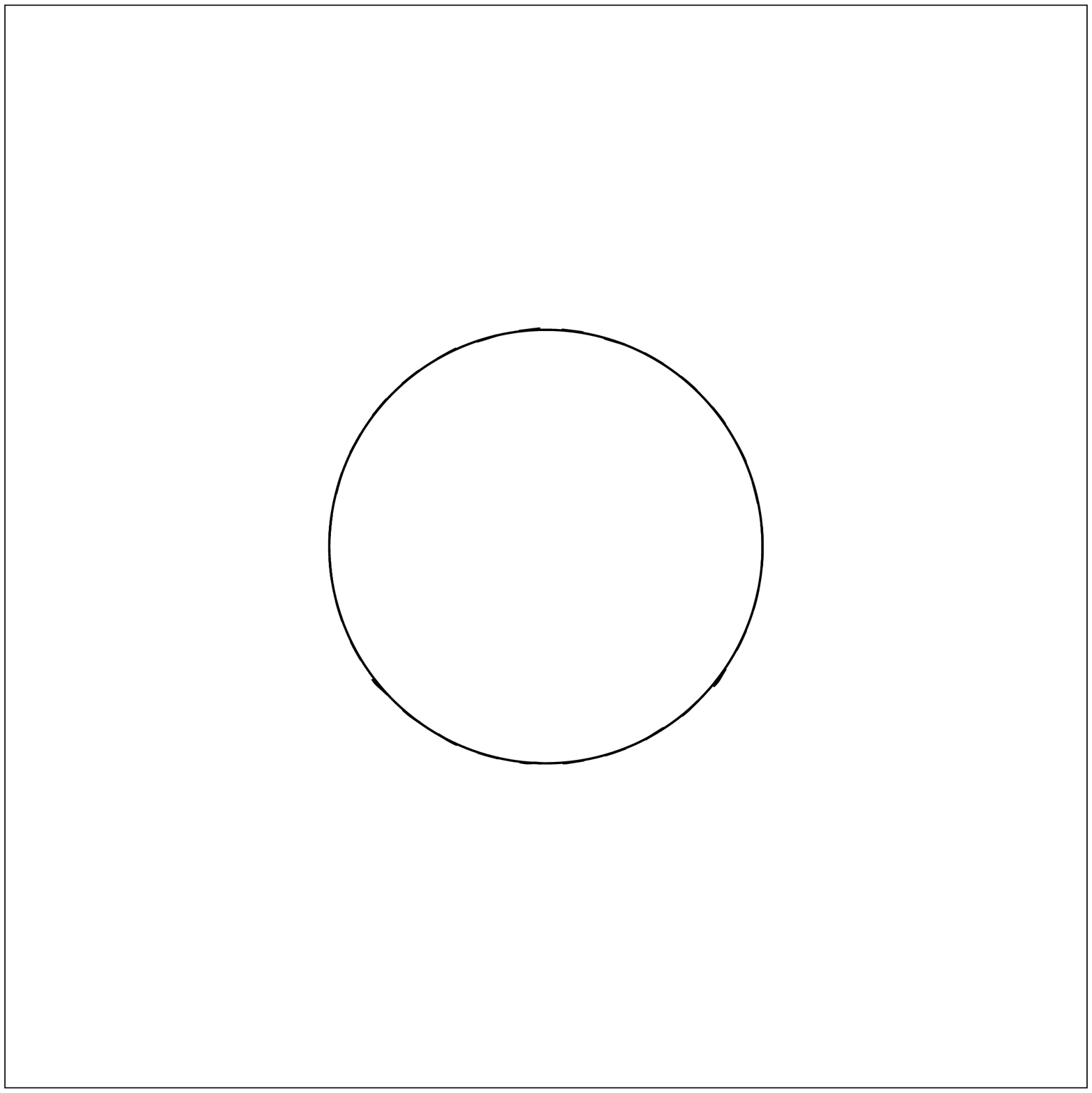}
      \caption[]{DUGKS-T3S3, $C = 0.5$}
      \label{FIG:SD:DUGKS-T3S3-4-4T}
    \end{subfigure}
\caption{Interface deformation in a smoothed shear flow with $Pe = 2048$, $Cn = 4/512$, $M_\phi = 0.02$.}
\label{FIG:SD:Scheme-4-4T}
\end{figure}

\begin{figure}[htbp]
    \centering
    \begin{minipage}[b]{0.9 \columnwidth}
    \begin{subfigure}{0.45 \columnwidth}
      \centering
      \includegraphics[width=1\linewidth]{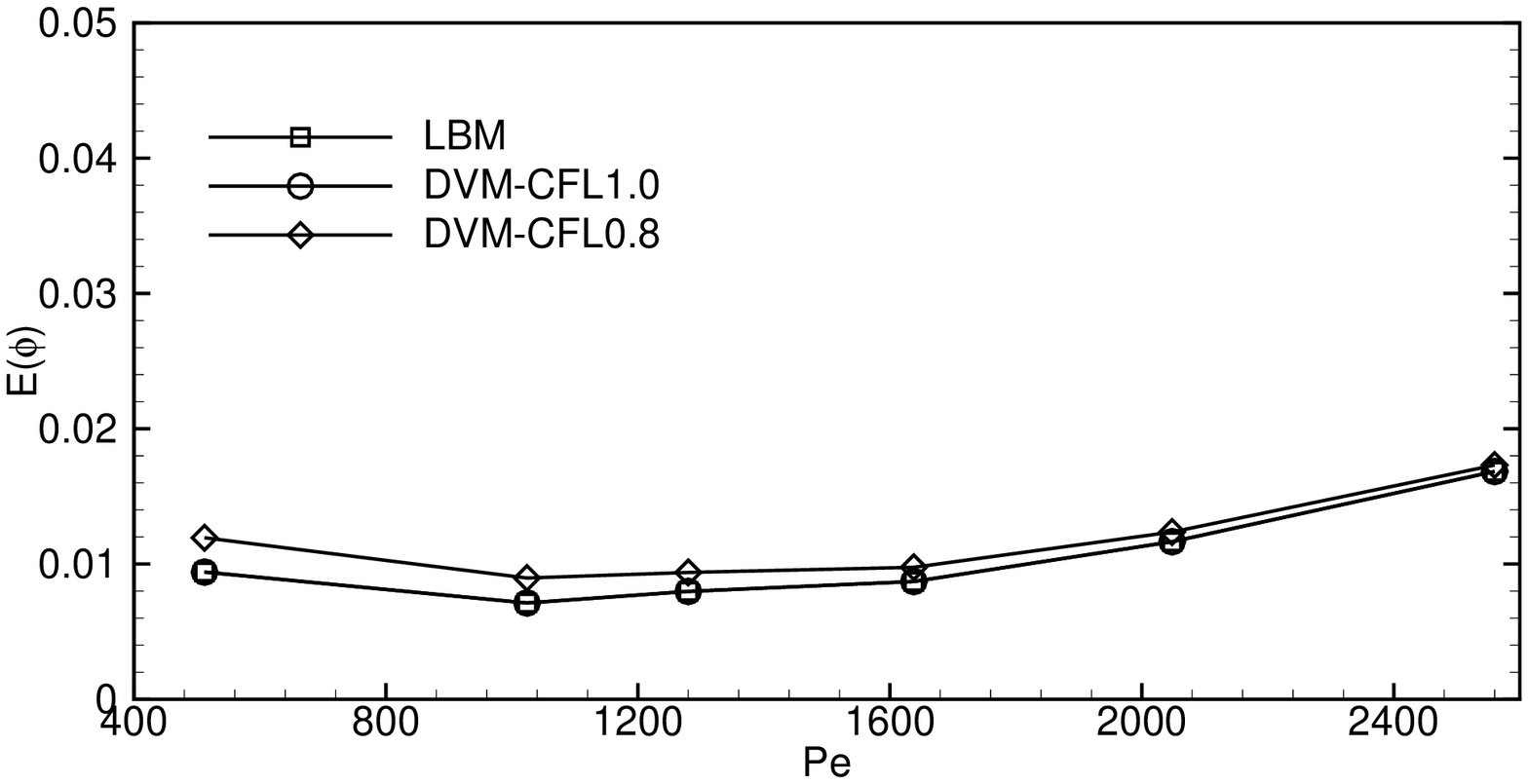}
      \caption{LBM, DVM}
      \label{FIG:SD:LBMDVM-Pe-L2Phi}
    \end{subfigure}
    \end{minipage}
    \begin{minipage}[b]{0.9 \columnwidth}
    \begin{subfigure}{0.45 \columnwidth}
      \centering
      \includegraphics[width=1\linewidth]{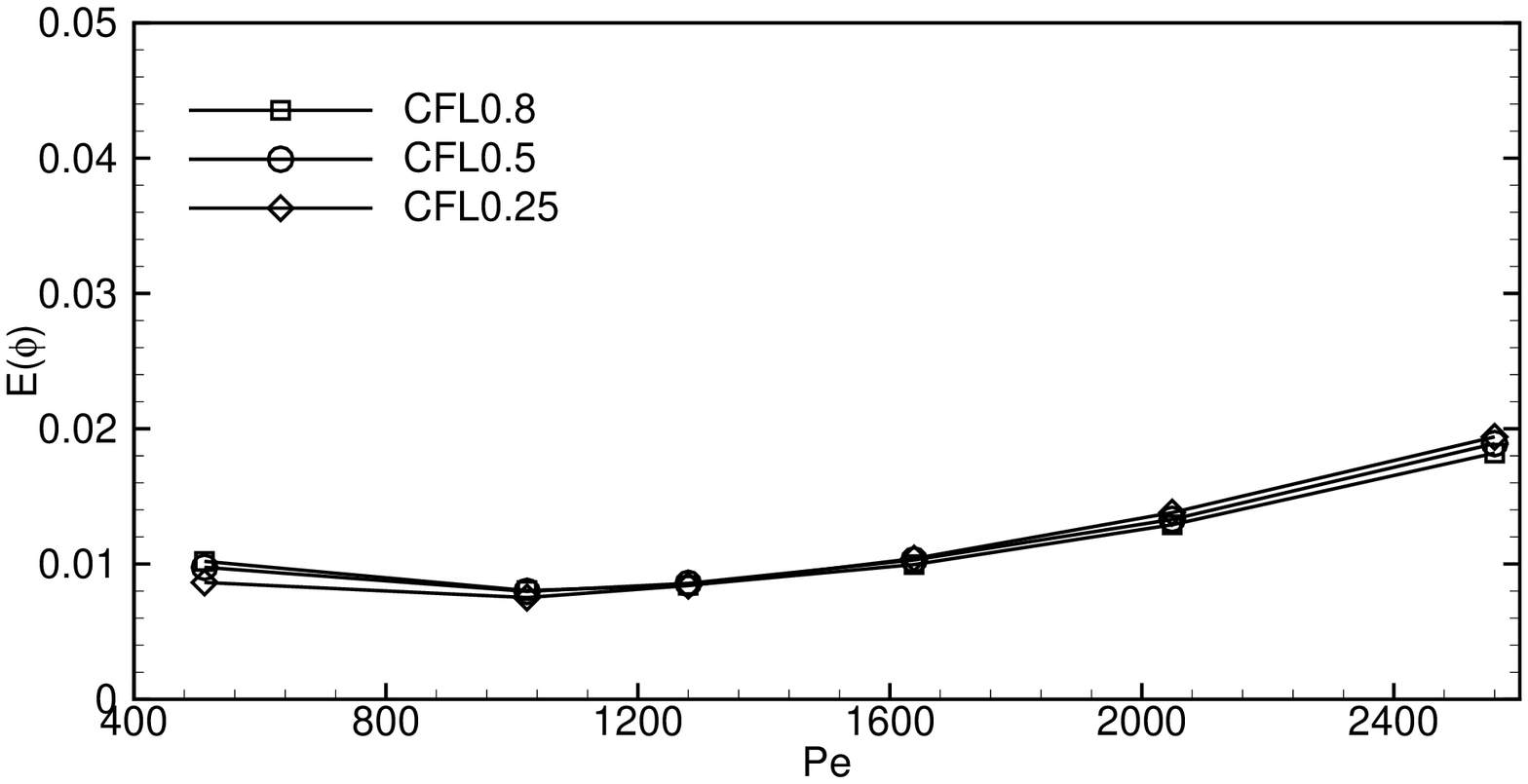}
      \caption{DUGKS-T2S2CD}
      \label{FIG:SD:DUGKST2S2CD-Pe-L2Phi}
    \end{subfigure}
    \end{minipage}
    \begin{minipage}[b]{0.9 \columnwidth}
    \begin{subfigure}{0.45 \columnwidth}
      \centering
      \includegraphics[width=1\linewidth]{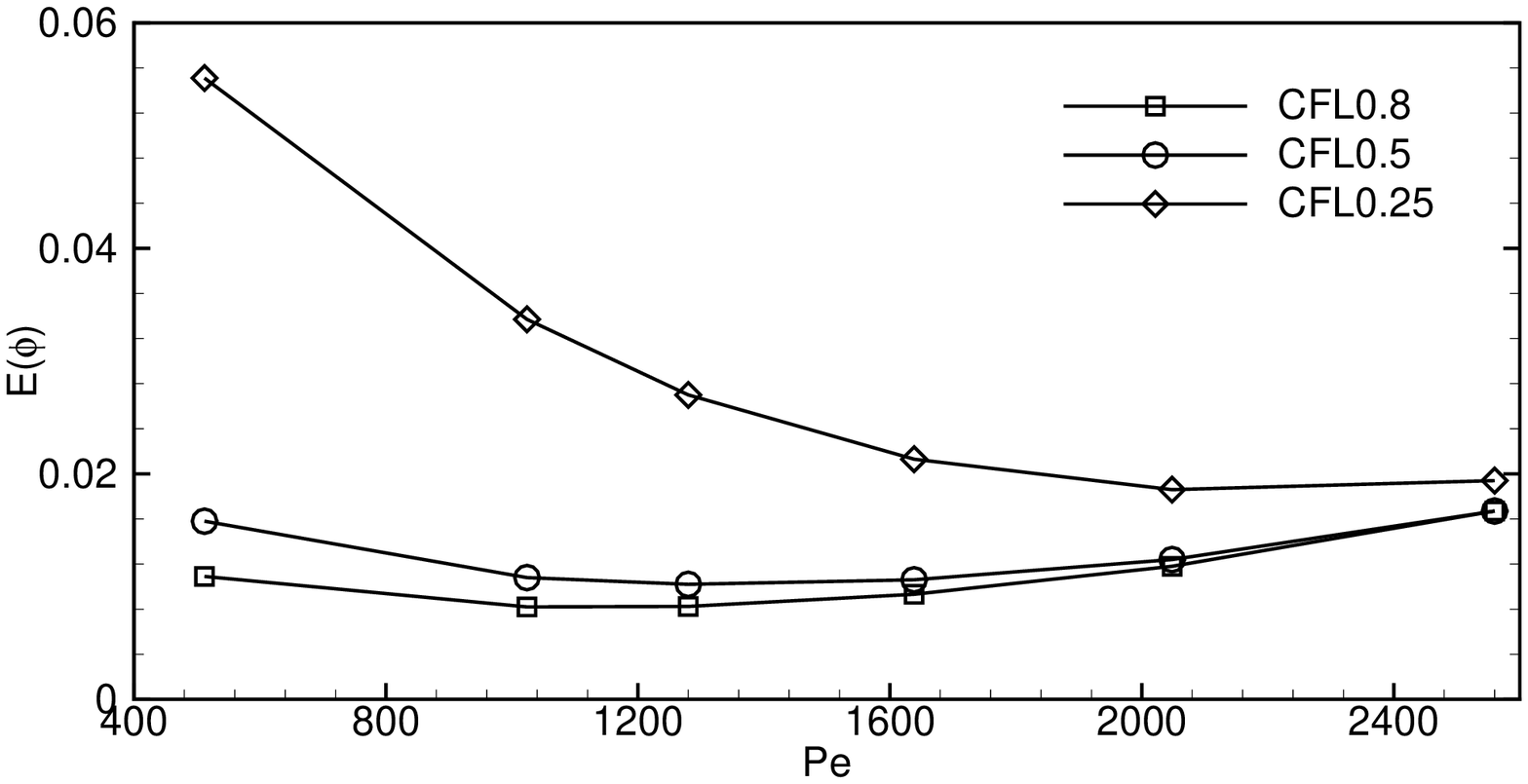}
      \caption{DUGKS-T2S3}
      \label{FIG:SD:DUGKST2S3-Pe-L2Phi}
    \end{subfigure}
    \end{minipage}
    \begin{minipage}[b]{0.9 \columnwidth}
    \begin{subfigure}{0.45 \columnwidth}
      \centering
      \includegraphics[width=1\linewidth]{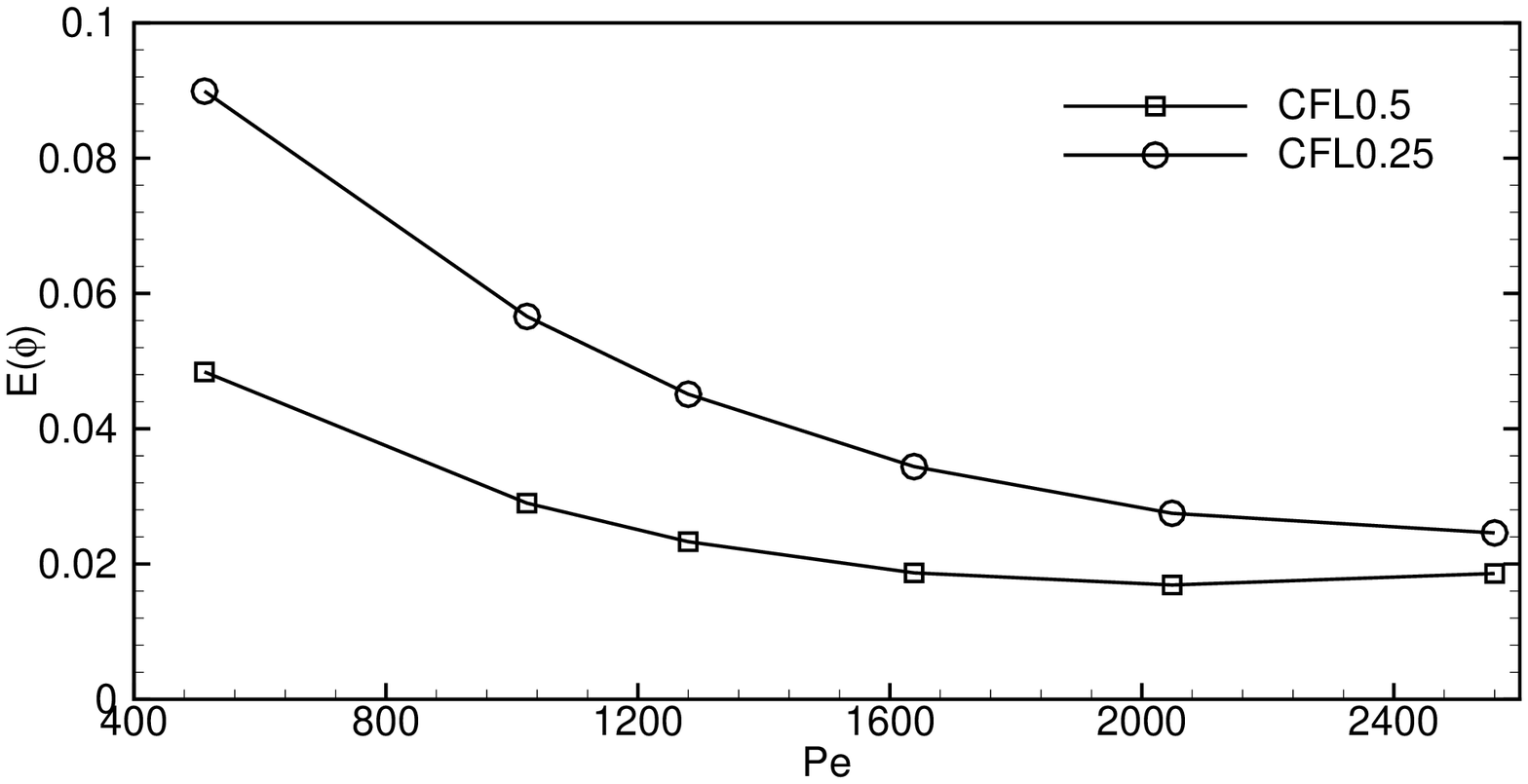}
      \caption{DUGKS-T3S3}
      \label{FIG:SD:DUGKST3S3-Pe-L2Phi}
    \end{subfigure}
    \end{minipage}
    \begin{minipage}[b]{0.9 \columnwidth}
    \begin{subfigure}{0.45 \columnwidth}
      \centering
      \includegraphics[width=1\linewidth]{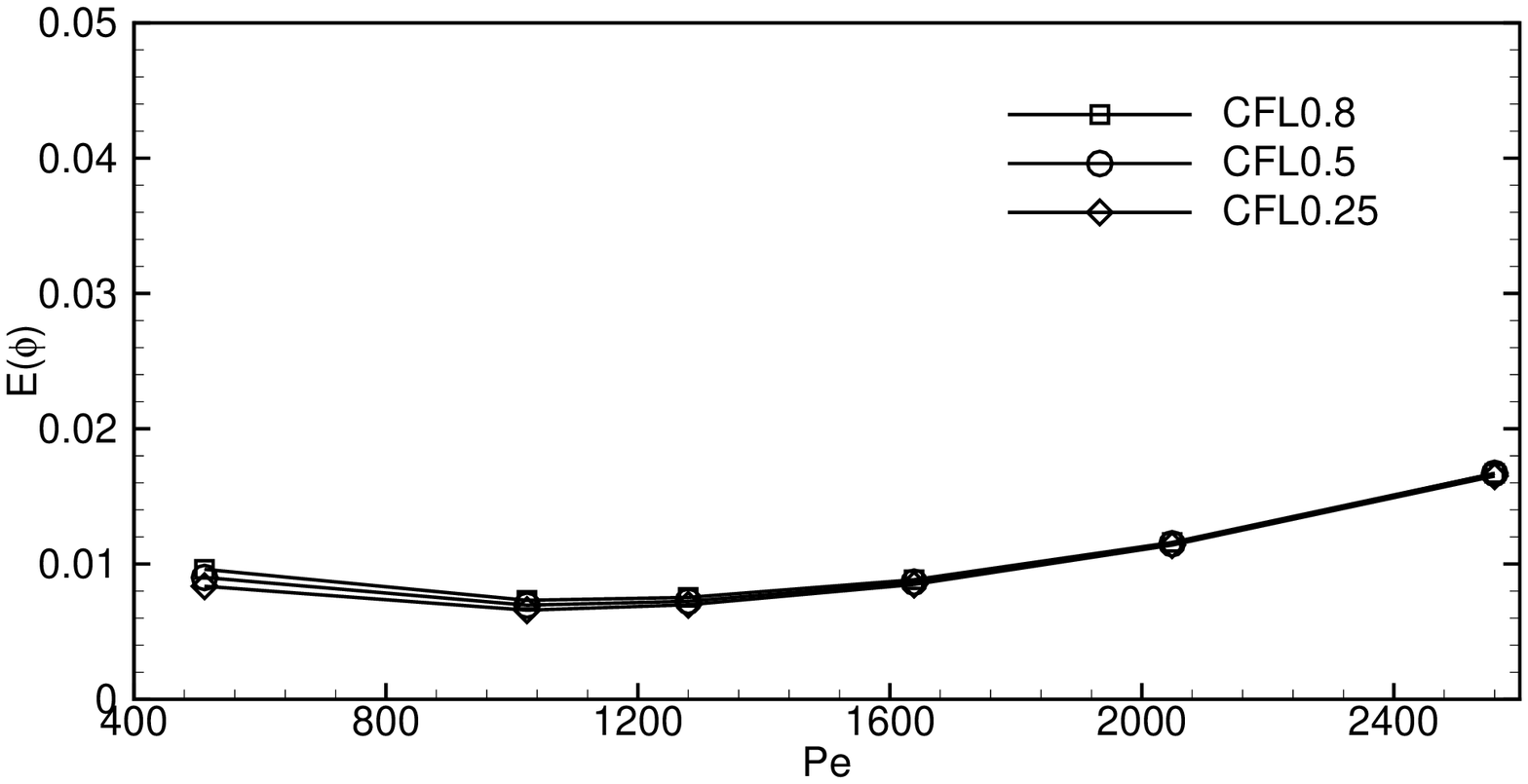}
      \caption{DUGKS-T2S5}
      \label{FIG:SD:DUGKST2S5-Pe-L2Phi}
    \end{subfigure}
    \end{minipage}
\caption{$L_2$-norm error of $\phi$ for smoothed deformation obtained by multiple methods with various $Pe$, $Cn = 4/512$, $M_\phi = 0.02$.}
\label{FIG:SD:Pe-L2Phi}
\end{figure}

\begin{figure}[htbp]
    \centering
    \begin{minipage}[b]{0.9 \columnwidth}
    \begin{subfigure}{0.45 \columnwidth}
      \centering
      \includegraphics[width=1\linewidth]{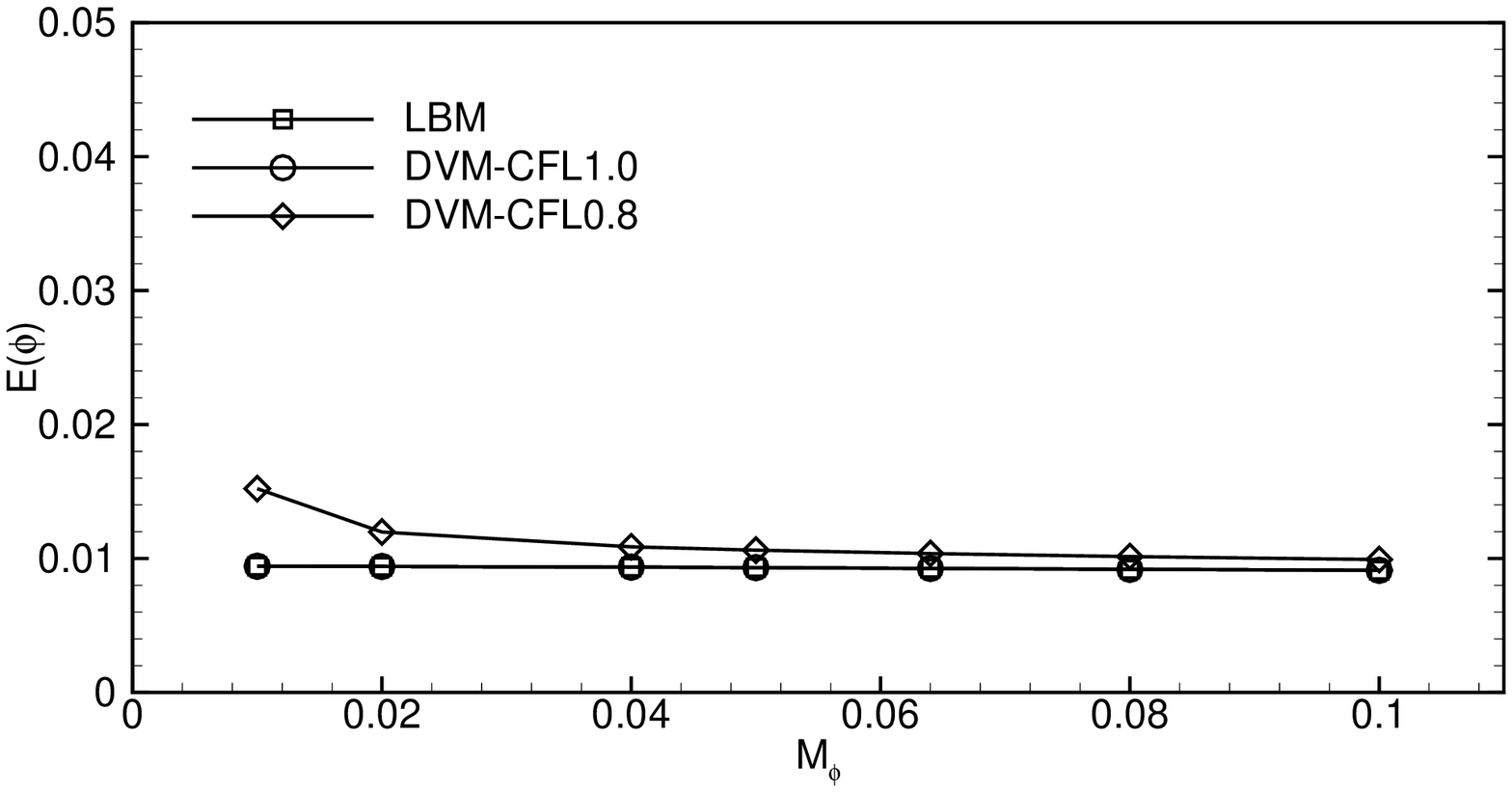}
      \caption{LBM, DVM}
      \label{FIG:SD:LBMDVM-M-L2Phi}
    \end{subfigure}
    \end{minipage}
    \begin{minipage}[b]{0.9 \columnwidth}
    \begin{subfigure}{0.45 \columnwidth}
      \centering
      \includegraphics[width=1\linewidth]{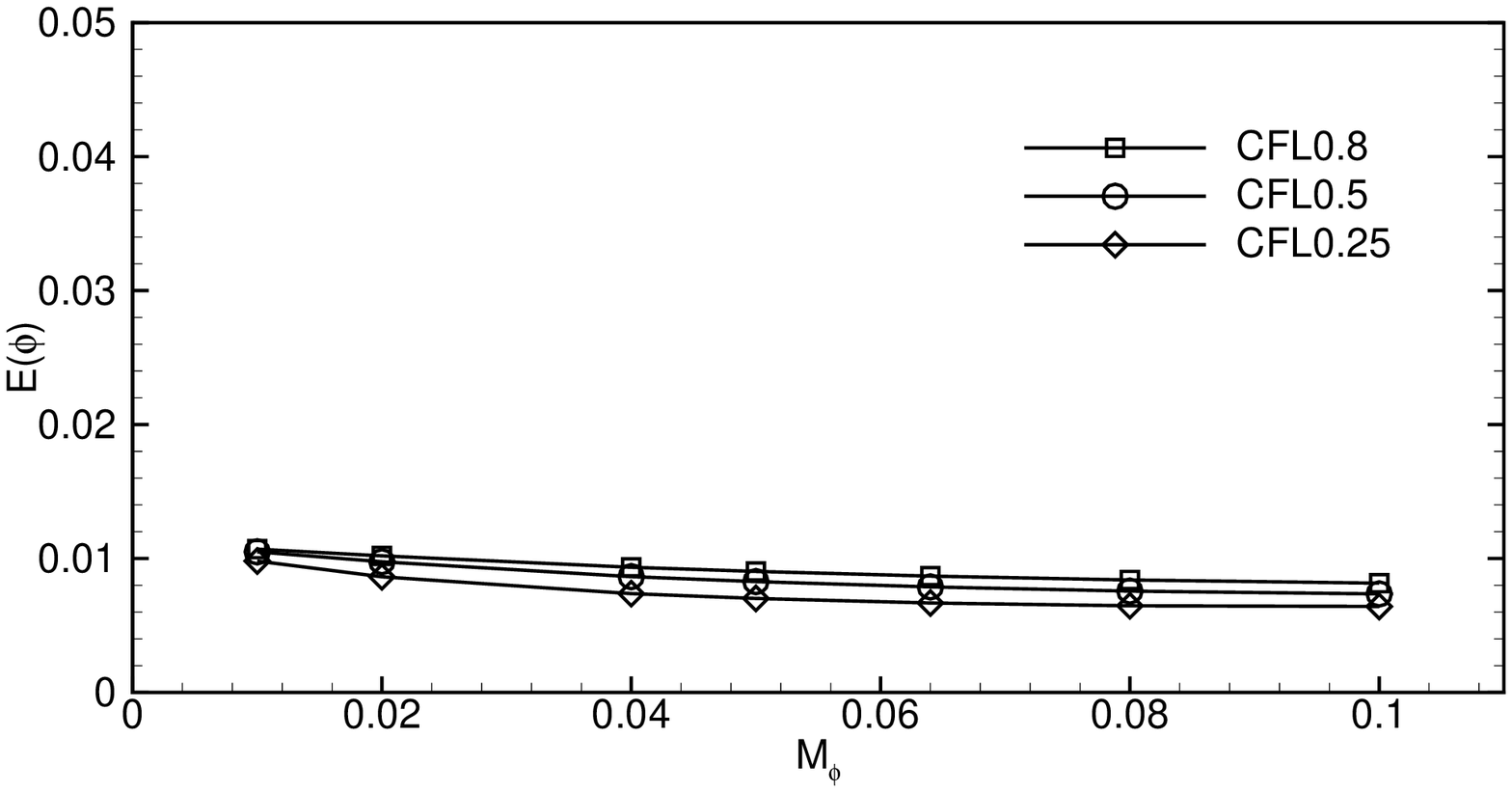}
      \caption{DUGKS-T2S2CD}
      \label{FIG:SD:DUGKST2S2CD-M-L2Phi}
    \end{subfigure}
    \end{minipage}
    \begin{minipage}[b]{0.9 \columnwidth}
    \begin{subfigure}{0.45 \columnwidth}
      \centering
      \includegraphics[width=1\linewidth]{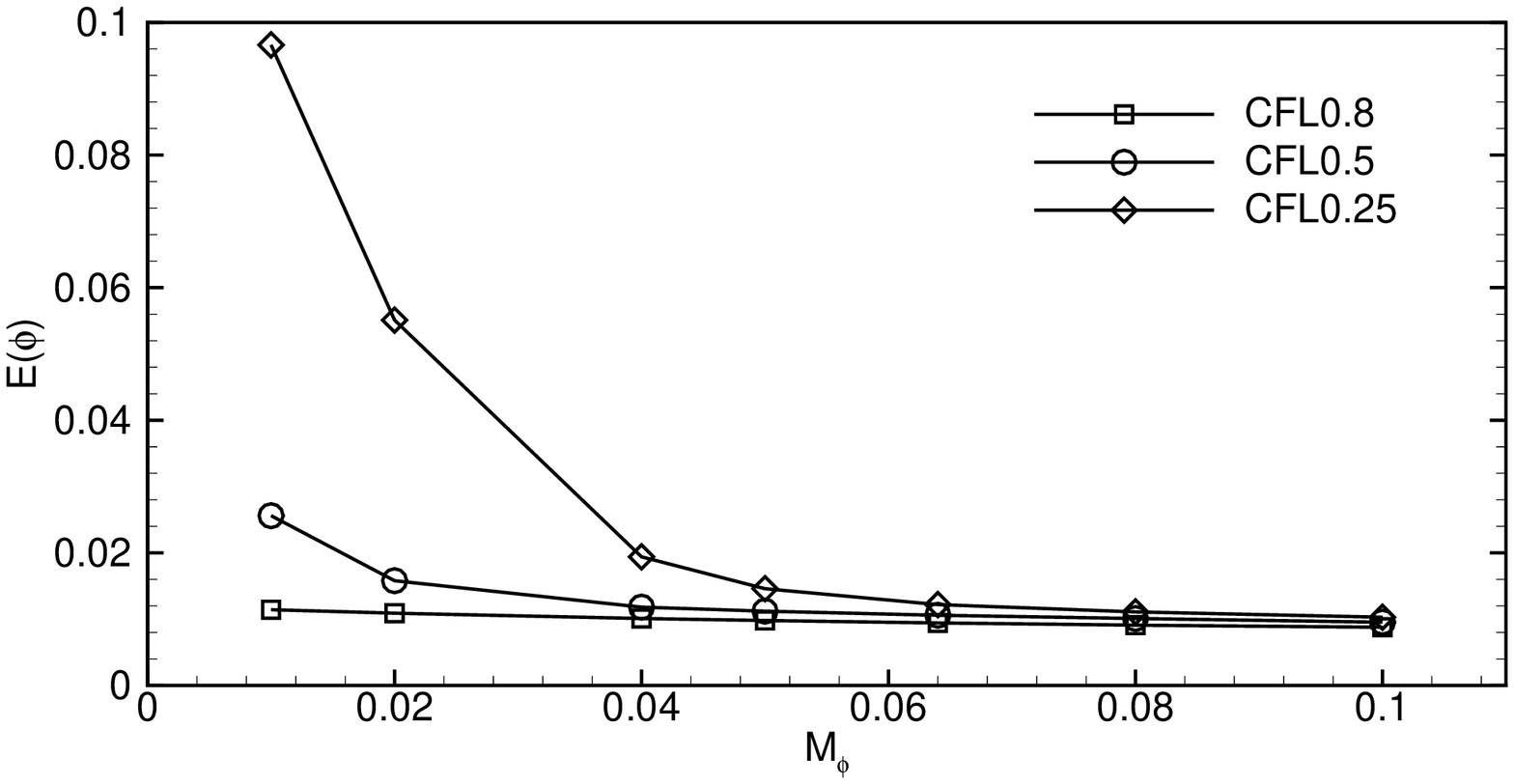}
      \caption{DUGKS-T2S3}
      \label{FIG:SD:DUGKST2S3-M-L2Phi}
    \end{subfigure}
    \end{minipage}
    \begin{minipage}[b]{0.9 \columnwidth}
    \begin{subfigure}{0.45 \columnwidth}
      \centering
      \includegraphics[width=1\linewidth]{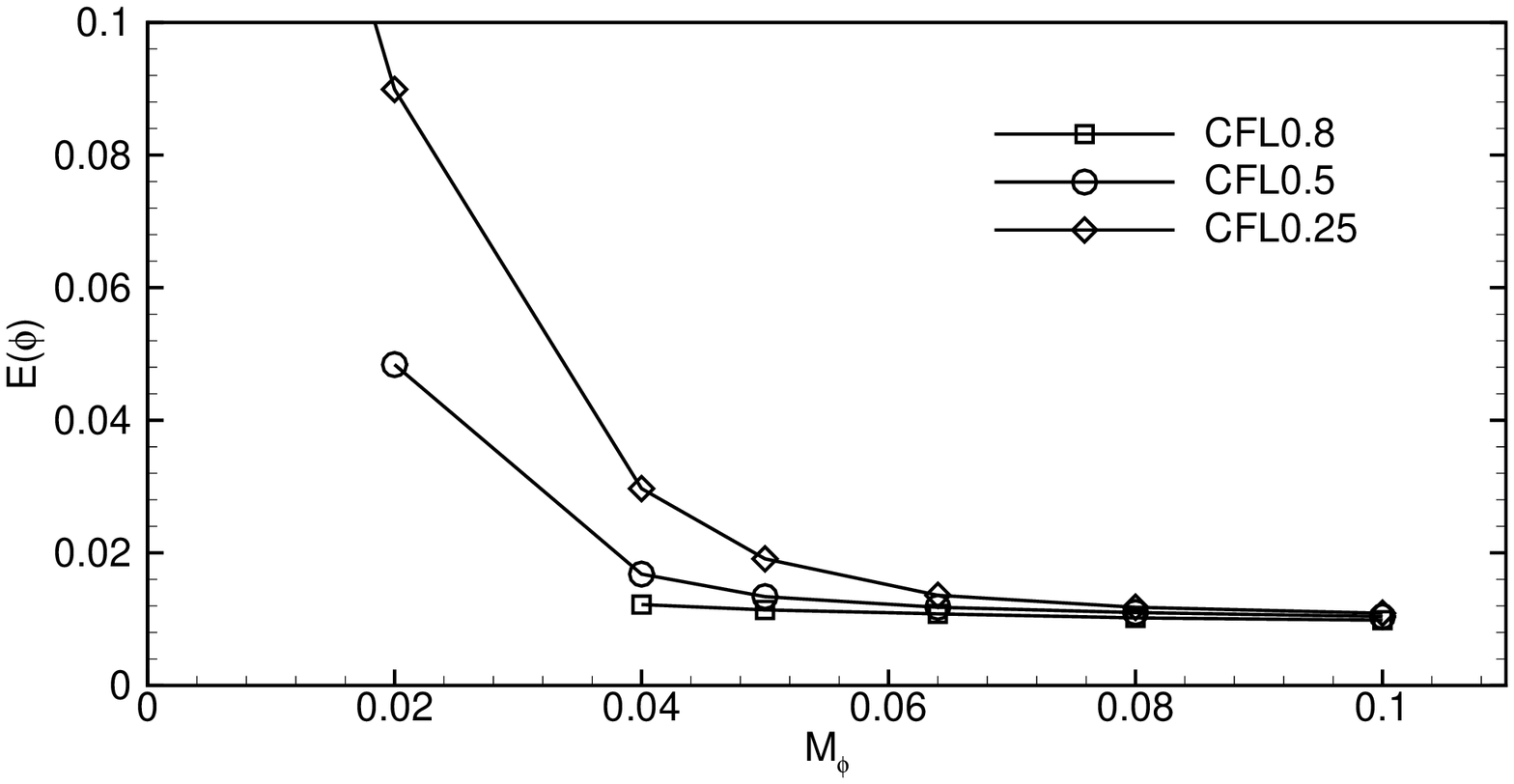}
      \caption{DUGKS-T3S3}
      \label{FIG:SD:DUGKST3S3-M-L2Phi}
    \end{subfigure}
    \end{minipage}
    \begin{minipage}[b]{0.9 \columnwidth}
    \begin{subfigure}{0.45 \columnwidth}
      \centering
      \includegraphics[width=1\linewidth]{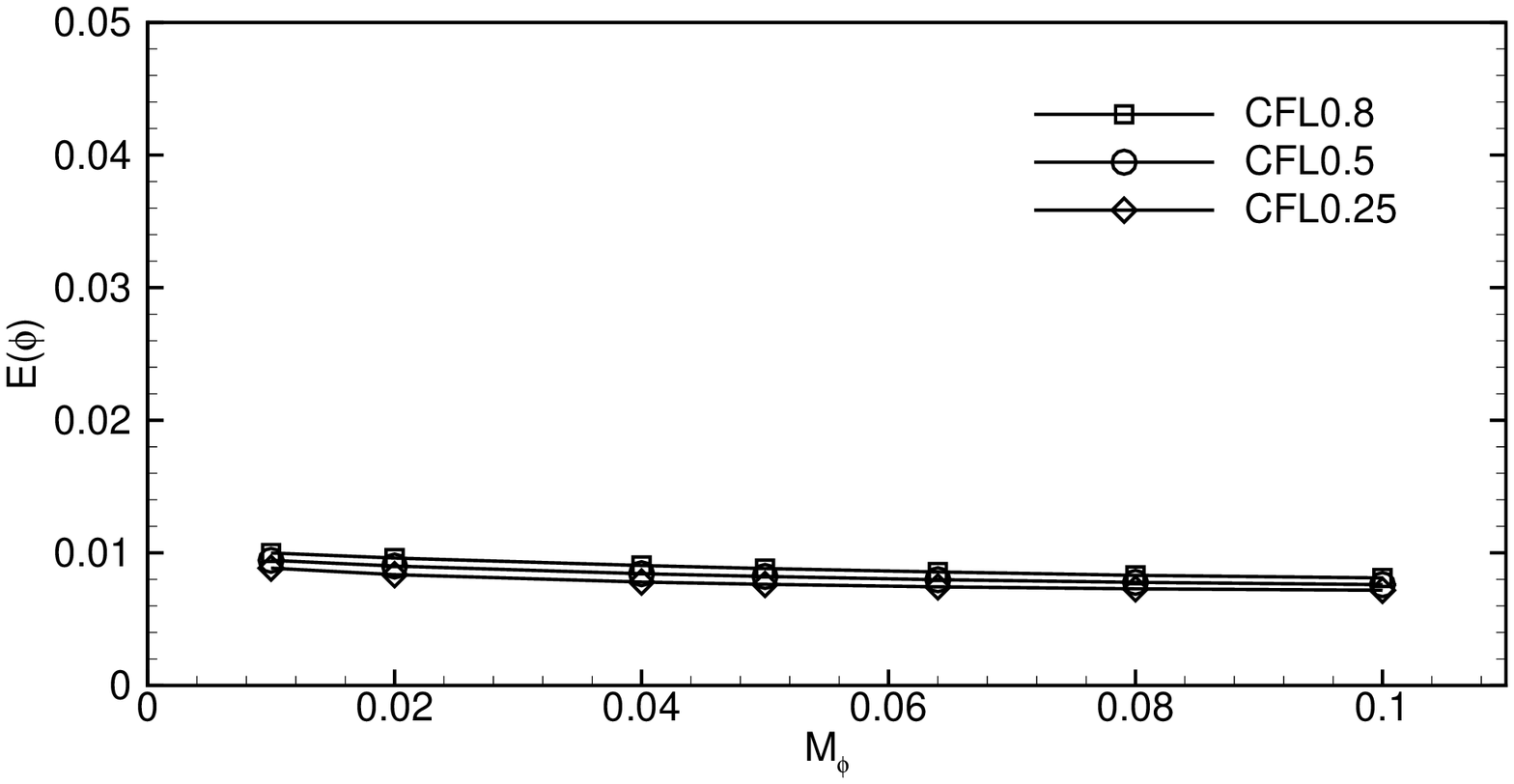}
      \caption{DUGKS-T2S5}
      \label{FIG:SD:DUGKST2S5-M-L2Phi}
    \end{subfigure}
    \end{minipage}
\caption{$L_2$-norm error of $\phi$ for smoothed deformation obtained by multiple methods with various $M_\phi$, $Cn = 4/512$, $Pe = 512$.}
\label{FIG:SD:M-L2Phi}
\end{figure}

As the sudden reverse in velocity field has an effect on the results obtained by various methods, the test of interface deformation in a smoothed shear flow is conducted in this subsection. A circular interface is placed in the center of a square domain at initial state. The velocity field is controlled by
\begin{equation}
\begin{aligned}
&u(x,y)=-U_0\textrm{sin}\big(\frac{4{\pi}x}{L_0}\big)\textrm{sin}\big(\frac{4{\pi}y}{L_0}\big)\textrm{cos}\big(\frac{{\pi}t}{T_f}\big),
\\
&v(x,y)=-U_0\textrm{cos}\big(\frac{4{\pi}x}{L_0}\big)\textrm{cos}\big(\frac{4{\pi}y}{L_0}\big)\textrm{cos}\big(\frac{{\pi}t}{T_f}\big).
\end{aligned}
\end{equation}
The interface will evolve into filamentary structures in the first half period and restore to its initial shape gradually in the second half period. Due to the complicated structures that the interface forms, the reference length $L_0$ is increased to 512 and the $Cn$ number is fixed at $4/512$. Again, the discrepancy between initial and final interface is regarded as the evaluation criteria of performance for various methods.
\\
Fig.~\ref{FIG:SD:Scheme-4-4T} illustrates the differences between the restored interface (solid line) and the initial interface (dash line) at $Pe = 2048, M_{\phi} = 0.02$ provided by various methods. It can be observed that the differences are almost indistinguishable to the naked eyes. The corresponding data are described in Table~\ref{tab:SD:Pe-L2Phi} of the sixth column. The $L_2$ norm error obtained by various methods is no larger than $1.3{\times}10^{-2}$, which is so tiny that the differences illustrated in Fig.~\ref{FIG:SD:Scheme-4-4T} are barely noticeable. In this example, the best result is provided by the method of DUGKS-T2S5 with a CFL number of 0.25. LBM and DVM-CFL1.0 offer the next-best results, which are nearly identical to that obtained by DUGKS-T2S5. The absolute difference between the result of DUGKS-T2S3 with a CFL number of 0.8 and the result of LBM is 0.01, which means that the performance of those two methods shown in this test is comparable. DUGKS-T2S2CD offers a result that is not that comparable to the previous results. The worst-performing method is DUGKS-T3S3 due to its intrinsic limitation of the ratio of time step to relaxation time.
\\
Fig.~\ref{FIG:SD:Pe-L2Phi} illustrates the results of $L_2$ norm error obtained by multiple methods at various P\'{e}clet number for the smoothed deformation test. The corresponding data are described in Table~\ref{tab:SD:Pe-L2Phi}. It can be observed that the performance of DUGKS-T2S5 with a CFL number of 0.25 is most excellent. The result of DUGKS-T2S5 with a CFL number of 0.5 is still better than that provided by the methods of LBM and DVM-CFL1.0. DUGKS-T2S2CD with a CFL number of 0.25 performs well when the P\'{e}clet number is small whereas DUGKS-T2S3 with a CFL number of 0.8 shows an advantage when the P\'{e}clet number is relatively large. This conclusion is far different from the previous one that the performance of DUGKS-T2S3 is always better than that of DUGKS-T2S2CD. This phenomenon can be mainly attributed to the enhanced mesh resolution in this test. As is mentioned before, the spatial dissipation of DUGKS-T2S2CD is so large that the accumulated error would rise if the number of iteration steps is increased. Since the mesh resolution is this test is enhanced, the spatial dissipation of all the methods would be reduced. In this way, DUGKS-T2S2CD with a low CFL number would be able to show its advantage when the P\'{e}clet number is small. However, due to the central scheme utilized in the evaluation of meso-flux, its ability would not be that good when the flow is dominated by advection. Opposite trend can be observed in the results produced by DUGKS-T2S3 since upwind scheme is employed in the evaluation of its meso-flux. As for the influence of CFL number, let us take a look at the comparative results illustrated in Fig.~\ref{FIG:SD:Pe-L2Phi}. DVM-CFL0.8 performs worse than DVM-CFL1.0 and LBM because of the spatial dissipation introduced in the reconstruction procedure. DUGKS-T2S2CD with a smaller CFL number has shown its advantage when the P\'{e}clet number is small. As the P\'{e}clet number increases, the evaluation of meso-flux would lose accuracy due to the central scheme utilized, which would in turn lead to an increase in the accumulated error. In such a condition, fewer iteration steps would be able to reduce the impacts of the low precision meso-flux on the accumulated error. That is why DUGKS-T2S2CD with a larger CFL number performs better when the P\'{e}clet number is large. As for the method of DUGKS-T2S3, it can be concluded that the larger the CFL number is, the better the performance would be. The reason is that an increase in CFL number would lead to a decrease in the spatial numerical dissipation, which thereby restrain the error accumulation. The effects of CFL number on the performance of DUGKS-T2S5 are almost indistinguishable, which is similar to the phenomenon shown in Fig.~\ref{FIG:IE:DUGKST2S5-Pe-L2Phi}. This observation confirms again that the temporal dissipation has limited effects on the accuracy of those methods.
\\
Fig.~\ref{FIG:SD:M-L2Phi} illustrates the results of $L_2$ norm error obtained by multiple methods at various mobility coefficients for the smoothed deformation test and the corresponding data are presented in Table~\ref{tab:SD:M-L2Phi}. It can be observed that the best results are produced by DUGKS-T2S2CD with a CFL number of 0.25. The finer mesh resolution in this test alleviates the spatial dissipation and thus the error accumulation would be suppressed. Since this test is conducted at a small P\'{e}clet number of 512, there is no wonder that the performance of DUGKS-T2S2CD is the most excellent. The results provided by DUGKS-T2S5 with a CFL number of 0.25 are comparable to that obtained by DUGKS-T2S2CD with the identical CFL number. LBM and DVM-CFL1.0 fail to outperform the two methods discussed above and yet they do not show evident superiority over DUGKS-T2S3, which indicates that the performance of DUGKS would be comparable to that of LBM if the mesh resolution is fine enough. As for the effects of CFL number, similar conclusions can be reached. The results of DVM-CFL0.8 are a little worse than the results obtained by DVM-CFL1.0 due to the reconstruction procedure introduced. DUGKS-T2S2CD has shown a better performance as the CFL number is smaller. Since the P\'{e}clet number in current case is small, the central scheme utilized in the evaluation of meso-flux would be able to take the full advantage. And because the mesh resolution is improved, the spatial dissipation has been suppressed. What is more, the smaller the CFL number is, the lower the spatial dissipation would be for DUGKS-T2S2CD method. With the combined effects of these three facts, there is no doubt that DUGKS-T2S2CD offer better results when the CFL number is smaller. As for the DUGKS-T2S3 method, it can be concluded that an increase in CFL number would lead to improvements in the results, which is mainly caused by the spatial dissipation mechanism behind the evaluation of meso-flux. The same trend can be observed in the results provided by DUGKS-T3S3. However, DUGKS-T3S3 always performs worse than DUGKS-T2S3 due to the extra flux reconstruction procedure in each evolution step. DUGKS-T2S3 and DUGKS-T3S3 with small CFL numbers have both shown bad performances when the mobility coefficient is small. Although DUGKS-T2S5 provides the best results with a CFL number of 0.25, the differences among those results obtained with various CFL numbers are barely distinguishable, which confirms the previous conclusion that the temporal dissipation contributes little to the accumulated error.
\\
\begin{table}[htbp]
\caption
{
  \label{tab:SD:Pe-L2Phi}
  $L_2$-norm error of $\phi$ for smoothed deformation obtained by multiple methods with various $Pe$, $Cn = 4/512$, $M_\phi = 0.02$.
}
	\begin{minipage}{1.0\textwidth}
	\begin{subtable}{1.0\textwidth}
	\caption{\label{tab:SD:LBMDVM-Pe-L2Phi}LBM, DVM}
		\begin{ruledtabular}
		\begin{tabular}{ccccccc}
		Pe&512&1024&1280&1638.4&2048&2560\\
		\colrule
		LBM & $9.41\times10^{-3}$ & $7.12\times10^{-3}$ & $7.98\times10^{-3}$ & $8.71\times10^{-3}$ & $1.17\times10^{-2}$ & $1.68\times10^{-2}$\\
		DVM-CFL1.0 & $9.40\times10^{-3}$ & $7.12\times10^{-3}$ & $7.98\times10^{-3}$ & $8.71\times10^{-3}$ & $1.17\times10^{-2}$ & $1.68\times10^{-2}$\\
		DVM-CFL0.8 & $1.19\times10^{-2}$ & $8.97\times10^{-3}$ & $9.38\times10^{-3}$ & $9.76\times10^{-3}$ & $1.24\times10^{-2}$ & $1.73\times10^{-2}$\\
		\end{tabular}
		\end{ruledtabular}
	\end{subtable}%
	\end{minipage}
	\par\medskip
	\begin{minipage}{1.0\textwidth}
	\begin{subtable}{1.0\textwidth}
	\caption{\label{tab:SD:DUGKS-T2S2CD-Pe-L2Phi}DUGKS-T2S2CD}
		\begin{ruledtabular}
		\begin{tabular}{ccccccc}
		Pe&512&1024&1280&1638.4&2048&2560\\
		\colrule		
		CFL0.25 & $8.64\times10^{-3}$ & $7.52\times10^{-3}$ & $8.41\times10^{-3}$ & $1.04\times10^{-2}$ & $1.38\times10^{-2}$ & $1.94\times10^{-2}$\\
		CFL0.5 & $9.76\times10^{-3}$ & $7.98\times10^{-3}$ & $8.57\times10^{-3}$ & $1.03\times10^{-2}$ & $1.33\times10^{-2}$ & $1.89\times10^{-2}$\\
		CFL0.8 & $1.02\times10^{-2}$ & $8.04\times10^{-3}$ & $8.43\times10^{-3}$ & $9.96\times10^{-3}$ & $1.29\times10^{-2}$ & $1.82\times10^{-2}$\\
		\end{tabular}
		\end{ruledtabular}
	\end{subtable}%
	\end{minipage}
	\par\medskip
	\begin{minipage}{1.0\textwidth}
	\begin{subtable}{1.0\textwidth}
	\caption{\label{tab:SD:DUGKS-T2S3-Pe-L2Phi}DUGKS-T2S3}
		\begin{ruledtabular}
		\begin{tabular}{ccccccc}
		Pe&512&1024&1280&1638.4&2048&2560\\
		\colrule
		CFL0.25 & $5.51\times10^{-2}$ & $3.37\times10^{-2}$ & $2.70\times10^{-2}$ & $2.13\times10^{-2}$ & $1.86\times10^{-2}$ & $1.94\times10^{-2}$\\
		CFL0.5 & $1.58\times10^{-2}$ & $1.08\times10^{-2}$ & $1.02\times10^{-2}$ & $1.06\times10^{-2}$ & $1.24\times10^{-2}$ & $1.67\times10^{-2}$\\
		CFL0.8 & $1.09\times10^{-2}$ & $8.20\times10^{-3}$ & $8.24\times10^{-3}$ & $9.32\times10^{-3}$ & $1.18\times10^{-2}$ & $1.67\times10^{-2}$\\
		\end{tabular}
		\end{ruledtabular}
	\end{subtable}%
	\end{minipage}
	\par\medskip
	\begin{minipage}{1.0\textwidth}
	\begin{subtable}{1.0\textwidth}
	\caption{\label{tab:SD:DUGKS-T3S3-Pe-L2Phi}DUGKS-T3S3}
		\begin{ruledtabular}
		\begin{tabular}{ccccccc}
		Pe&512&1024&1280&1638.4&2048&2560\\
		\colrule
		CFL0.25 & $8.99\times10^{-2}$ & $5.66\times10^{-2}$ & $4.51\times10^{-2}$ & $3.44\times10^{-2}$ & $2.75\times10^{-2}$ & $2.46\times10^{-2}$\\
		CFL0.5 & $4.84\times10^{-2}$ & $2.90\times10^{-2}$ & $2.33\times10^{-2}$ & $1.87\times10^{-2}$ & $1.69\times10^{-2}$ & $1.86\times10^{-2}$
		\end{tabular}
		\end{ruledtabular}
	\end{subtable}%
	\end{minipage}
	\par\medskip
	\begin{minipage}{1.0\textwidth}
	\begin{subtable}{1.0\textwidth}
	\caption{\label{tab:SD:DUGKS-T2S5-Pe-L2Phi}DUGKS-T2S5}
		\begin{ruledtabular}
		\begin{tabular}{ccccccc}
		Pe&512&1024&1280&1638.4&2048&2560\\
		\colrule
		CFL0.25 & $8.36\times10^{-3}$ & $6.59\times10^{-3}$ & $6.99\times10^{-2}$ & $8.49\times10^{-2}$ & $1.14\times10^{-2}$ & $1.65\times10^{-2}$\\
		CFL0.5 & $9.01\times10^{-3}$ & $6.96\times10^{-3}$ & $7.25\times10^{-3}$ & $8.66\times10^{-3}$ & $1.15\times10^{-2}$ & $1.67\times10^{-2}$\\
		CFL0.8 & $9.61\times10^{-3}$ & $7.32\times10^{-3}$ & $7.53\times10^{-3}$ & $8.84\times10^{-3}$ & $1.16\times10^{-2}$ & $1.67\times10^{-2}$\\
		\end{tabular}
		\end{ruledtabular}
	\end{subtable}%
	\end{minipage}
\end{table}
\begin{table}[htbp]
\caption
{
  \label{tab:SD:M-L2Phi}
  $L_2$-norm error of $\phi$ for smoothed deformation obtained by multiple methods with various $M_\phi$, 
}
	\begin{minipage}{1.0\textwidth}
	\begin{subtable}{1.0\textwidth}
	\caption{\label{tab:SD:LBMDVM-M-L2Phi}LBM, DVM}
		\begin{ruledtabular}
		\begin{tabular}{cccccccc}
		M$_\phi$&0.01&0.02&0.04&0.05&0.064&0.08&0.1\\
		\colrule
		LBM & $9.42\times10^{-3}$ & $9.41\times10^{-3}$ & $9.36\times10^{-3}$ & $9.32\times10^{-3}$ & $9.26\times10^{-3}$ & $9.19\times10^{-3}$ & $9.12\times10^{-3}$\\
		DVM-CFL1.0 & $9.42\times10^{-3}$ & $9.41\times10^{-3}$ & $9.36\times10^{-3}$ & $9.32\times10^{-3}$ & $9.26\times10^{-3}$ & $9.19\times10^{-3}$ & $9.12\times10^{-3}$\\
		DVM-CFL0.8 & $1.52\times10^{-2}$ & $1.20\times10^{-2}$ & $1.09\times10^{-2}$ & $1.06\times10^{-2}$ & $1.04\times10^{-2}$ & $1.01\times10^{-2}$ & $9.93\times10^{-3}$\\
		\end{tabular}
		\end{ruledtabular}
	\end{subtable}%
	\end{minipage}
	\par\medskip
	\begin{minipage}{1.0\textwidth}
	\begin{subtable}{1.0\textwidth}
	\caption{\label{tab:SD:DUGKS-T2S2CD-M-L2Phi}DUGKS-T2S2CD}
		\begin{ruledtabular}
		\begin{tabular}{cccccccc}
		M$_\phi$&0.01&0.02&0.04&0.05&0.064&0.08&0.1\\
		\colrule
		CFL0.25 & $9.81\times10^{-3}$ & $8.64\times10^{-3}$ & $7.38\times10^{-3}$ & $7.02\times10^{-3}$ & $6.68\times10^{-3}$ & $6.47\times10^{-3}$ & $6.42\times10^{-3}$
		\\
		CFL0.5 & $1.05\times10^{-2}$ & $9.76\times10^{-3}$ & $8.65\times10^{-3}$ & $8.28\times10^{-3}$ & $7.88\times10^{-3}$ & $7.57\times10^{-3}$ & $7.36\times10^{-3}$
		\\
		CFL0.8 & $1.07\times10^{-2}$ & $1.02\times10^{-2}$ & $9.35\times10^{-3}$ & $9.04\times10^{-3}$ & $8.69\times10^{-3}$ & $8.40\times10^{-3}$ & $8.16\times10^{-3}$
		\\
		\end{tabular}
		\end{ruledtabular}
	\end{subtable}%
	\end{minipage}
	\par\medskip
	\begin{minipage}{1.0\textwidth}
	\begin{subtable}{1.0\textwidth}
	\caption{\label{tab:SD:DUGKS-T2S3-M-L2Phi}DUGKS-T2S3}
		\begin{ruledtabular}
		\begin{tabular}{cccccccc}
		M$_\phi$&0.01&0.02&0.04&0.05&0.064&0.08&0.1\\
		\colrule
		CFL0.25 & $9.66\times10^{-2}$ & $5.51\times10^{-2}$ & $1.94\times10^{-2}$ & $1.46\times10^{-2}$ & $1.22\times10^{-2}$ & $1.11\times10^{-2}$ & $1.03\times10^{-2}$\\
		CFL0.5 & $2.56\times10^{-2}$ & $1.58\times10^{-2}$ & $1.18\times10^{-2}$ & $1.12\times10^{-2}$ & $1.06\times10^{-2}$ & $1.01\times10^{-2}$ &$9.55\times10^{-2}$\\
		CFL0.8 & $1.14\times10^{-2}$ & $1.09\times10^{-2}$ & $1.01\times10^{-2}$ & $9.80\times10^{-3}$ & $9.43\times10^{-3}$ & $9.10\times10^{-3}$ & $8.79\times10^{-3}$\\
		\end{tabular}
		\end{ruledtabular}
	\end{subtable}%
	\end{minipage}
	\par\medskip
	\begin{minipage}{1.0\textwidth}
	\begin{subtable}{1.0\textwidth}
	\caption{\label{tab:SD:DUGKS-T3S3-M-L2Phi}DUGKS-T3S3}
		\begin{ruledtabular}
		\begin{tabular}{cccccccc}
		M$_\phi$&0.01&0.02&0.04&0.05&0.064&0.08&0.1\\
		\colrule
		CFL0.25 & $1.53\times10^{-1}$ & $8.99\times10^{-2}$ & $2.97\times10^{-2}$ & $1.91\times10^{-2}$ & $1.36\times10^{-2}$ & $1.18\times10^{-2}$ & $1.09\times10^{-2}$\\
		CFL0.5 & - & $4.84\times10^{-2}$ & $1.68\times10^{-2}$ & $1.34\times10^{-2}$ & $1.18\times10^{-2}$ & $1.10\times10^{-2}$ &$1.04\times10^{-2}$\\
		CFL0.8 & - & - & $1.22\times10^{-2}$ & $1.14\times10^{-2}$ & $1.08\times10^{-2}$ & $1.02\times10^{-2}$ & $9.85\times10^{-3}$\\
		\end{tabular}
		\end{ruledtabular}
	\end{subtable}%
	\end{minipage}
	\par\medskip
	\begin{minipage}{1.0\textwidth}
	\begin{subtable}{1.0\textwidth}
	\caption{\label{tab:SD:DUGKS-T2S5-M-L2Phi}DUGKS-T2S5}
		\begin{ruledtabular}
		\begin{tabular}{cccccccc}
		M$_\phi$&0.01&0.02&0.04&0.05&0.064&0.08&0.1\\
		\colrule
		CFL0.25 & $8.85\times10^{-3}$ & $8.36\times10^{-3}$ & $7.80\times10^{-3}$ & $7.63\times10^{-3}$ & $7.44\times10^{-3}$ & $7.29\times10^{-3}$ & $7.18\times10^{-3}$\\
		CFL0.5 & $9.44\times10^{-3}$ & $9.01\times10^{-3}$ & $8.43\times10^{-3}$ & $8.22\times10^{-3}$ & $7.98\times10^{-3}$ & $7.78\times10^{-3}$ & $7.61\times10^{-3}$\\
		CFL0.8 & $9.99\times10^{-3}$ & $9.61\times10^{-3}$ & $9.05\times10^{-3}$ & $8.82\times10^{-3}$ & $8.56\times10^{-3}$ & $8.31\times10^{-3}$ & $8.11\times10^{-3}$\\
		\end{tabular}
		\end{ruledtabular}
	\end{subtable}%
	\end{minipage}
\end{table}
\section{\label{sec:sec4}DISCUSSIONS AND CONCLUSIONS}
In this paper, several high-order approaches have been utilized in the solution reconstruction of discrete unified gas-kinetic scheme and the performance on capturing dynamic interfaces for each approach has been examined. The lattice Boltzmann method and discrete velocity method have been introduced as references to facilitate comparative analysis. Detailed results have been provided by the three kinetic methods and performance distinctions among them are explained from an informed perspective.
\\
Results produced by LBM are in full agreement with the previous results presented in the literature\cite{Yang2019PRE} and second order accuracy can be observed in the interface diagonal translation test. Across all the results obtained from various interface-tracking tests, it has been validated that variations in P\'{e}clet number and mobility coefficient have limited effects on the performance of LBM. Although the comparative results provided by those three kinetic methods has demonstrated the  superiority of LBM in the condition of uniform grids, its advantages will disappear when non-uniform grids, say block-structured multi-grid, are applied due to the destruction of perfect streaming process.
\\
The streaming and collision DVM with a unit time step has predicted identical results as the numerical solution of LBM. The reason comes from the reconstruction procedure performed in the evolution process. As the time step of DVM is tuned to unit, the particles will migrate precisely from the center of its neighbor cell, which implies that the derivative information in Eq.~(\ref{Eq:DVMDerivatives}) is not utilized and thus no interpolation error is introduced. In such a condition, DVM will turn into a pure streaming and collision method which shares the majority properties with LBM. Therefore, the results produced by DVM with a unit time step are identical to the results provided by LBM. When the time step of DVM is tuned to 0.8, the position from which particles migrate do not coincide with the cell center. Hence, the derivative information in Eq.~(\ref{Eq:DVMDerivatives}) is used and the interpolation error generated in the calculation of derivatives will reduce the accuracy of simulation results. Although second order accuracy has been verified in the interface diagonal test, it can be observed clearly that the dissipation of DVM implemented with a time step of 0.8 is a bit larger than LBM or DVM with a time step of unit. The performance distinction can be observed clearly in the condition of small mobility coefficient. Since the spatial dissipation of DVM is inversely proportional to the time step, it is recommended to maximize the time step when applying this method to numerical simulations.
\\
The performance of discrete unified gas-kinetics scheme implemented by four different types of reconstruction methods has been studied with the same series of tests. DUGKS-T2S2CD, in which the evaluation of meso-flux is implemented by central scheme, has proven its second order accuracy in the interface diagonal translation test, which is in accordance with the fact that the truncation error appeared in the reconstruction of distribution function and source terms on mesh interface has a magnitude of $O(\Delta{x}^2)$. Due to the central scheme utilized in the evaluation of meso-flux, DUGKS-T2S2CD is capable to provide results with enough accuracy when the P\'{e}clet number is relatively small. That is to say, this method performs well when the flow is dominated by diffusion. In the smoothed deformation test, DUGKS-T2S2CD can even offer results better than any other ones with a small P\'{e}clet number of 512. However, when it comes to the flow identified by advection, the results produced by DUGKS-T2S2CD would lose accuracy. Those results obtained by DUGKS-T2S2CD at large P\'{e}clet number are generally the worst ones among all of the results produced by different methods. Also it is worth noting that the requirements imposed by DUGKS-T2S2CD on the mesh resolution is more stringent than other methods being tested. The suitable condition for the employment of DUGKS-T2S2CD is mainly determined by the advection transport rate. DUGKS-T2S3, in which the evaluation of meso-flux is implemented by upwind scheme and the spatial derivatives are calculated by the third order isotropic finite-difference scheme, has shown a second order accuracy in the test of interface diagonal translation. Although the spatial derivatives are updated by the third order finite difference scheme, the local accuracy of the source terms on mesh interface is second order. Hence, the overall accuracy of this method is limited to second order. Since the truncation error introduced in the estimation of distribution function on mesh interface measures a magnitude of $O(\vert{\Delta{x}-\bm{\xi}_i\Delta{t}}\vert^3)$, which maintains a direct relationship with the time step $\Delta{t}$, the results provided by DUGKS-T2S3 are generally sensitive to the variation of time step. An increase in time step would lead to a reduction in the truncation error, which in turn results in a decrease in the spatial dissipation. Hence, the results produced by DUGKS-T2S3 with a large time step are usually better than those obtained with small time steps. As the upwind scheme is utilized in the evaluation of meso-flux, the performance of DUGKS-T2S3 in the condition of high P\'{e}clet number is outstanding. What is particularly exciting is DUGKS-T2S3 performs satisfactorily even when it comes to the circumstance of low P\'{e}clet number. The results provided by DUGKS-T2S3 with the maximum time step are generally more accurate than the results obtained by DUGKS-T2S2CD. The comparative results have also verified that the performance differences between DUGKS-T2S3 and DVM at the time step of 0.8 are nearly indistinguishable. One thing to note in terms of the results from DUGKS-T2S3 is that it performs poorly when the mobility coefficient and time step are both set to small values. It is recommended to maximize the time step as far as possible in the simulations undertaken with DUGKS-T2S3. An improved version of DUGKS-T2S3 is DUGKS-T3S3, which is said to guarantee third order accuracy in both space and time. However, the real accuracy of this method revealed in the interface diagonal test is only second order. This is mainly caused by the precision lose in the evaluation of source terms on mesh interface. As is mentioned above, the overall accuracy is limited to second order due to the second order scheme utilized in the estimation of source terms. It has been verified that DUGKS-T3S3 and DUGKS-T2S3 share many common features. Nevertheless, the time step of DUGKS-T3S3 is severely constrained by the particle collision time due to the limitation introduced in the evaluation of intermediate collision term. This is in direct contradiction to the dissipation mechanism behind the construction of meso-flux. The comparative results obtained in the same conditions have also proven that the performance of DUGKS-T3S3 is inferior to that of DUGKS-T2S3. Therefore, there is no necessity to apply DUGKS-T3S3 in the simulations governed by the conservative Allen-Cahn equation. The other improved version of DUGKS is named DUGKS-T2S5 due to the fifth order finite difference scheme utilized in the calculation of spatial derivatives. Even if the truncation error of the spatial derivatives has a magnitude of $O(\Delta{x}^5)$, the accuracy of this method manifested in the interface diagonal translation test maintains second order. Similarly, this is attributed to the precision lose in the evaluation of source terms on mesh interface. Nevertheless, results provided by DUGKS-T2S5 are always superior to that obtained by DUGKS-T2S3. The fact that DUGKS-T2S5 does have an absolute advantage over DUGKS-T2S3 should not be ignored. However, when the algorithm complexity and time consumption are taken into consideration, the superiority of DUGKS-T2S5 becomes less evident. The various schemes utilized in DUGKS have proved that without the implementation of high order scheme in the estimation of source terms, it is impossible to construct global high order methods by merely applying high order difference schemes to the evaluation of spatial and temporal derivatives. Taking everything into consideration, it can be concluded that DUGKS-T2S3 with a large time step is the most cost-effective method amongst various kinds of DUGKS implemented by different schemes for solving the conservative Allen-Cahn equation.
\\
The current research conducts a comprehensive assessment on discrete unified gas-kinetic scheme implemented with several high-order reconstruction methods and provides a basis for further applications of DUGKS in studying multi-phase flows. Since DUGKS is not confined to uniform mesh, performance of DUGKS implemented with adaptive mesh refinement technique is worth expecting.

\begin{acknowledgments}
This study is supported by the National Natural Science Foundation of China (Grant No. 11472219, 11902264), the 111 Project of China (B17037), and the Natural Science Basic Research Program of Shaanxi (Program No. 2019JQ-315).
\end{acknowledgments}
\section*{DATA AVAILABILITY}
The data that support the findings of this study are available from the corresponding author upon reasonable request.
\clearpage
\bibliographystyle{elsarticle-num}
\bibliography{LBMDVMDUGKS}
\end{document}